\documentclass[14pt]{article}
\usepackage{amssymb,amsmath}
\usepackage{helvet,amsmath,amssymb,amsfonts} 
\usepackage[english]{babel}

\def\un{{\rm 1\mkern-4mu I}}

\def\cQ{ {\cal Q}  }
\def\cC{ {\cal C}  }
\def\cS{ {\cal S}  }

\def\px{\varpi^{23}}
\def\py{\varpi^{13}}
\def\pz{\varpi^{12}}
\def\Sx{S^{[23]}}
\def\Sy{S^{[13]}}
\def\Sz{S^{[12]}}
\def\pX{\varpi^1}
\def\pY{\varpi^{2}}
\def\pZ{\varpi^3}
\def\nz{\nu_{12}}
\def\ny{\nu_{13}}
\def\nx{\nu_{23}}
 
 \title{Quantum Einstein-Dirac Bianchi Universes}
 
\author{  Thibault Damour$\,^a$, Philippe Spindel$\,^b$
\\(a) Institut des Hautes 
\'Etudes Scientifiques\\ 
Bures sur Yvette, F-91440, France\\
(b) M\'ecanique et gravitation, 
Universit\'e de Mons,\\
7000 Mons, Belgique}
\date{Version from \today} 
\def\mettresous#1\sous#2{\mathrel{\mathop{\kern0pt #2}\limits_{#1}}}                                          \newcommand{\PG}{\mettresous{G}\sous{\Pi}}
\newcommand{\Psib}{\overline \Psi}
\begin{document}
\maketitle
\section*{ }{\bf Abstract:} 
We study the mini--superspace quantization of spatially homogeneous (Bianchi) 
cosmological universes sourced by a Dirac spinor field. The quantization of the homogeneous
spinor leads to a finite-dimensional fermionic Hilbert space and thereby to a multi-component
Wheeler-DeWitt equation whose main features are: (i) the presence of spin-dependent 
Morse-type potentials, and (ii) the appearance of a q-number squared-mass term, which
is of order ${\cal O}(\hbar^2)$, and which is affected by ordering ambiguities. We give the
exact quantum solution of the Bianchi type-II system (which contains both scattering states
and bound states), and discuss the main qualitative
features of the quantum dynamics of the (classically chaotic) 
Bianchi type-IX system. We compare the exact
quantum dynamics of fermionic cosmological billiards to previous works that described
the spinor field as being either  classical or  Grassmann-valued.

\section{Introduction}
The main aim of this work is to clarify the physical structure of the coupled quantum Einstein-Dirac
system within a minisuperspace cosmological setting, and, in particular, its dynamics
in the vicinity of a cosmological 
singularity. Let us recall that the seminal work of Belinsky, Khalatnikov and Lifshitz (BKL) \cite{BKL} (see also Misner \cite{Mi}) has brought 
into light the chaotically `` oscillatory" behaviour, near the singularity,  of the diagonal components of the metric both
in  homogeneous Bianchi IX cosmological models, and in the ``general (classical) solution'' of Einstein equations. 
Recently, this chaotic BKL behaviour acquired a new significance through the discovery of its unexpected link with infinite-dimensional
Kac-Moody algebras \cite{DaHe2,Damour:2001sa,DHN}  (for reviews, see \cite{DHN2,Henneaux:2007ej}). 
The presence of such a chaotic behaviour depends both on the spacetime dimension, $D$,
and on the matter content of the considered cosmological model. For instance, pure gravity (i.e. the vacuum Einstein equations) has a
chaotic, BKL-like behaviour in spacetime dimension $ D \leq 10$, but a monotonic, Kasner-like behaviour in spacetime dimensions $ D \geq 11$ \cite{Demaret:1986ys}.

The effect, near a singularity, of a general {\it bosonic} matter content (scalar field and $p-$forms), in any spacetime dimension $D$, 
has been studied in Ref. \cite{DHN2}. The main result is that
almost all the  bosonic  degrees of freedom ``freeze" near the singularity (i.e. admit limits, at each spatial point, as $ t \to 0$), except for the 
diagonal part of the spatial metric, parametrized as $g_{a a} = \exp (- 2 \beta_a)$
(together with any scalar field $\varphi = \beta_0$, if present). The dynamics of the $\beta_a$'s can be represented as that of a 
"ball" (of position $ \beta_a$) moving on a Lorentzian (or hyperbolic) billiard. The latter billiard motion
can then either be chaotic, or non-chaotic (i.e. ultimately monotonic), depending on the geometry of the `` billiard table",
which is a polyhedron domain in  hyperbolic space. 
The quantum mini--superspace versions of some of these bosonic  cosmological
billiards have been studied in several works \cite{Misner:1969ae,Ryan72,RyanShepley75,Montani:2007vu}

While the effect of {\it bosonic} matter on the dynamics near a singularity is well understood, the effect of {\it fermionic} matter
is less well understood. This difference in understanding has both  technical and physical roots. From the technical point of view,
the Hamiltonian description of spinor fields, coupled to gravity, is rather subtle and complex. The Hamiltonian description
of a Dirac (spin $\frac{1}{2}$) field, coupled to gravity (notably in Bianchi spacetimes), has been clarified in several papers,
see Refs. \cite{NelsonTeitelboim,IsNe,HenPRD,HenIHP,ObRy,Jantzen:1982je,Christodoulakis:1983qb,D'Eath:1986mc}. Here, we shall not consider the case of the gravity coupling to
a Rarita-Schwinger (spin $\frac{3}{2}$) field, i.e. the case of ``supersymmetric (quantum) cosmology", though we view the Einstein-Dirac system as a toy model for the
Einstein-Rarita-Schwinger case.  See, e.g., \cite{D'Eath:1996at} for a entry in the literature on
supersymmetric quantum cosmology.
On the other hand, the physical meaning of having spinorial sources in cosmology has remained somewhat obscure. Various (conflicting)
approaches to this issue have been assumed in the literature. We shall distinguish three different approaches to the treatment of
spinorial fields, say $\psi$: (i) bilinears\footnote{Some authors even replace the $\psi$ variables themselves by real numbers.} 
in $\psi$ (in the source terms, $T^{\mu \nu} $, for gravity) are replaced by real numbers (or ``c-numbers"); 
(ii) the $\psi$ variables are treated as Grassmann-valued (or ``G-numbers");
or (iii) the $\psi$'s are treated as quantum operators (or ``q-numbers"). 

When treating the $\psi$'s as G-numbers, they do not affect the BKL chaos (if the latter is present in absence of fermions).
Indeed,  $T^{\mu \nu} \propto \psi^2$ only modifies the ``soul" $g_2 + g_4 + \cdots$ of the metric $g$, i.e. the part of the Grassmann expansion
$ g = g_0 + g_2 + g_4 \cdots$ of $g$ which contains an even (non zero) number of Grassmann generators. By contrast, the ``body" $g_0$ of $g$ (i.e. the
part of $g$ which does not contain any Grassmann generator) is unaffected by the $\psi$'s, and is entirely determined by the bosonic
sector of the theory. This approach has been recently used \cite{DaHi}  to discuss how the chaotic behaviour of the body $g_0$ of the metric induces a
corresponding chaotic dynamics in the (Grassmann-level-one) fermions $ \psi= \psi_1 + \psi_3 + \cdots$.

By contrast, if one replaces the $ \sim \psi^2$ source terms by c-numbers (that react back on the body $g_0$ of the metric),
they can drastically modify the asymptotic behaviour near the singularity. Indeed, it has been argued by Belinsky and Khalatnikov \cite{BK}
that the presence of a Dirac field would, if so treated, ultimately destroy the BKL chaos. This result has been confirmed,
and streamlined, by the Hamiltonian treatment of \cite{dBHP} which showed that the $ \sim \psi^2$ source terms modify the billiard
dynamics of the (logarithmic) cosmological scale factors $\beta_a$ by adding a (positive)  ``squared-mass term" $\mu^2$ to the effective
(Lorentzian-signature) Hamiltonian describing the dynamics of $\beta_a$. Indeed, such a squared-mass term (in Lorentzian $\beta$ space)
slows down the motion of the $\beta$ particle and ultimately prevent collisions on the cushions of the billiard table.

In the present work we shall treat the fermions as q-numbers and face the problem of discussing the meaning of the back reaction of fermions
within a quantum framework. This is a notoriously difficult problem, but we shall be able to get answers by restricting ourselves
to the mini-superspace framework of homogeneous cosmological spacetimes. 
In that framework, the quantization of the spinor $\psi$ is equivalent to
considering that the wavefunction $\Phi$ of the universe is a multi-component object,
say $\Phi_\sigma$,  where the discrete index $\sigma$ labels the finite number of independent
states allowed by the anticommutation relations of $\psi$. In particular, as we shall see in 
detail below, when $\psi$ is
a Majorana spinor the discrete index $\sigma$ takes 4 values and can be identified with the
spinor index of an auxiliary $SO(4)$ Clifford algebra defined by the quantum operators
$\Gamma_\alpha= 2 g^{\frac14} \psi_\alpha$, $\alpha=1,2,3,4$. The appearance 
in the wavefunction of such
a discrete spinor-type label when considering zero-mode (spatially independent)
Fermionic operators is well known from the Ramond string case \cite{Ramond:1971gb}
, and has also been
used in some studies of quantum supersymmetric minisuperspace cosmologies,
see, e.g., \cite{Obregon:1998hb}.

Most of our discussion will allow for a general (class A) Bianchi model,
but we shall discuss the quantum dynamics in details only for two cases:
the  Bianchi IX (or `mixmaster') model, and the (non-chaotic)   Bianchi II model.
We shall also compare the quantum solution to the various ways of discussing its classical analogs.

Our paper is organized as follows. Section 2 discusses the classical Lagrangian formulation
(with a specific fixing of the vielbein in terms of the metric) of an homogeneous spinor
field coupled to a Bianchi metric. This is followed in Sec. 3 by the corresponding classical
Hamiltonian formulation. The quantization of this system, and the discussion of the quantum
dynamics of the Bianchi-IX and Bianchi-II Einstein-Dirac systems is presented in Sec. 4.  The comparison of the q-number and G-number approaches
is discussed in Sec. 5, while our main conclusions are presented in Sec. 6.  Finally,
three appendices present complementary material, namely: Appendix A : alternative
Hamiltonian approach; Appendix B: alternative fixing of the dreibein; and Appendix C: 
classical dynamics of the Bianchi-II case.

\section{Classical Lagrangian formulation of a homogeneous spinor field coupled to a Bianchi cosmological metric}\label{sec2}
\setcounter{equation}{0}

\subsection{Choice of approach}\label{ssec2.1}

The coupling of a spinor field to gravity poses special problems in view of the need to use a vielbein, say $h_{\ \mu}^{\hat\alpha}$, in addition to the metric $g_{\mu\nu}$, to describe the spinor degrees of freedom. The metric and vielbein components are related by
\begin{equation}
\label{eq2.1}
g_{\mu\nu} = \eta_{\hat\alpha\hat\beta} \, h_{\ \mu}^{\hat\alpha} \, h_{\ \nu}^{\hat\beta} \, .
\end{equation}
Here, and in the following, we shall use hatted indices to denote ``flat indices'' referring to a local orthonormal frame where $\eta_{\hat\alpha\hat\beta} = {\rm diag} (-1,+1,+1,+1)$. [We restrict ourselves to a four-dimensional spacetime, and use greek indices to denote spacetime indices, and latin indices to denote spatial ones.]

\smallskip

There are basically two different approaches to the description of the coupled gravity-spinor (or Einstein-Dirac) dynamics. Either: (i) the gravity degrees of freedom are only described by means of the metric components $g_{\mu\nu}$, the vielbein ones $h_{\ \mu}^{\hat\alpha}$ being (locally) determined in terms of $g_{\mu\nu}$ by some suitable gauge-fixing of the local Lorentz symmetry $SO(3,1)$, or (ii) one does not break the local Lorentz symmetry, and describes the gravity degrees of freedom by means of the redundant vielbein variables $h_{\ \mu}^{\hat\alpha}$. The approach (i) is technically simpler, but depends on the choice of a specific gauge-fixing of the local $SO(3,1)$ symmetry. The approach (ii) does not depend (until one discusses explicit solutions) on the choice of a gauge-fixing of $SO(3,1)$, but involves more constraints than the first approach, namely the constraints linked to the local $SO(3,1)$ gauge symmetry. We shall use the approach (i) in the text, and discuss the approach (ii) in an Appendix. Note that the approach
(i) has been advocated in Ref. \cite{ObRy}, but that its implementation in that reference 
differs from the one we shall use (and does not take advantage of the useful automorphic
potentialities of the matrix $S$ discussed below).

\smallskip

Having decided on the approach (i), we need to choose a specific way of fixing the local $SO(3,1)$ gauge symmetry, {\it i.e.} of determining a specific vielbein $h_{\ \mu}^{\hat\alpha}$, when given the metric components $g_{\mu\nu}$ (in some coordinate system, or, more generally, some
non-holonomic frame). Here, the (assumed) symmetry properties of Bianchi models come to our help. Let us recall that the geometry of homogeneous spacetimes admits the special form
\begin{eqnarray}
\label{eq2.2}
ds^2 &= &g_{\mu\nu} (x) \, dx^{\mu} \, dx^{\nu} \nonumber \\
&= &-N^2 (t) \, dt^2 + g_{ab} (t) (N^a(t) \, dt + \tau^a (x)) (N^b (t) \, dt + \tau^b (x))
\end{eqnarray}
where the one-forms $\tau^a (x) = \tau^a_i (x) \, dx^i$ only depend on the spatial coordinates used in the $t = {\rm const.}$ slices. In the case (considered here) of Bianchi geometries, {\it i.e.} such that the spatial slices admit a simply transitive Lie group $G$ preserving the geometry, the one-forms $\tau^a$ can be chosen to be invariant under the group $G$, {\it i.e.} $\pounds_{\xi_b} \tau^a = 0$, where $\pounds$ denotes a Lie derivative and $\xi_b = \xi_b^i (x) \, \partial / \partial x^i$, with $b = 1,2,3$, a basis of three infinitesimal generators of $G$. With a suitable choice of $\tau^a$, the structure constants $C^c \, _{ab}$ of (the Lie algebra of) $G$, which enter the Lie brackets of the Killing vectors, $[\xi_a , \xi_b] = + \, C^c \, _{ab} \ \xi_c$, also enter the Cartan differential of the forms $\tau^a$, namely
\begin{equation}
\label{eq2.3}
d\tau^a = + \, \frac{1}{2} \, C^a \, _{bc} \ \tau^b \wedge \tau^c \, .
\end{equation}
Note that the constants $C^a \, _{bc}$ enter with the {\it same} sign in $[\xi_b , \xi_c]$ and in $d \tau^a$, because there are two flips of sign when going from the $\xi_a$'s to the $\tau^a$'s: one flip between the bracket of $[\xi , \xi]$ and the $d$ of the co-frame dual to the $\xi$'s, and a second flip between the structure constants of the left action of $G$ (defined, say, by the $\xi$'s) and those of its right action (which commutes with the $\xi$'s, and corresponds to the invariant (co)-frame linked to the $\tau$'s).

\smallskip

Apart from the $\tau^a$'s (which depend on the spatial coordinates only), the other objects entering the homogeneous metric (\ref{eq2.2}) only depend on the (coordinate) time, $t$. In addition, by using some time-dependent change of coordinates, one can ensure that the components of the shift vector $N^a(t)$ vanish for all times. We shall generally assume that this is the case, but we have left them in Eq.~(\ref{eq2.2}) as a reminder that, in the Hamiltonian formalism, the $N^a$'s enter the action as Lagrange multipliers of the three diffeomorphism constraints ${\mathcal H}_a \approx 0$. Similarly, the ``lapse'' $N(t)$ enters the action as a Lagrange multiplier for the Hamiltonian constraint ${\mathcal H} \approx 0$. Finally, the dynamical variables of Bianchi geometries are the six
functions of time $g_{ab} (t)$.

\subsection{Fixing the local Lorentz gauge symmetry}\label{ssec2.2}

In view of the special structure (\ref{eq2.2}) of the geometry it is natural to choose a vielbein co-frame $\theta^{\hat\alpha} = h_{\ \mu}^{\hat\alpha} \, dx^{\mu}$ of the form
\begin{eqnarray}
\label{eq2.4}
\theta^{\hat 0} &= &N(t) \, dt \, , \\
\label{eq2.5}
\theta^{\hat a} &= &\sum_b e^{-\beta_a(t)} \, S^{\hat a} \, _b \ (t) (\tau^b + N^b(t) \, dt) \, ,
\end{eqnarray}
where the matrix $S^{\hat a} \, _b$ satisfies\footnote{We generally use Einstein's summation convention, except when there are ambiguities, as in Eqs.~(\ref{eq2.5}) or (\ref{eq2.6}).}
\begin{equation}
\label{eq2.6}
\sum_a e^{-2\beta_a} \, S^{\hat a} \, _b \ S^{\hat a} \, _c = g_{bc} \, .
\end{equation}
In other words, the matrix $S^{\hat a} \, _b$, or rather its inverse $S^b \, _{\hat a}$, such that
\begin{equation}
\label{eq2.7}
S^{\hat a} \, _c \ S^c \, _{\hat b} = \delta_{\hat b}^{\hat a} \, , \quad S^a \, _{\hat c} \ S^{\hat c} \, _b = \delta_b^a \, ,
\end{equation}
transforms the quadratic form $g_{bc}$ into a diagonal form
\begin{equation}
\label{eq2.8}
g_{ab} \, S^a \, _{\hat c} \ S^b \, _{\hat d} = [{\rm diag} (e^{- 2\beta_1} , e^{-2\beta_2} , e^{-2\beta_3})]_{\hat c \hat d} \, .
\end{equation}

The idea behind the representation (\ref{eq2.6}) is to encode the six independent components of $g_{ab}$ into two sets of three variables: (1) three ``diagonal'' degrees of freedom $\beta_1 , \beta_2 , \beta_3$; and (2) three ``off diagonal'' ones, parametrizing the ``diagonalizing'' matrix $S$. For such a decomposition to be uniquely defined, one needs to restrict the structure of the ($3$ by $3$) matrix $S$ by six conditions. This can be achieved in several different ways. For instance, one could require the lower-than-diagonal elements of the matrix $S^{\hat a} \, _b$ to vanish, and its diagonal elements to be equal to 1; this would correspond to the so-called Iwasawa decomposition of $h^{\hat a} \, _b$ (and, correspondingly, of $g_{cd} = \delta_{\hat a \hat b} \ h^{\hat a} \, _c \ h^{\hat b} \, _d)$, which is unique, and which was found to be useful in recent work on the hidden presence of Kac-Moody symmetries in gravity theories \cite{DHN2}. We shall discuss in Appendix B the use of this decomposition in the study of the dynamics of type~II Bianchi cosmologies. However, in the case of the most generic (class~A) Bianchi models, namely type~IX and type~VIII, the Iwasawa decomposition is rather inconvenient. 

As was emphasized by R.T.~Jantzen \cite{Jantzen:1982je,Jantzen79,Jantzen:2001me}
it is quite advantageous to use decompositions (\ref{eq2.5}), (\ref{eq2.6}) with a matrix $S^{\hat a} \, _b$ restricted to belonging to the {\it automorphism group}, say ${\mathcal A}$, of the Lie algebra ${\frak G}$ of $G$. Explicitly, this means that ${\mathcal A}$ is the group of linear transformations which leave invariant the structure constants $C^a \, _{bc}$.
The reason why the choice of a ``diagonalizing'' matrix $S$ belonging to the automorphism group ${\mathcal A}$ is advantageous is that (as we shall see explicitly below) the potential terms in the Hamiltonian can be expressed in terms of the $\beta$'s and of the components, say $\overline C^{\overline a} \, _{\overline b \overline c}$, of the structure constants w.r.t. the intermediate, co-frame $\overline\tau^{\overline a} = S^{\hat a} \, _b \ \tau^b$. These components are given by
\begin{equation}
\label{eq2.9}
\overline C^{\overline a} \, _{\overline b \overline c} = S^{\hat a} \, _{a'} \ S^{b'} \, _{\hat b} \ S^{c'} \, _{\hat c} \ C^{a'} \, _{b'c'} \, .
\end{equation}
For a general choice of $S^{\hat a} \, _b$ the components $\overline C^{\overline a} \, _{\overline b \overline c}$ would depend on the off-diagonal variables entering $S^{\hat a} \, _b$. For instance, when using an Iwasawa decomposition of $g_{ab}$ (with an upper diagonal matrix $S$), the $\overline C^{\overline a} \, _{\overline b \overline c}$ explicitly depend on the off-diagonal variables $\nu_{12} , \nu_{23} , \nu_{13}$ entering Eq.~(\ref{Iwasrep}), so that the potential terms in the Hamiltonian also depend on these off-diagonal variables. By contrast, by definition of the automorphism group ${\mathcal A}$ (as fixing ${\frak G}$ and therefore the $C$'s) when $S \in {\mathcal A}$ the components $\overline C^{\overline a} \, _{\overline b \overline c}$ are simply equal to the original $C^a \, _{bc}$, and thereby do not introduce any dependence on the off-diagonal metric variables.

\smallskip

When $G$ is a simple group (and when its Dynkin diagram has no symmetries), the automorphism group ${\mathcal A}$ of ${\frak G}$ is the adjoint group of $G$. In the case of Bianchi type~IX (where $G = SU(2)$), this means that ${\mathcal A}$ is $SO(3)$ (which is the
quotient  $SU(2)/Z_2$). In that case, $S^{\hat a} \, _b$ is an orthogonal matrix, and the decomposition (\ref{eq2.6}) is the so-called ``Gauss decomposition'', corresponding to the diagonalization of the quadratic form $g_{ab}$ w.r.t. a given Euclidean metric $\delta_{ab}$. [Such
a Gauss decomposition was advocated by M. Ryan \cite{Ryan72,RyanShepley75}.]
In the case of Bianchi type~VIII, the automorphism group is $SO(1,2)$, which means that one should use an ``hyperbolic'' generalization of the Gauss decomposition of $g_{ab}$ w.r.t. a given Lorentzian metric $\eta_{ab} = {\rm diag} (-1 , -1 , +1)$. The special metrics, $\delta_{ab}$ or $\eta_{ab}$, that enter here are simply (modulo a suitable normalization) the Cartan-Killing metrics $k_{ab}$ associated to the Lie algebra ${\frak G}$, say
\begin{equation}
\label{eq2.10}
k_{ab} = - \, \frac{1}{2} \, C^c \, _{ad} \ C^d \, _{bc} \, .
\end{equation}
For instance, in the usual basis for type~IX we have $C^a \, _{bc} = \varepsilon_{abc}$, so that $k_{ab} = + \, \delta_{ab}$.

\smallskip

Summarizing: In the Bianchi type~IX case, we parametrize the six metric degrees of freedom contained in $g_{ab} (t)$ by means of the three diagonal variables $\beta_1 (t) , \beta_2 (t) , \beta_3 (t)$ and the three Euler angles $\theta_1 (t) , \theta_2 (t) , \theta_3 (t)$ parametrizing the {\it orthogonal} metric $S^{\hat a} \, _b \, (\theta_1 , \theta_2 , \theta_3)$ entering the Gauss decomposition (\ref{eq2.6}) of the quadratic form $g_{ab}$ (``transformation to principal axes''). On the other hand, in the Bianchi type~VIII case, the parametrization of $g_{ab} (t)$ by three diagonal variables $\beta_1 (t) , \beta_2 (t) , \beta_3 (t)$ and three ``diagonalizing angles'' $\theta_1 (t) , \theta_2 (t) , \theta_3 (t)$ should be done by an hyperbolic $(S \in SO (1,2))$ generalization of the Gauss decomposition, {\it i.e.} a transformation of $g_{ab}$ to principal axes w.r.t. a Lorentzian metric $\eta_{ab} = {\rm diag} (-1 , -1 , +1)$. [Such a transformation is always possible. It can be built from the eigenvectors of the  non positive-definite quadratic form $\eta_{ab}$ w.r.t. the  positive-definite one $g_{ab}$ .]

\smallskip

Then, in terms of such a parametrization, $g_{ab} \leftrightarrow (\beta_1 , \beta_2 , \beta_3 \, , \ S^{\hat a} \, _b \ (\theta_1 , \theta_2 , \theta_3))$ of the metric, we gauge-fix the local Lorentz symmetry by defining the specific vielbein $\theta^{\hat\alpha} (\beta_a , \theta_a) = h_{\mu}^{\hat\alpha} (\beta_a , \theta_a) \, dx^{\mu}$ by means of Eqs.~(\ref{eq2.4}), (\ref{eq2.5}).

\subsection{Lagrangian formulation of the Einstein-Dirac system}\label{ssec2.3}

Having uniquely determined (for types~IX and VIII) a vielbein $\theta^{\hat\alpha} \, _{\mu} \ dx^{\mu}$ in terms of the usual metric degrees of freedom $g_{ab} (t)$, $N^a (t)$, $N(t)$, we can now consider the general Einstein(-Hilbert)-Dirac Lagrangian density,
\begin{equation}
\label{eq2.11}
L = L_{EH} + L_D \, ,
\end{equation}
where\footnote{We use units such that $16 \pi \, G = 1 = c$.}
\begin{eqnarray}
\label{eq2.12}
L_{EH} &= &\sqrt{- \, ^4g} \ ^4R \, , \\
\label{eq2.13}
L_D &= &\sqrt{- \, ^4g} \ \left( \overline\Psi \, \gamma^{\hat\alpha} \, \nabla_{\hat\alpha} \, \Psi - m \, \overline\Psi \, \Psi \right) \, ,
\end{eqnarray}
as a functional of the metric $g_{\mu\nu}$, and of the spinor field $\Psi$ (w.r.t. the gauge-fixed vielbein $h_{\mu}^{\hat\alpha} (g_{\nu\lambda})$). We use gamma matrices adapted to our mostly plus signature $-+++$, namely
\begin{equation}
\label{eq2.14}
\gamma^{\hat\alpha} \, \gamma^{\hat\beta} + \gamma^{\hat\beta} \, \gamma^{\hat\alpha} = 2 \, \eta^{\hat\alpha\hat\beta} \, \un \, ,
\end{equation}
with an anti-hermitian $\gamma^{\hat 0}$ (satisfying $(\gamma^{\hat 0})^2 = -\un$), and three hermitian $\gamma^{\hat 1} , \gamma^{\hat 2} , \gamma^{\hat 3}$. We then choose
\begin{equation}
\label{eq2.15}
\beta := i \, \gamma_{\hat 0} = - \, i \, \gamma^{\hat 0} \, ,
\end{equation}
with the involutive $\beta$ ($\beta^2 = + \un$) being used to define the usual Dirac conjugate (as defined in the mostly minus signature)
\begin{equation}
\label{eq2.16}
\overline\Psi := \Psi^{\dagger} \, \beta \, .
\end{equation}

The frame covariant derivative of the spinor entering the Dirac action Eq.~(\ref{eq2.13}) is $\nabla_{\hat\alpha} \, \Psi \equiv h^{\mu} \, _{\hat\alpha} \ \nabla_{\mu} \, \Psi$, where $h^{\mu} \, _{\hat\alpha}$ is the vielbein frame dual to the vielbein co-frame $h^{\hat\alpha} \, _{\mu}$ ({\it i.e.} $h^{\hat\alpha} \, _{\mu} \ h^{\mu} \, _{\hat\beta} = \delta^{\hat\alpha} \, _{\hat\beta}$), and where the world-index covariant derivative $\nabla_{\mu} \, \Psi$ is given  by
\begin{equation}
\label{eq2.17}
\nabla_{\mu} \, \Psi = \partial_{\mu} \, \Psi + \frac{1}{4} \, \omega_{\hat\alpha\hat\beta \, \mu} \, \gamma^{\hat\alpha\hat\beta} \, \Psi \, ,
\end{equation}
where (denoting $h_{\hat\alpha  \mu}\equiv \eta_{\hat\alpha \hat\beta} h^{\hat\beta}\,_{\mu}$)
$$
\omega_{\hat\alpha\hat\beta \, \mu} := h_{\hat\alpha \nu} \, \nabla_{\mu} \, h^{\nu}\,_{\hat\beta}
$$
are the connection components of the vielbein (with last index taken as world index), and 
where\footnote{Everywhere we use brackets $[\cdots]$ around indices to denote antisymmetrization with weight one.}
\begin{equation}
\label{eq2.18}
\gamma^{\hat\alpha\hat\beta} := \frac{1}{2} \, (\gamma^{\hat\alpha} \, \gamma^{\hat\beta} - \gamma^{\hat\beta} \, \gamma^{\hat\alpha}) \equiv \gamma  ^{[\hat\alpha} \gamma^{\hat\beta]} \, .
\end{equation}
We shall only consider here class~A Bianchi models, {\it i.e.} models satisfying $C^a \, _{ac} = 0$, which is equivalent to saying that the dualization of the structure constants w.r.t. the antisymmetric lower indices, $n^{ad} := \frac{1}{2} \, \varepsilon^{bcd} \, C^a \, _{bc}$, yields a symmetric  
tensor density. In other words
\begin{equation}
\label{eq2.19}
C^a \, _{bc} = \varepsilon_{bcd} \, n^{ad} \, ,
\end{equation}
where $\varepsilon_{abc} = \varepsilon_{[abc]}$ (with $\varepsilon_{123} = + \, 1$), and $n^{ab} = n^{ba}$. For type~IX, one has $n^{ab} = \delta^{ab}$ in the usual basis, while, for type~VIII $n^{ab} = {\rm diag} (+1,+1,-1)$. [Note that the Cartan-Killing metric $k_{ab}$, Eq.~(\ref{eq2.10}), associated to the $C$'s is quadratic in $n^{ab}$. In type~IX $k_{ab} = \delta_{ab} = {\rm diag} (+1,+1,+1)$ is numerically equal to $n^{ab}$, while in type~VIII $k_{ab} = {\rm diag} (-1,-1,+1)$ is of signature $--+$, independently of whether one chooses a basis where $n^{ab} = {\rm diag} (+1,+1,-1)$ or $n^{ab} = {\rm diag} (-1,-1,+1)$.]

\smallskip

It is well-known that the dynamics of class~A Bianchi models derives from a Lagrangian which is obtained simply by substituting in the general action (\ref{eq2.11}) the symmetry-reduced form of the metric, {\it i.e.} Eq.~(\ref{eq2.2}). This is also true when the metric is coupled to a homogeneous spinor. Here, we shall define the spatial homogeneity of a spinor $\Psi$ simply as meaning that the components of the spinor w.r.t. any frame $h_{\hat\alpha} := h_{\hat\alpha}^{\mu} \, \partial_{\mu}$ which is invariant under the homogeneity group $G$, {\it i.e.} $\pounds_{\xi_a} \, h_{\hat\alpha} = 0$, depend only on time\footnote{In some cases, some spatial variation of $\Psi$, of the type $\pounds_{\xi_a} \, \Psi = i \, \lambda_a \, \Psi$, with real quantities $\lambda_a$ subject to the integrability constraint $\lambda_a \, C^a \, _{bc} = 0$, is compatible with the homogeneity of the geometry \cite{HenIHP}. However, such a generalization is allowed neither in the case (we shall focus on) of a Majorana spinor, nor in the case of  simple Lie algebras, such as type~IX or type~VIII, which are of most physical interest.}.

\smallskip

We know on general grounds \cite{ADM}  that the lapse, $N$, and the shift vector, $N^a$, will enter the final Hamiltonian action $S = \int (pdq - H(q,p) \, dt)$ as Lagrange multipliers of, respectively, the Hamiltonian constraint, ${\mathcal H}$, and the diffeomorphism (or momentum) constraints, ${\mathcal H}_a$. Namely, the Hamiltonian is of the general form
\begin{equation}
\label{eq2.20}
H = \int (N {\mathcal H} + N^a \, {\mathcal H}_a) \, \mu
\end{equation}
where $\mu \equiv \tau^1 \wedge \tau^2 \wedge \tau^3$ is the spatial volume density [in the co-frame $(dt,\tau^1 , \tau^2 , \tau^3)$], and where
\begin{eqnarray}
\label{eq2.21}
{\mathcal H} &= &\sqrt g \, (2 \ ^4G_0^0 - T_0^0) = \sqrt g \, (^4R_0^0 - \, ^4R_a^a - T_0^0) \, , \\
\label{eq2.22}
{\mathcal H}_a &= &\sqrt g \, (2 \ ^4G_a^0 - T_a^0) = \sqrt g \, (2 \ ^4R_a^0 - T_a^0) \, .
\end{eqnarray}

Here, $g$ denotes the determinant of the spatial metric $g_{ab}$ (w.r.t. the spatial-coframe $\tau^a$, see Eq.~(\ref{eq2.2})), $^4G_{\beta}^{\alpha} \equiv \, ^4R_{\beta}^{\alpha} - \frac{1}{2} \, ^4R \, \delta_{\beta}^{\alpha}$ the spacetime Einstein tensor, and $T_{\beta}^{\alpha}$ the matter stress-energy tensor. The factor 2 multiplying $^4G_{\alpha}^0$ in the equations above represents $(8\pi G)^{-1}$ in the units we use where $16\pi G = 1$.

\smallskip

Knowing in advance the structure (\ref{eq2.20}), one can simplify the computation of the Hamiltonian by working in the special quasi-Gaussian gauge where
\begin{equation} \label{gauge}
N = \sqrt g \, , \ {\rm and} \ N^a = 0 \,.
\end{equation}
[This gauge was found useful in many previous cosmological studies, see
e.g. \cite{BKL,Ryan72,DHN2}.]
In addition, we shall assume (for notational simplicity) that we consider the dynamics of a (comoving) piece of a homogeneous universe which has a unit (comoving) volume $1 = \int \mu$. This allows one to identify the total Lagrangian with the Lagrangian density: 
$$
S = \int \mu \int dt \, L(q,\dot q) = \int dt \, L (q,\dot q) \, .
$$

\subsection{Gravity part of the Lagrangian}\label{ssec2.4}

The gravity part of the Lagrangian, which generically reads (modulo a total divergence)
\begin{equation}
\label{eq2.23}
L_{EH} = N \, \sqrt g \, [g^{ac} \, g^{bd} \, K_{ab} \, K_{cd} - (g^{ab} \, K_{ab})^2 + R(g)] \, ,
\end{equation}
in terms of the spatial scalar curvature $R(g) \equiv \, ^3R (g)$, and of the second fundamental form\footnote{Note that Ref. \cite{ADM} defines $K_{ab}$ with the opposite sign.},
\begin{equation}
\label{eq2.24}
K_{ab} := \frac{1}{2N} \, (\partial_t \, g_{ab} - D_a \, N_b - D_b \, N_a)
\end{equation}
($D$ denoting the $3$-dimensional covariant derivative), reads, when working in the gauge (\ref{gauge}) 
\begin{equation}
\label{eq2.25}
L_{EH} = T_g (g,\dot g) - V_g (g) \, .
\end{equation}
Here $T_g$ denotes the ``kinetic-energy'' part of the gravity Lagrangian $(\dot g \equiv \partial_t \, g)$,
\begin{equation}
\label{eq2.26}
T_g (g,\dot g) = \frac{1}{4} \, g^{ac} \, g^{bd} \, \dot g_{ab} \, \dot g_{cd} - \frac{1}{4} \, (g^{ab} \, \dot g_{ab})^2 \, ,
\end{equation}
while $V_g$ denotes its ``potential'' part,
\begin{equation}
\label{eq2.27}
V_g (g) = - \, g \, R(g) \, .
\end{equation}

Using the decomposition (\ref{eq2.6}) of $g_{ab}$ into the three diagonal variables $\beta_1$, $\beta_2$, $\beta_3$, and the three angles $\theta_1 , \theta_2 , \theta_3$ parametrizing the $SO(3)$ [respectively $SO(1,2)$] matrix $S^{\hat a} \, _b$ in the type~IX (resp. type~VIII) case, we can express the gravity part (\ref{eq2.25}) of the Lagrangian in terms of $\beta_a , \theta_a$ and $\dot\beta_a , \dot\theta_a$.

\smallskip

The spatial scalar curvature of a general homogeneous metric $g_{ab} (t) \, \tau^a (x)$ $\tau^b (x)$ is expressible in terms of the structure constants of Eq.~(\ref{eq2.3}), namely
(see, e.g., \cite{SpindelGBS})
\begin{equation}
\label{eq2.28}
R(g) = - \frac{1}{4} \, C^a \, _{bc} \ C^{\underline a} \, _{\underline b \underline c} - \frac{1}{2} \, C^a \, _{bc} \ C^b \, _{a \underline c} - C^a \, _{ac} \ C^a \, _{a \underline c}
\end{equation}
where it is understood (for notational transparence) that a summation over, say, $a$ and $\underline a$ denotes an appropriate contraction by means of $g_{ab}$ or its inverse $g^{ab}$ ({\it e.g.} $A_c \, B_{\underline c} \equiv g^{cc'} \, A_c \, B_{c'}$). In the class~A case the last term in Eq.~(\ref{eq2.28}) vanishes, while the other ones can be expressed in terms of the (symmetric) dual, $n^{ad}$ of $C^a \, _{bc}$ (see Eq.~(\ref{eq2.19}). This leads to the following simple expression for the ``gravity potential'' $V_g$, Eq.~(\ref{eq2.27}),
\begin{equation}
\label{eq2.29}
V_g (g) = n^{ab} \, n^{\underline a \underline b} - \frac{1}{2} \, (n^{a \underline a})^2 \equiv g_{aa'} \, g_{bb'} \, n^{ab} \, n^{a'b'} - \frac{1}{2} \, (g_{ab} \, n^{ab})^2 \, .
\end{equation}
Inserting the decomposition (\ref{eq2.6}) into this result, then yields the expression of $V_g$ in terms of $\beta_a$ and the matrix $S$. As announced above, the fact that $S^{\hat a} \, _b$ was chosen to belong to the automorphism group ${\mathcal A}$ (leaving the structure constants $C^a \, _{bc}$ invariant) implies that the right-hand-side (r.h.s.) of Eq.~(\ref{eq2.28}) is independent of $S$, and only depends on the diagonal variables $\beta_a$. This is also true for the potential $V_g = - \, g \, R(g)$ if $S$ belongs to the ``special'' subgroup of ${\mathcal A}$ having $\det S = 1$. In the cases we consider here (types IX and VIII) this is automatically the case as ${\mathcal A} = SO(3)$ or $SO(1,2)$. Finally, we conclude that $V_g (g)$ is given by the same expression that it would have if $g_{ab}$ had been assumed to be diagonal, namely
\begin{eqnarray}
\label{eq2.30}
V_g (g) = V_g (\beta) &\equiv &n_1^2 e^{-4\beta_1} + n_2^2 e^{-4\beta_2} + n_3^2 e^{-4 \beta_3} - \frac{1}{2} \, (n_1 e^{-2 \beta_1} + n_2 e^{-2 \beta_2} + n_3 e^{-2 \beta_3})^2 \nonumber \\
&= &\frac{1}{2} \, [n_1^2 e^{-4 \beta_1} + n_2^2 e^{-4\beta_2} + n_3^2 e^{-4 \beta_3}] \nonumber \\
&&- \, [n_1n_2e^{-2(\beta_1 + \beta_2)} + n_2n_3e^{-2(\beta_2 + \beta_3)} +n_3n_1 e^{-2(\beta_3 + \beta_1)}] \, ,
\end{eqnarray}
where $n_a$ denote the diagonal components of $n^{ab} ={\rm diag} (n_1,n_2,n_3)$.

Turning now to the kinetic part $T_g$, Eq.~(\ref{eq2.26}), we need to evaluate it in terms of the $\beta - S$ parametrization (\ref{eq2.6}) of $g_{ab}$. To be explicit we should, at this stage, choose a specific convention for the definition of the three (possibly generalized) Euler angles $\theta_1 , \theta_2 ,\theta_3$ parametrizing the (pseudo-)orthogonal matrix $S$. It is, however, better to introduce a notation for the ``angular velocity'', say $w$, of the matrix $S$, before specifying its expression in terms of $\dot\theta_1 , \dot\theta_2 , \dot\theta_3$ and the $\theta$'s. We define $w^{\hat a} \, _{\hat b}$ by writing
\begin{equation}
\label{eq2.31}
\dot S^{\hat a} \, _c = \sum_{\hat b} w^{\hat a} \, _{\hat b} \ S^{\hat b} \ _c \, ,
\end{equation}
or $\dot S = w \, S$, if we consider $S \equiv S^{\hat a} \, _b$ and $w \equiv w^{\hat a} \, _{\hat b}$ as matrices. In this matrix notation, the decomposition (\ref{eq2.6}) reads $g = S^T \, e^{-2\beta} \, S$, where $g$ denotes here the matrix $g_{ab}$, $\beta$ the diagonal matrix $\beta_a \, \delta_{ab}$, and the superscript $T$ the transposition of a matrix. Differentiating the matrix $g$ then yields
\begin{equation}
\label{eq2.32}
\dot g = S^T (-2 \dot\beta e^{-2\beta} + e^{-2\beta} \, w + w^T \, e^{-2\beta}) \, S \, .
\end{equation}
At this stage, the calculation depends on whether the matrix $S$ is orthogonal ($SO(3)$; type~IX) or pseudo-orthogonal ($SO(1,2)$; type~VIII). We shall henceforth focus on the type~IX case, giving only some indications of the differences that arise in the type~VIII case. In the type~IX case we have 
(when using the usual basis where $n^{ab}=\delta^{ab}$) $S^T S = SS^T = \un$ so that the ``matrix angular velocity'' $w$ defined by Eq.~(\ref{eq2.31}) is antisymmetric in the usual sense: $w^T = -w$. [In the type~VIII case $w$ would be antisymmetric in the Lorentzian sense, {\it i.e.} after considering $w_{\hat a \hat b} : = \eta_{\hat a \hat a'} \, w^{\hat a'} \, _{\hat b}$ where $\eta_{\hat a \hat b} = {\rm diag} (-1,-1,+1)$.] Inserting this knowledge in Eq.~(\ref{eq2.32}) then yields an explicit expression for $\dot g_{ab}$ of the form (denoting $w^{\hat a \hat b} \equiv \delta^{\hat b} \, _{\hat c} \ w^{\hat a} \, _{\hat c} = w^{\hat a} \, _{\hat b}$)
\begin{equation}
\label{eq2.33}
\dot g_{cd} = S^{\hat a} \, _c \ S^{\hat b} \, _d \ \overline k_{\overline a \overline b} \, ,
\end{equation}
\begin{equation}
\label{eq2.34}
\overline k_{\overline a \overline b} \equiv - \, 2 \, \dot\beta_a \, e^{-2\beta_a} + (e^{-2\beta_a} - e^{-2\beta_b}) \, w^{\hat a \hat b} \, .
\end{equation}
Here, we can think of $\overline k_{\overline a \overline b}$ as the components of the covariant tensor $k_{ab} := \dot g_{ab}$ w.r.t. to the ``rotated'' co-frame
\begin{equation}
\label{eq2.35}
\overline\tau^{\overline a} := S^{\hat a} \, _b \ \tau^b \, .
\end{equation}
The co-frame (\ref{eq2.35}) is intermediate between the basic co-frame $\tau^a$ (w.r.t. which $ds^2 = -N^2 \, dt^2 + g_{ab} \, \tau^a \, \tau^b$), and the orthonormal frame $\theta^{\hat\alpha}$ (w.r.t. which $ds^2 = \eta_{\hat\alpha\hat\beta} \, \theta^{\hat\alpha} \, \theta^{\hat\beta}$). Indeed, in the co-frame $(dt , \overline\tau^{\overline a})$ we have a non-Minkowskian, but {\it diagonal} form of the spacetime metric, namely:  $ds^2 = -N^2 \, dt^2 + \underset{a}{\sum} \ e^{-2\beta_a} (\overline\tau^a)^2$, with $N = \sqrt{\det g_{ab}} = e^{-(\beta_1 + \beta_2 + \beta_3)}$.

\smallskip

As the gravitational kinetic-energy term (\ref{eq2.26}) is manifestly invariant under any linear change of basis of the co-frame $\tau^a$, it can be rewritten as
\begin{equation}
\label{eq2.36}
T_g = \frac{1}{4} \, \overline g^{\overline a \overline c} \ \overline g^{\overline b \overline d} \ \overline k_{\overline a \overline b} \ \overline k_{\overline c \overline d} - \frac{1}{4} \, (\overline g^{\overline a \overline b} \ \overline k_{\overline a \overline b})^2 \, ,
\end{equation}
where $\overline g^{\overline a \overline b}$ are the components of the contravariant metric w.r.t. the intermediate frame $\overline\tau^{\overline a}$, namely:  $ \overline g^{\overline a \overline b} = e^{+2\beta_a} \, \delta_{ab}$. This yields the explicit result
\begin{equation}
\label{eq2.37}
T_g = T_{\beta} (\dot\beta) + T_w (w,\beta) \, ,
\end{equation}
where
\begin{equation}
\label{eq2.38}
T_{\beta} (\dot\beta) = \sum_a \dot\beta_a^2 - \left( \sum_a \dot\beta_a \right)^2 = -2 \, (\dot\beta_1 \, \dot\beta_2 + \dot\beta_2 \, \dot\beta_3 + \dot\beta_3 \, \dot\beta_1) \, ,
\end{equation}
and (in the type IX case)
\begin{eqnarray}
\label{eq2.39}
T_w^{\rm IX} (w,\beta) &= &2 \sinh^2 (\beta_1 - \beta_2) (w^{\hat 1 \hat 2})^2 + 2 \sinh^2 (\beta_2 - \beta_3)(w^{\hat 2 \hat 3})^2 \nonumber \\
&+ &2 \sinh^2 (\beta_3 - \beta_1)(w^{\hat 3 \hat 1})^2 \, .
\end{eqnarray}
Summarizing so far: the gravity part of the Lagrangian has the form
\begin{equation}
\label{eq2.40}
L_{EH} (g,\dot g) = T_{\beta} (\dot\beta) + T_w (w,\beta) - V_g (\beta) \, ,
\end{equation}
with a $\beta$-kinetic energy given by (\ref{eq2.38}), a rotational kinetic energy linked to the dynamics of the Euler angles entering $S(\theta_a)$ given by (\ref{eq2.39}) and a potential energy given by (\ref{eq2.30}) (with $n^{ab}=\delta^{ab}$). Note that the matrix $S$, {\it i.e.} the Euler angles $\theta_1 , \theta_2 , \theta_3$, do not explicitly enter the result (\ref{eq2.40}). As explained above, they do not enter the potential term $V_g$ because $S$ was chosen to belong to the automorphism group of the Lie algebra ${\mathfrak G}$. However, they do implicitly enter the rotational kinetic energy $T_w$, as the rotational angular velocity $w^{\hat a \hat b}$ do depend both on $\theta_a$ and $\dot\theta_a$. For instance, if we define the Euler angles as in the standard references \cite{Landau-Lifchitz,Goldstein}, namely using the $z-x-z$ convention, {\it i.e.} a matrix
\begin{equation}
\label{eq2.41}
S = \begin{pmatrix}
\cos\psi &\sin\psi &0 \\ -\sin \psi &\cos \psi &0 \\ 0 &0 &1 
\end{pmatrix} \begin{pmatrix}
1 &0 &0 \\ 0 &\cos \theta &\sin\theta \\ 0 &-\sin\theta &\cos\theta
\end{pmatrix} \begin{pmatrix}
\cos\varphi &\sin\varphi &0 \\ -\sin \varphi &\cos \varphi &0 \\ 0 &0 &1 
\end{pmatrix} ,
\end{equation}
the angular velocities $w^{\hat a \hat b}$ (with $w = \dot S \, S^{-1}$) are given by
\begin{eqnarray}
\label{eq2.42}
w^{\hat 1 \hat 2} &= &\dot\varphi \cos\theta + \dot\psi \, , \nonumber \\
w^{\hat 2 \hat 3} &= &\dot\varphi \sin\theta \sin\psi + \dot\theta \cos\psi \, , \nonumber \\
w^{\hat 3 \hat 1} &= &\dot\varphi \sin\theta \cos\psi - \dot\theta \sin\psi \, .
\end{eqnarray}
As several authors (see, e.g., \cite{Ryan72}) have previously remarked, the rotational part $T_w$ of the gravity Lagrangian is analogous to the kinetic energy of a rotating body (``asymmetric top''), namely $T_w = \frac{1}{2} (I_1 \, \Omega_1^2 + I_2 \, \Omega_2^2 + I_3 \, \Omega_3^2)$, where $\Omega_1 \equiv w^{\hat 2 \hat 3}$, $\Omega_2 \equiv w^{\hat 3 \hat 1}$, $\Omega_3 \equiv w^{\hat 1 \hat 2}$ are the body-frame components of the angular velocity and $I_3 = 4 \sinh^2 (\beta_1 - \beta_2)$, {\it etc.} the (body-frame) moments of inertia. Note, however, that, contrary to a usual rigid solid, the body-frame moments of inertia are time-dependent. Indeed, they depend on the $\beta$'s, which have a (coupled) dynamics determined by their kinetic energy $T_{\beta}$, their potential energy $V_{\beta}$, and their couplings to the other variables [and notably the Euler angles themselves, through the term $T_w (\dot\theta_a , \theta_a , \beta_a)$].

\subsection{Spinor part of the Lagrangian}\label{ssec2.5}

Let us now derive the explicit form of the spinor part, $L_D$, Eq.~(\ref{eq2.13}), of the Lagrangian. To do this, we need the explicit expression of the connection coefficients, say $\omega_{\hat\alpha\hat\beta\hat\gamma} \equiv  \omega_{\hat\alpha\hat\beta\mu} \,  h^{\mu}_{\ \hat\gamma} \,$, of our specifically chosen vielbein $\theta^{\hat\alpha} =h^{\hat\alpha} \,_{\mu} dx^{\mu}$, defined in Eq.~(\ref{eq2.5}) in terms of the metric degrees of freedom $\beta_a$, $S (\theta_a)$ (parametrizing $g_{ab}$ {\it via} Eq.~(\ref{eq2.6})). [In this
subsection we take as above a vanishing shift vector $N^a=0$, but
allow for an arbitrary lapse $N(t)$.] If we consider the ``structure constants'', say ${\mathcal C}^{\hat\alpha} \, _{\hat\beta\hat\gamma}$, of the orthonormal co-frame $\theta^{\hat\alpha}$, defined as
\begin{equation}
\label{eq2.43}
d \theta^{\hat\alpha} = \frac{1}{2} \, {\mathcal C}^{\hat\alpha} \, _{\hat\beta \hat\gamma} \ \theta^{\hat\beta} \wedge \theta^{\hat\gamma} \, ,
\end{equation}
the (frame) connection coefficients $\omega_{\hat\alpha\hat\beta\hat\gamma}$ can be expressed as
\begin{equation}
\label{eq2.44}
\omega_{\hat\alpha\hat\beta\hat\gamma} = \frac{1}{2} \, ({\mathcal C}_{\hat\alpha\hat\beta\hat\gamma} + {\mathcal C}_{\hat\beta\hat\gamma\hat\alpha} - {\mathcal C}_{\hat\gamma\hat\alpha\hat\beta}) \, ,
\end{equation}
where we denoted ${\mathcal C}_{\hat\alpha\hat\beta\hat\gamma} := \eta_{\hat\alpha\hat\sigma} \, {\mathcal C}^{\hat\sigma} \, _{\hat\beta\hat\gamma}$.

\smallskip

As $\theta^{\hat 0} = N(t) \, dt$ only involves $t$, we have $d \theta^{\hat 0} = 0$, so that
\begin{equation}
\label{eq2.45}
{\mathcal C}^{\hat 0} \, _{\hat\beta\hat\gamma} = 0 \qquad (\mbox{and} \quad {\mathcal C}_{\hat 0 \hat\beta \hat\gamma} = 0) \, .
\end{equation}
On the other hand $\theta^{\hat a}$ involves both $t$ and the spatial coordinates (that are implicit in $\tau^a (x)$). Using Eq.~(\ref{eq2.3}), one finds that the Cartan differential of $\theta^{\hat a}$ reads (no summation on $a$, but summation on $b,c,d$))
\begin{equation}
\label{eq2.46}
d \theta^{\hat a} = \partial_t (e^{-\beta_a} \, S^{\hat a} \, _{b}) \, dt \wedge \tau^b + \frac{1}{2} \, e^{-\beta_a} \, S^{\hat a} \, _d \ C^d \, _{bc} \ \tau^b \wedge \tau^c \, .
\end{equation}
Rewriting the r.h.s. in terms of $\theta^{\hat 0} = N  dt$ and $\theta^{\hat a} = e^{-\beta_a} \, S^{\hat a} \, _b \ \tau^b$ yields the structure constants ${\mathcal C}^{\hat a} \, _{\hat 0 \hat b}$ and ${\mathcal C}^{\hat a} \, _{\hat b \hat c}$, namely (no summation on $a,b,c$)
\begin{equation}
\label{eq2.47}
{\mathcal C}^{\hat a} \, _{\hat 0 \hat b} \equiv - {\mathcal C}^{\hat a} \, _{\hat b \hat 0} = - \frac{1}{N} \, \dot\beta_a \, \delta_{ab} + e^{-\beta_a + \beta_b} \, \frac{w^{\hat a}\,_{ \hat b}}{N}
\end{equation}
\begin{equation}
\label{eq2.48}
{\mathcal C}^{\hat a} \, _{\hat b \hat c} = e^{-\beta_a + \beta_b + \beta_c} \, \overline C^{\overline a} \, _{\overline b \overline c} \, ,
\end{equation}
where the $\overline C$ coefficients are the structure constants w.r.t. the intermediate frame $\overline\tau^{\overline a} \equiv S^{\hat a} \, _b \ \tau^b$, as defined in Eq.~(\ref{eq2.9}) above. Again, the choice of the matrix $S$ as belonging to the automorphism group of the Lie algebra ${\mathfrak G}$ implies that the $\overline C$ coefficients are simply equal to the original structure constants $C^a \, _{bc}$.

\smallskip

Inserting the results (\ref{eq2.45}), (\ref{eq2.47}), (\ref{eq2.48}) in Eq.~(\ref{eq2.44}) 
(and considering the special case of type IX, i.e. $w^T =-w$)
yields the explicit expressions of the connection coefficients, namely
\begin{equation}
\label{eq2.49}
\omega_{\hat 0 \hat b \hat c} = - \frac{1}{2} \, \left[ {\mathcal C}_{\hat b \hat 0 \hat c} + {\mathcal C}_{\hat c \hat 0 \hat b} \right] = \frac{1}{N} \, \dot\beta_b \, \delta_{bc} + \sinh (\beta_b - \beta_c) \, \frac{w^{\hat b \hat c}}{N}
\end{equation}
\begin{equation}
\label{eq2.50}
\omega_{\hat a \hat b \hat 0} = \frac{1}{2} \, \left[ - {\mathcal C}_{\hat a \hat 0 \hat b} + {\mathcal C}_{\hat b \hat 0 \hat a} \right] = -\cosh (\beta_a - \beta_b) \, \frac{w^{\hat a \hat b}}{N} \, ,
\end{equation}
\begin{equation}
\label{eq2.51}
\omega_{\hat a \hat b \hat c} = \frac{1}{2} \, \left( e^{-\beta_a + \beta_b + \beta_c} \, C^a \, _{bc} + e^{-\beta_b + \beta_c + \beta_a} \, C^b \, _{ca} - e^{-\beta_c + \beta_a + \beta_b} \, C^c \, _{ab} \right) \, .
\end{equation}
In the last expressions, one has simply $C^a \, _{bc} = \varepsilon_{abc}$ when using the usual basis $\tau_a$ for type~IX, so that
\begin{equation}
\label{eq2.52}
\omega_{\hat a \hat b \hat c} = \frac{1}{2} \, e^{\beta_1 + \beta_2 + \beta_3} \, (e^{-2\beta_a} + e^{-2\beta_b} - e^{-2\beta_c}) \,  \varepsilon_{abc} \, .
\end{equation}
When inserting these results into the Dirac Lagrangian (\ref{eq2.13}), {\it i.e.}
\begin{equation}
\label{eq2.53}
L_D = N \, \sqrt g \left( \overline\Psi \, \frac{\gamma^{\hat 0}}{N} \, \partial_t \, \Psi + \overline\Psi \, \gamma^{\hat\gamma} \, \omega_{\hat\alpha\hat\beta\hat\gamma} \, \frac{\gamma^{\hat\alpha\hat\beta}}{4} \, \Psi - m \, \overline \Psi \Psi \right) \, ,
\end{equation}
there arise several types of terms: (a) a term involving $\partial_t \, \Psi$; (b) some terms involving $\partial_t \, \beta$; (c) terms involving the rotational velocities $w \sim \dot\theta$; (d) terms involving the purely spatial components $\omega_{\hat a \hat b \hat c}$ of the connection; and (e) the term involving the mass $m$ of the spinor field. Let us first note that the lapse cancels out in all the terms involving one time derivative ($\partial_t \, \Psi$, $\partial_t \, \beta$ or $w$), while it contributes a factor $N^{+1}$ in the ``potential'' terms (d) and (e). Let us first focus on the terms (a) and (b), {\it i.e.} those involving either $\partial_t \, \Psi$ or $\partial_t \, \beta$. They are easily found to be
\begin{eqnarray}
\label{eq2.54}
L_D &= &\sqrt g \, \overline\Psi \, \gamma^{\hat 0} \left( \dot\Psi - \frac{1}{2} \, (\dot\beta_1 + \dot\beta_2 + \dot\beta_3) \, \Psi \right) + \ldots \nonumber \\
&= &g^{\frac{1}{2}} \, \overline\Psi \, \gamma^{\hat 0} \left( \dot\Psi + \frac{\partial_t \, g^{1/4}}{g^{1/4}} \, \Psi \right) + \ldots \, ,
\end{eqnarray}
where we introduced the logarithmic derivative of $g^{\frac{1}{4}} = e^{-\frac{1}{2} (\beta_1 + \beta_2 + \beta_3)}$. This shows (as had been used in many previous works), that the replacement of the original spinor variable $\Psi$ by the rescaled spinor
\begin{equation}
\label{eq2.55}
\chi : = g^{\frac{1}{4}} \, \Psi
\end{equation}
disposes of the coupling to the $\dot\beta_a$'s. This rescaling has also the effect of absorbing the prefactor $\sqrt g$ in Eq.~(\ref{eq2.53}) into the various spinor bilinears:  $\sqrt g \, \overline\Psi (\ldots) \Psi \equiv \overline\chi (\ldots) \chi$. Note, however, that the lapse prefactor $N$ remains in factor of the bilinears of types (d) and (e), {\it i.e.} those involving no time derivatives.

\smallskip

Finally, after using the rescaling (\ref{eq2.55}) (and remembering the convention (\ref{eq2.15})), we end up with a spinor part of the Lagrangian of the form
\begin{eqnarray}
\label{eq2.56}
L_D &= &i \, \chi^{\dagger} \, \dot\chi - \cosh (\beta_1 - \beta_2) \, w^{\hat 1 \hat 2} \, \Sigma^{\hat 1 \hat 2} - \cosh (\beta_2 - \beta_3) \, w^{\hat 2 \hat 3} \, \Sigma^{\hat 2 \hat 3} \nonumber \\
&&- \, \cosh (\beta_3 - \beta_1) \, w^{\hat 3 \hat 1} \, \Sigma^{\hat 3 \hat 1} - V_{s \, {\rm grav}} - V_{s \, {\rm mass}} \, ,
\end{eqnarray}
where we have introduced the short-hand notation
\begin{equation}
\label{sigma}
\Sigma^{\hat a \hat b} := \frac{1}{2} \, \overline\chi \, \gamma^{\hat 0} \, \gamma^{\hat a \hat b} \, \chi = \frac{i}{2} \, \chi^{\dagger} \, \gamma^{\hat a \hat b} \, \chi
\end{equation}
for the spinor bilinears that couple to the rotational velocities $w^{\hat a \hat b}$ (we used (\ref{eq2.15}) in the last equation). We are including a factor $\frac{1}{2}$ in the definition (\ref{sigma}) so that the hermitian operator $\Sigma^{\hat a \hat b}$ (in the quantum theory) measures the (second quantized) spin of the spinor field $\chi$ (with eigenvalues $\pm \, \frac{1}{2}$ or $0$; see below).

\smallskip

In addition to the (body-frame) ``spin-angular-velocity'' coupling terms $\propto w^{\hat a \hat b} \, \Sigma^{\hat a \hat b}$ in the spinor Lagrangian, there are also (velocity-independent) ``spinor potential terms''\footnote{The classical analogs of these terms were discussed in Ref. \cite{ObRy}, where 
$\chi$ was treated as a c-number.},  that are naturally divided into two separate contributions:  (i) the spinor potentials $V_{s \, {\rm grav}}$ coming from the coupling to the spatial connection coefficients $\omega_{\hat a \hat b \hat c}$; and (ii) the spinor potential $V_{s \, {\rm mass}}$ coming from the Dirac mass term $m \, \overline\Psi \Psi$. The original expression (from (\ref{eq2.53})) of the gravitational-spinor potential is
\begin{equation}
\label{eq2.57}
V_{s \, {\rm grav}} = - \frac{N}{4} \, \overline\chi \, \gamma^{\hat c} \, \omega_{\hat a \hat b \hat c} \, \gamma^{\hat a \hat b} \, \chi \, .
\end{equation}

Using the gamma identity (where $\gamma^{\hat a \hat b \hat c} \equiv \gamma^{[\hat a}\gamma^{\hat b} \gamma^{\hat c]}$)
\begin{equation}
\label{eq2.58}
\gamma^{\hat c} \, \gamma^{\hat a \hat b} = \gamma^{\hat c \hat a \hat b} + \eta^{\hat c \hat a} \, \gamma^{\hat b} - \eta^{\hat c \hat b} \, \gamma^{\hat a} \, ,
\end{equation}
and the fact that all the traces of $\omega_{\hat a \hat b \hat c}$ vanish (because of the vanishing of $C^a \, _{ab}$), the potential (\ref{eq2.57}) can be rewritten, using (\ref{eq2.44}), as
\begin{equation}
\label{eq2.59}
V_{s \, {\rm grav}} = - \sum_{a,b,c} \, \frac{N}{8} \, \overline\chi \ {\mathcal C}^{\hat a} \, _{\hat b \hat c} \ \gamma^{\hat a \hat b \hat c} \, \chi \, .
\end{equation}

Using $N = \sqrt g = e^{-(\beta_1 + \beta_2 + \beta_3)}$ and (\ref{eq2.48}), this yields,
in any (class A) Bianchi type (with $n^{ab} = {\rm diag} (n_1,n_2,n_3)$)
\begin{equation}
\label{eq2.60bis}
V_{s \, {\rm grav}} = - \frac{1}{4} \, n_1 e^{-2\beta_1} \, \overline\chi \, \gamma^{\hat 1 \hat 2 \hat 3} \, \chi - \frac{1}{4} \, n_2 e^{-2\beta_2} \, \overline\chi \, \gamma^{\hat 2 \hat 3 \hat 1} \, \chi - \frac{1}{4} \, n_3 e^{-2\beta_3} \, \overline\chi \, \gamma^{\hat 3 \hat 1 \hat 2} \, \chi,
\end{equation}
where, for instance, the first term comes from the $C^1 \, _{23}$ structure constant, and is associated with the corresponding $(1;2,3)$ gravitational wall in $\beta$-space, namely $w^g_{1;23} = \beta_1 - \beta_2 - \beta_3 + \underset{a}{\sum} \, \beta_a = 2 \, \beta_1$ (when considering, as we do here, the $3$-dimensional case).
We see (in agreement with Refs.~\cite{dBHP,DaHi}) that the corresponding spinor coupling involves
\begin{equation}
\label{eq2.61}
\overline\chi \, \gamma^{\hat 1 \hat 2 \hat 3} \, \chi = i \, \chi^{\dagger} \, \gamma_{\hat 0} \, \gamma^{\hat 1 \hat 2 \hat 3} \, \chi \, .
\end{equation}
Another peculiar feature of $3$ dimensions is that all the different gravitational walls, $w^g_{1;23}$, $w^g_{2;31}$ and $w^g_{3;12}$, involve the same spinor bilinear $\overline\chi \, \gamma^{\hat 1 \hat 2 \hat 3} \, \chi = \overline\chi \, \gamma^{\hat 2 \hat 3 \hat 1} = \overline\chi \, \gamma^{\hat 3 \hat 1 \hat 2}$. This implies that, for instance in the type~IX case, the spinorial-gravitational coupling explicitly reads
\begin{equation}
\label{eq2.60}
V_{s \, {\rm grav}}^{\rm IX } = - \frac{1}{4} \, ( e^{-2\beta_1} \, +  e^{-2\beta_2} \, +e^{-2\beta_3} ) \
\overline\chi \, \gamma^{\hat 1 \hat 2 \hat 3} \, \chi ,
\end{equation}

Finally, the last term in the spinor Lagrangian (\ref{eq2.56}) reads
\begin{equation}
\label{eq2.62}
V_{s \, {\rm mass}} = + \, m \, N \, \sqrt g \ \overline\Psi \, \Psi = m \, N \, \overline\chi \, \chi = m \, e^{-(\beta_1 + \beta_2 + \beta_3)} \, \overline\chi \, \chi \, ,
\end{equation}
with 
\begin{equation}
\label{eq2.63}
\overline\chi \, \chi = i \, \chi^{\dagger} \, \gamma_{\hat 0} \, \chi \, .
\end{equation}
Contrary to what happened for $V_{s \, {\rm grav}}$, this term does not correspond to one of the gravitational walls entering $V_g (\beta)$. It would correspond to the wall form 
$w_{\Lambda} (\beta)$ associated to a cosmological constant $\Lambda$. Indeed, in the $N = \sqrt g$ gauge, $\Lambda$ generates a term $\propto \Lambda \, N \, \sqrt g = \Lambda \, e^{-2w_{\Lambda} (\beta)}$ with $w_{\Lambda} (\beta) = \underset{a}{\sum} \, \beta_a$. As usual (see, {\it e.g.}, Ref.~\cite{dBHP}) one expects the corresponding ``spinor wall'' to have a halved exponent, {\it i.e.} $\propto e^{-w_{\Lambda} (\beta)}$, which is the case of the spinor mass term (\ref{eq2.62}).

\section{Classical Hamiltonian formulation of the Einstein-Dirac Bianchi system.}

Having obtained a gauge-fixed Lagrangian formulation of the Einstein-Dirac system, let us
now show how to pass to a Hamiltonian formalism.

\subsection{From the Lagrangian to the Hamiltonian in presence of derivative couplings}\label{ssec2.6}

A (well known) peculiarity of the Einstein-Dirac Lagrangian is the presence of {\it derivative couplings} between gravity and the spinor. In our homogeneous-cosmology context (and after the rescaling (\ref{eq2.55}) of the spinor) these derivative couplings are the terms $\propto w^{\hat a \hat b} \, \Sigma^{\hat a \hat b}$ in (\ref{eq2.56}), where $w^{\hat a \hat b}$ are linear in the time derivatives of the Euler angles (see Eq.~(\ref{eq2.42})). A well-known instance of such derivative couplings is given by the coupling of a charged particle to a magnetic field say (suppressing indices)
\begin{equation}
\label{eq2.64}
L(q,\dot q) = \frac{1}{2} \, m \, \dot q^2 + e \, \dot q \, A - e \, V \, .
\end{equation}
Let us recall the effect of the derivative coupling $ e \, \dot q \, A$, when going from the Lagrangian to the Hamiltonian formalism. First, it modifies the relation between the (canonical) momentum and the velocity, namely
\begin{equation}
\label{eq2.65}
p = \frac{\partial L}{\partial \dot q} = m \, \dot q + e \, A \, .
\end{equation}
Second, $e \, A$ cancels out when computing the energy (because ``a magnetic force does no work'')
\begin{equation}
\label{eq2.66}
E(q,\dot q) := p \, \dot q - L = (m \, \dot q + e \, A) \, \dot q - \frac{1}{2} \, m \, \dot q^2 - e \, A \, \dot q + e \, V = \frac{1}{2} \, m \, \dot q^2 + e \, V \, .
\end{equation}
However, third, the $e \, A$ coupling reappears in the Hamiltonian because one must replace $\dot q$ by its expression in terms of $p$: 
\begin{equation}
\label{eq2.67}
H(q,p) = [E(q,\dot q)]_{\dot q(p)} = \left[ \frac{1}{2} \, m \, \dot q^2 + e \, V \right]_{\dot q (p)} = \frac{(p-e \, A)^2}{2m} + e \, V \, .
\end{equation}
This mechanism is easily seen to hold for any derivative coupling which is linear in time derivatives. [In a more general case the mass $m$ in (\ref{eq2.64}) becomes some $q$-dependent quadratic form, and the $m^{-1}$ factor in $H(q,p)$ becomes the inverse quadratic form.] The crucial end result is that the Hamiltonian is numerically equal to the sum of the original kinetic energy terms (without the velocity coupling term $e \, \dot q \, A$) and of the potential energy, but with the replacement of the concerned velocities by their expressions in terms of their ($e \, A$-shifted) canonical momenta.

\subsection{Hamiltonian formulation of the Einstein-Dirac system}\label{ssec2.7}

We can now write down the explicit Hamiltonian of the Einstein-Dirac system, for homogeneous configurations, when using the gauge-fixed vielbein (\ref{eq2.4}), (\ref{eq2.5}) (with zero shift vector). More precisely, the Hamiltonian action density has the form
\begin{eqnarray}
\label{eq2.68}
L_{\rm Ham} &= &\sum_a \, \pi_{\beta_a} \, \dot\beta_a + \sum_a p_{\theta_a} \, \dot\theta_a + i \, \chi^{\dagger} \, \dot\chi \nonumber \\
&&- \, \widetilde N \, \widetilde{\mathcal H} (\beta , \pi_{\beta} , \theta , p_{\theta} , \chi^{\dagger} , \chi) - N^a \, {\mathcal H}_a (\beta , \pi_{\beta} , \theta , p_{\theta} , \chi^{\dagger} , \chi) \, ,
\end{eqnarray}
where we have introduced the canonical momenta $\pi_{\beta_a} , p_{\theta_a}$, respectively conjugated to the three $\beta$'s, and to the three Euler angles $\theta_a$, and where $\widetilde N$ denotes the rescaled lapse $\widetilde N := N/\sqrt g$. We shall work in a gauge where $\widetilde N = 1$ and $N^a = 0$, which makes the Hamiltonian in the Hamiltonian action, $L_{\rm Ham} = p \, \dot q - H$, simply equal to $\widetilde{\mathcal H}$. Note that $\widetilde{\mathcal H} \equiv \sqrt g \, {\mathcal H}$ where ${\mathcal H}$ is the usual Arnowitt-Deser-Misner Hamiltonian density entering Eq.~(\ref{eq2.21}) (${\mathcal H}$ is a spatial density of weight $+1$, while $\widetilde{\mathcal H}$ has weight $+2$). Note also that we did not introduce any notation for the conjugate momentum of $\chi$ as the Dirac action is first order, {\it i.e.} already in $p \, \dot q - H$ form (with $p = i \, \chi^{\dagger}$ when $q=\chi$). Using the results above (notably in the last Section) the value of $\widetilde{\mathcal H}$ is the sum of kinetic-energy and potential-energy terms: 
\begin{eqnarray}
\label{eq2.69}
\widetilde{\mathcal H} &= &T_{\beta} (\pi_{\beta}) + T_w (\beta , \pi_w , \chi , \chi^{\dagger}) + V_g (\beta) \nonumber \\
&&+ \, V_{s \, {\rm grav}} (\beta , \chi , \chi^{\dagger}) + V_{s \, {\rm mass}} (\beta , \chi , \chi^{\dagger}) \, .
\end{eqnarray}
The various potential terms are the same as written above:  $V_g (\beta)$ is given by Eq.~(\ref{eq2.30}), $V_{s \, {\rm grav}} (\beta , \chi , \chi^{\dagger})$ by Eq.~(\ref{eq2.60bis}), and $V_{s \, {\rm mass}} (\beta , \chi , \chi^{\dagger})$ by Eq.~(\ref{eq2.62}). [One should use $\overline\chi := i \, \chi^{\dagger} \, \gamma_{\hat 0}$ to replace $\overline \chi$ in terms of $\chi^{\dagger}$.]

\smallskip

The $\beta$-kinetic-term\footnote{When discussing the $\beta$-kinetic-term in the Lagrangian ($L= T-V$), we do not take out a factor $\widetilde N$,
as we do in discussing the corresponding term in the Hamiltonian.} is originally given (in any dimension, and for any time gauge, i.e. any value of
$\widetilde N := N/\sqrt g$) by
\begin{equation}
\label{eq2.70}
T_{\beta} (\dot\beta) = \frac{1}{\widetilde N} \, G_{ab} \, \dot\beta_a \, \dot\beta_b \equiv \frac{1}{\widetilde N} \,\left[ \sum_a \dot\beta_a^2 - \left( \sum_a \dot\beta_a \right)^2 \right]\, ,
\end{equation}
which defines the $\beta$-space metric $G_{ab}$. [Though we put the $a$ index on $\beta$
as a subscript, one should think of it as a contravariant index $\beta^a$.]
After the rescaling (\ref{eq2.55}) of the spinor field, $\dot\beta$ appears only in the $\beta$ kinetic term, so that the conjugate momenta $\pi_{\beta_a}$ to the $\beta_a$'s are given by
\begin{equation}
\label{eq2.71}
\pi_{\beta_a} = \frac{2}{\widetilde N} \, G_{ab} \, \dot\beta_b \, .
\end{equation}
This leads to the following value for the $\beta$ kinetic energy expressed in terms of $\pi_{\beta}$
\begin{equation}
\label{eq2.72}
T_{\beta} (\pi_{\beta}) = \frac{1}{4} \, G^{ab} \, \pi_{\beta_a} \, \pi_{\beta_b} \, ,
\end{equation}
where $G^{ab}$ is the inverse of $G_{ab}$, {\it i.e.} (in our $3+1$ dimensional case)
\begin{equation}
\label{eq2.73}
G^{ ab} \, \pi_{\beta_a} \, \pi_{\beta_b} = \sum_a \pi_{\beta_a}^2 - \frac{1}{2} \left( \sum_a \pi_{\beta_a}\right)^2 \, .
\end{equation}
Note that, if we were in a space dimension $d \ne 3$, the coefficient of the second term on the r.h.s. of (\ref{eq2.73}) would be $\frac{1}{d-1}$ instead of $\frac{1}{2}$.

\smallskip

It remains to discuss the rotational kinetic term $T_w$ linked to the angular motion parametrized by the three Euler angles $\theta_a$. To simplify
this discussion, we shall henceforth come back to using the time gauge (\ref{gauge}), as we did in Section 2. 
[The formulas above for the $\beta$ kinetic terms 
were written in a general time gauge as a reminder of the various occurrences of $N$ and $\sqrt g$ in the
kinetic part of the action.]
As the rotational velocities $w^{\hat a \hat b} \sim \dot\theta_c$ are linearly coupled to $\chi$ {\it via} the spinor bilinears $\Sigma^{\hat a \hat b}$ (see Eq.~(\ref{eq2.56})), we must apply the result of the previous Section to this term. To ease the writing of the Hamiltonian version of the rotational kinetic term $T_w$ it is convenient to introduce the following momentum-like variables
\begin{equation}
\label{eq2.74}
\pi_{w^{ab}} := \frac{\partial L}{\partial \, w^{ab}}
\end{equation}
{\it i.e.} explicitly (in the type IX case)
\begin{equation}
\label{eq2.75}
\pi_{w^{ab}} = 4 \sinh^2 (\beta_a - \beta_b) \, w^{ab} - \cosh (\beta_a - \beta_b) \, \Sigma^{ab} \, ,
\end{equation}
as deduced from Eqs.~(\ref{eq2.39}) and (\ref{eq2.56}). Here, and henceforth, we have simplified the notation by dropping the carets over the (``flat'') indices $a,b$ on $w^{ab}$ and $\Sigma^{ab}$. [Note that the partial derivatives in Eq.~(\ref{eq2.74}) are done w.r.t. the three {\it restricted} independent components $w^{12}$, $w^{23}$, $w^{31}$, of $w$; the other ones being defined in terms of these by, {\it e.g.}, $w^{21} := -w^{12}$, {\it etc.}]

\smallskip

Solving Eq.~(\ref{eq2.75}) for the $w$'s as functions of the $\pi_w$'s, then yields the following Hamiltonian version $(\sim (p-eA)^2 / (2m))$ of the rotational kinetic term
\begin{eqnarray}
\label{eq2.76}
T_w (\beta , \pi_w , \chi , \chi^{\dagger}) &= &\frac{1}{8 \sinh^2 (\beta_1 - \beta_2)} \, [\pi_{w^{12}} + \cosh (\beta_1 - \beta_2) \, \Sigma^{12}]^2 \nonumber \\
&+ &\frac{1}{8 \sinh^2 (\beta_2 - \beta_3)} \, [\pi_{w^{23}} + \cosh (\beta_2 - \beta_3) \, \Sigma^{23}]^2 \nonumber \\
&+ &\frac{1}{8 \sinh^2 (\beta_3 - \beta_1)} \, [\pi_{w^{31}} + \cosh (\beta_3 - \beta_1) \, \Sigma^{31}]^2 \, .
\end{eqnarray}

If one wanted to express the Hamiltonian ${\mathcal H}$ completely in terms of canonically conjugated variables, one should replace the non-canonical, momentum-like variables $\pi_w$ by the linear combinations of the Euler-angle conjugate momenta $p_{\theta_a}$ (see Eq.~(\ref{eq2.68})) that they represent. As $\pi_w \, w = w \, \partial L / \partial \, w = \dot\theta \, \partial L / \partial \, \dot\theta$ $= p_{\theta} \, \dot\theta$, the transpose of the matrix $A$ appearing on the r.h.s. of (\ref{eq2.42}) yields the link $p_{\theta} = A^T \pi_w$, namely
\begin{eqnarray}
\label{eq2.77}
p_{\varphi} &= &\cos \theta \, \pi_{w^{12}} + \sin \theta \sin \psi \, \pi_{w^{23}} + \sin \theta \cos \psi \, \pi_{w^{31}} \, , \nonumber \\
p_{\theta} &= &\cos \psi \, \pi_{w^{23}} - \sin \psi \, \pi_{w^{31}} \, , \nonumber \\
p_{\psi} &= &\pi_{w^{12}} \, ,
\end{eqnarray}
whose inverse reads
\begin{eqnarray}
\label{eq2.78}
\pi_{w^{12}} &= &p_{\psi} \, , \nonumber \\
\pi_{w^{23}} &= &\sin \psi \, \frac{p_{\varphi} - \cos \theta \, p_{\psi}}{\sin \theta} + \cos \psi \, p_{\theta} \, , \nonumber \\
\pi_{w^{31}} &= &\cos \psi \, \frac{p_{\varphi} - \cos \theta \, p_{\psi}}{\sin \theta} - \sin \psi \, p_{\theta} \, .
\end{eqnarray}

The canonical Poisson brackets $\{ \theta_a , \theta_b \} = 0 = \{p_{\theta_a} , p_{\theta_b} \}$ and $\{ \theta_a , p_{\theta_b} \} = \delta_{ab}$ are easily found to imply the following brackets for the non-canonical variables $\pi_w$: 
\begin{eqnarray}
\label{eq2.79}
\{ \pi_{w^{12}} , \pi_{w^{23}} \} &= &- \, \pi_{w^{31}} \, , \nonumber \\
\{ \pi_{w^{23}} , \pi_{w^{31}} \} &= &- \, \pi_{w^{12}} \, , \nonumber \\
\{ \pi_{w^{31}} , \pi_{w^{12}} \} &= &- \, \pi_{w^{23}} \, .
\end{eqnarray}
Note that these are the {\it opposite-sign} brackets, compared to those of a usual angular momentum vector:  $\{ L_x , L_y \}$ $= + \, L_z$, {\it etc.} Indeed, in the analogy of (\ref{eq2.76}) (without the spin term) with the Hamiltonian of an asymmetric top, {\it i.e.}
\begin{equation}
\label{eq2.80}
H_{\rm top} = \frac{L_1^2}{2 I_1} + \frac{L_2^2}{2 I_2} + \frac{L_3^2}{2 I_3} \, ,
\end{equation}
the quantities $L_1 , L_2 , L_3$ (which are analogous to $\pi_{w^{23}} , \pi_{w^{31}} , \pi_{w^{12}}$) represent the (non-conserved) body-frame components of the angular momentum, which have {\it opposite-type} brackets, namely $\{ L_1 , L_2 \} = - L_3$, {\it etc.} This change of sign is linked to the fact that the body-frame $L_1 , L_2 , L_3$ are related to the (normal-bracket, and conserved) space-frame $L_x , L_y , L_z$ by the time-dependent matrix $S$ effecting the rotation between the space-frame and the body-frame. 

\smallskip

For most purposes, the expression (\ref{eq2.76}) of the rotational energy in terms of the $\pi_w$'s, together with the knowledge of the Poisson brackets (\ref{eq2.79}) between the $\pi_w$'s, suffices to discuss the Hamiltonian dynamics of the system.

\smallskip

Summarizing:  the total Hamiltonian of the gravity-spinor system is given by Eq.~(\ref{eq2.69}), together with Eq.~(\ref{eq2.72}) $[T_{\beta} (\pi_{\beta})]$, Eq.~(\ref{eq2.76}) $[T_w (\beta , \pi_w , \chi)]$, Eq.~(\ref{eq2.30}) $[V_g (\beta)]$, Eq.~(\ref{eq2.60}) $[V_{s \, {\rm grav}} (\beta , \chi)]$ and Eq.~(\ref{eq2.62}) $[V_{s \, {\rm mass}} (\beta , \chi)]$. The Poisson brackets between the various variables is given by canonical pairs appearing in the $p \dot q$ terms in Eq.~(\ref{eq2.68}). [See also (\ref{eq2.79}).] When considering ``classical'' spinor variables, one should use an odd, Grassmann-valued $\chi$, and define a graded (anti-) bracket under which $\chi$ is conjugated to $i \, \chi^{\dagger}$. The quantum case will be discussed in detail below.

\subsection{Diffeomorphism constraints}\label{ssec2.8}

For simplifying the calculations, we have assumed above that we were working in a gauge where the shift vector $N^a$ vanishes. However, as recalled in Eq.~(\ref{eq2.68}), when we relax this assumption we know that $N^a$ will simply enter as a Lagrange multiplier of the diffeomorphism constraint ${\mathcal H}_a \approx 0$, where (see Eq.~(\ref{eq2.22})
\begin{equation}
\label{eq2.81}
{\mathcal H}_a = \sqrt g \, (2 \ ^4R_a^0 - T_a^0) = -2 \, D^b \, \pi_{ab}^{\rm kin} - \sqrt g \, T_a^0 \, .
\end{equation}
Here, the tensorial objects refer to the spatial metric $g_{ab}$ ({\it e.g.} $D_c \, g_{ab} \equiv 0$), and all indices are projected on the co-frame $\tau^a$, and its dual. In addition, $\pi_{\rm kin}^{ab}$ denotes the kinematical part of the conjugate momentum $\pi^{ab}$ of the metric $g_{ab}$. In the electromagnetic model discussed above, this would be the $m \, \dot q$ part, without the $e \, A$ shift in Eq.~(\ref{eq2.65}). In our case, $\pi_{\rm kin}^{ab}$ denotes the part of $\pi^{ab}$ proportional to the second fundamental form $K$, {\it i.e.} (in covariant form) $\pi_{ab}^{\rm kin} =\sqrt{g} ( K_{ab} - g_{ab} \, K^c \, _c)$, when using the $+ \, \dot g_{ab}$ convention (\ref{eq2.24}) for the definition of $K_{ab}$. Finally, the matter term in (\ref{eq2.81}) comes, in our case, from the stress-energy tensor of the spinor field. From Ref.~\cite{HenIHP}, the explicit expression of the r.h.s. of (\ref{eq2.81}) for a Bianchi spacetime reads
\begin{equation}
\label{eq2.82}
{\mathcal H}_a = 2 \, \pi_{\rm kin}^{bd} \, g_{bc} \, C^c \, _{ad} + \frac{1}{2} \, \Sigma^d \, _c \ C^c \, _{ad} \, ,
\end{equation}
where $\Sigma^d \, _c := h^d \, _{\hat a} \ h_{\hat b c} \, \Sigma^{\hat a \hat b}$.

\smallskip

Inserting the decomposition $h^{\hat a} \, _b = e^{-\beta_a} \, S^{\hat a} \, _b$ of the dreibein (with $g_{cd} = \delta_{\hat a \hat b} \, h^{\hat a} \, _c \ h^{\hat b} \, _d$), and transforming all tensor indices to the intermediate frame $\overline\tau^{\overline a} = S^{\hat a} \, _b \ \tau^b$, except the indices on $\Sigma^{\hat a \hat b}$ which must remain ``flat'' ({\it i.e.} referring to $\theta^{\hat a} = e^{-\beta_a} \, \overline\tau^{\overline a}$), one finds that (\ref{eq2.82}) becomes
\begin{equation}
\label{eq2.83}
\overline{\mathcal H}_{\overline m} := S^a \, _{\hat m} \ {\mathcal H}_a = \sum_{\overline p , \overline q} \left( 2 \, \overline\pi_{\rm kin}^{\overline p \overline q} \, e^{-2 \beta_p} + \frac{1}{2} \, e^{\beta_q - \beta_p} \, \Sigma^{\hat q \hat p} \right) \overline C^{\overline p} \, _{\overline m \overline q} \, .
\end{equation}
Note that, because of the covectorial nature of ${\mathcal H}_a$, the definition of $\overline{\mathcal H}_{\overline m}$ involves the inverse of the matrix $S^{\hat a} \, _b$. The automorphism property of the matrix $S$ guarantees that the $\overline\tau$-frame structure constants $\overline C$ entering (\ref{eq2.83}) are numerically equal to the original, $\tau$-frame, structure constants $C$. In addition, the rewriting of the $\pi^{ab} \, \dot g_{ab}$ term in the Hamiltonian action as $\underset{a}{\sum} \, \pi_{\beta_a} \, \dot\beta_a + \pi_{w^{12}} \, w^{12} + \pi_{w^{23}} \, w^{23} + \pi_{w^{31}} \, w^{31}$, shows that the components $\overline\pi_{\rm kin}^{\overline 1 \, \overline 2} = \overline\pi_{\rm kin}^{\overline 2 \, \overline 1}$ are linked to $\pi_{w^{12}}^{\rm kin}$ {\it via}
\begin{equation}
\label{eq2.84}
\pi_{w^{12}}^{\rm kin} = 2 \, (e^{-2 \beta_1} - e^{-2 \beta_2}) \, \overline\pi_{\rm kin}^{\overline 1 \, \overline 2} \, ,
\end{equation}
and similarly for the cyclic permutations $23$ and $31$. This implies that Eq.~(\ref{eq2.83}) reads, say for $\overline m = \overline 3$,
\begin{equation}
\label{eq2.85}
\overline{\mathcal H}_{\overline 3} = - \, \pi_{w^{12}}^{\rm kin} + \cosh (\beta_1 - \beta_2) \, \Sigma^{\hat 1 \hat 2} \, .
\end{equation}

From Eq.~(\ref{eq2.75}), we see that the last, spin contribution in (\ref{eq2.85}) has the correct magnitude and sign to transform the kinematic contribution $\pi_{w^{12}}^{\rm kin} \equiv 4 \sinh^2 (\beta_1 - \beta_2) \, w^{12}$ into the full momentum-like variable $\pi_{w^{12}}$. Finally, this shows 
(in agreement with  \cite{Jantzen:1982je}) that $\overline{\mathcal H}_{\overline 3} = - \, \pi_{w^{12}}$, {\it i.e.}, more explicitly
\begin{equation}
\label{eq2.86}
S^a \, _{\hat m} \ {\mathcal H}_a = - \, \pi_{w^{\hat p \hat q}} \, ,
\end{equation}
where $\hat m \, \hat p \, \hat q$ is a cyclic permutation of $123$. In other words, in the analogy where $\pi_{w^{23}} = L_1$, {\it etc.} are the {\it body-frame} angular momenta, we have that ${\mathcal H}_1 = - \, L_x$, {\it etc.} are (minus), the {\it space-frame} angular momenta. As in the asymmetric top situation, the ${\mathcal H}_a$, contrary to the $\pi_w$, are conserved because of the first-class bracket $\{ {\mathcal H}_a , {\mathcal H} \} = 0$. Let us also note that the diffeomorphism constraints satisfy the Poisson-bracket algebra
\begin{equation}
\label{eq2.87}
\{ {\mathcal H}_a , {\mathcal H}_b \} = \, - \, C^c \, _{ab} \ {\mathcal H}_c \, .
\end{equation}
[Here, the minus sign comes from the minus sign in the relation ${\mathcal H}_1 = - \, L_x$.]
The simple link (\ref{eq2.86}) between the diffeomorphism constraints and the (body-frame) angular momenta $\pi_w$ is not an accident but derives from basic symmetry properties. Indeed, while the three quantities ${\mathcal H}_a$ generate (in phase space) adjoint transformations of the homogeneity group $G$ \cite{HenIHP}, the three angular momenta $\pi_w$ generate, by definition, the group of the matrices $S(\theta_1 , \theta_2 , \theta_3)$ used in the parametrization (\ref{eq2.5}), (\ref{eq2.6}). However, for simple groups, such as type~IX and type~VIII, the latter group coincides with the adjoint representation of $G$.

\subsection{Explicit forms of some (class A) Bianchi Hamiltonians}\label{ssec2.9}

The full Hamiltonian action (\ref{eq2.68}) implies two types of constraints: (i) the diffeomorphism constraints ${\mathcal H}_a \approx 0$ linked to the arbitrariness of the shift vector $N^a$, and (ii) the Hamiltonian constraint $\widetilde{\mathcal H} \approx 0$ linked to the arbitrariness of the rescaled lapse $\widetilde N = N/\sqrt g$. When considering general (class A) Bianchi models, the expression and the number of effective diffeomorphism constraints strongly depend on the structure constraints of the homogeneity group. Let us summarize the results for the various Bianchi types.

\smallskip

In the Bianchi type~I case, {\it i.e.} when $C^a \, _{bc} = 0$, we see on the general expression (\ref{eq2.82}) that ${\mathcal H}_a \equiv 0$, {\it i.e.} that there are no diffeomorphism constraints. In that case, one can still conventionally decide to use the decomposition (\ref{eq2.6}) with a matrix $S (\theta_1 , \theta_2 , \theta_3) \in SO(3)$. One then ends up with an Hamiltonian action of the form (\ref{eq2.68}), except for the last term $N^a \, {\mathcal H}_a$ which is absent. This type~I action describes the coupled dynamics of the variables $\beta_1 , \beta_2 , \beta_3$, $\theta_1 , \theta_2 , \theta_3$, and $\chi$. The explicit form of the Hamiltonian is given (when $\widetilde N = N/\sqrt g$) by the r.h.s. of (\ref{eq2.69}), where, however, several terms vanish because of the vanishing of the $C$'s. Specifically, the two potential terms in (\ref{eq2.69}) linked to gravitational walls, namely $V_g (\beta)$ and $V_{s \, {\rm grav}} (\beta , \chi)$, are identically zero in type~I. Finally, the type~I Hamiltonian has the form
\begin{equation}
\label{eq2.88}
\widetilde{\mathcal H}^I = T_{\beta} (\pi_{\beta}) + T_w (\beta , \pi_w , \chi , \chi^{\dagger}) + V_{s \, {\rm mass}} (\beta , \chi , \chi^{\dagger})
\end{equation}
where $T_{\beta} (\pi_{\beta})$ is (universally) given by Eq.~(\ref{eq2.72}), $V_{s \, {\rm mass}} (\beta , \chi)$ by Eq.~(\ref{eq2.62}), and where $T_w (\beta , \pi_w , \chi)$ is given by the full expression (\ref{eq2.76}). Note that this Hamiltonian (which must be submitted to the single constraint $\widetilde{\mathcal H} \approx 0$) describes the coupled dynamics of the variables $\beta_1 , \beta_2 ,\beta_3$; $\varphi , \theta , \psi$, and $\chi$. It is a generalized asymmetric top dynamics (for the rotational degrees of freedom $\varphi , \theta , \psi$), where the moments of inertia $I_3 (\beta) = 4 \sinh^2 (\beta_1 - \beta_2)$, {\it etc.} have their own (coupled) dynamics, and which includes a coupling between the (bosonic) rotor and spinor degrees of freedom. As our main physical focus here concerns type~IX, we shall not discuss in detail the type~I dynamics. Let us only note that Ref.~\cite{BK} has explicitly solved the equations of motion of type~I dynamics, when using a different way of fixing the rotational state of the dreibein, and treating $\chi$ as a classical quantity.

\smallskip

The Bianchi type~II case cannot be directly described by the expressions derived above from the $SO(3)$ parametrization (\ref{eq2.6}) because (contrary to the type~I and type~IX cases) its structure constants are not invariant under the full $SO(3)$ group. Refs.~\cite{Jantzen:1982je,Jantzen79,Jantzen:2001me} has indicated other useful choices of three-dimensional matrix groups leaving invariant the $C$'s. To give a different perspective on the type~II case, and relate it to recent work on cosmological billiards, we treat the type~II case in an Appendix by using an Iwasawa decomposition of the metric, {\it i.e.} an upper triangular matrix $S$ in Eqs.~(\ref{eq2.5}) and (\ref{eq2.6}).

\smallskip

Leaving aside a discussion of the other {\it degenerate} class-A Bianchi types ({\it i.e.} VI$_0$ and VII$_0$), let us come back to the main focus of this work, namely the non-degenerate types~IX and VIII. In both cases, the link (\ref{eq2.86}) between the three diffeomorphism constraints, and the three body-frame angular momenta $\pi_w$, holds. [As mentioned above, in type~VIII the rotational angles and momenta refer to $SO(1,2)$ with metric ${\rm diag} (+1,-1,-1)$, rather than to $SO(3)$ with ${\rm diag} (+1,+1,+1)$.] We can therefore greatly simplify the dynamics by reducing the Hamiltonian by imposing the (first class) constraints ${\mathcal H}_a = 0$, {\it i.e.} $\pi_w = 0$, thereby eliminating both the $p_{\theta} \, \dot\theta$ terms in Eq.~(\ref{eq2.68}) and the $N^a \, {\mathcal H}_a$ contribution. This leads to a reduced Hamiltonian action of the simpler form
\begin{equation}
\label{eq2.89}
L_{\rm Ham}^{\rm VIII,IX} = \sum_a \pi_{\beta_a} \, \dot\beta_a + i \, \chi^{\dagger} \, \dot\chi - \widetilde N \, \widetilde{\mathcal H}^{\rm VIII,IX} (\beta , \pi_{\beta} , \chi^{\dagger} , \chi) \, ,
\end{equation}
where ($Z = {\rm VIII}, {\rm IX}$ being a label for the Bianchi type)
\begin{equation}
\label{eq2.90}
\widetilde{\mathcal H}^Z := T_{\beta} (\pi_{\beta}) + T_w^{(0)} (\beta , \chi) + V_g^Z (\beta) + V_{s \, {\rm grav}}^Z (\beta , \chi) + V_{s \, {\rm mass}} (\beta , \chi) \, .
\end{equation}
Here: $T_{\beta} (\pi_{\beta})$ and $V_{s \, {\rm mass}} (\beta , \chi)$ have the universal structure given above; $V_g^Z (\beta)$ is given for any Bianchi type (including degenerate ones) by Eq.~(\ref{eq2.29}); $V_{s \, {\rm grav}}^Z$ is given for any Bianchi type by Eq.~(\ref{eq2.59}), with ${\mathcal C}^{\hat a} \, _{\hat b \hat c}$ given by Eq.~(\ref{eq2.48}) (with $\overline C = C$ when $S$ leaves the $C$'s invariant, as is the case for $Z = {\rm VIII}$ and IX); and, finally, the rotational energy term takes, in type IX, the simplified form obtained by replacing $\pi_w \to 0$ in (\ref{eq2.76})
\begin{eqnarray}
\label{eq2.91}
T_w^{(0){\rm IX}} (\beta , \chi) &= &\frac{1}{8} \, \coth^2 (\beta_1 - \beta_2) (\Sigma^{12})^2 + \frac{1}{8} \coth^2 (\beta_2 - \beta_3)(\Sigma^{23})^2 \nonumber \\
&&+ \, \frac{1}{8} \coth^2 (\beta_3 - \beta_1) (\Sigma^{31})^2 \, .
\end{eqnarray}
The type VIII case is obtained by particularizing to the case $ (n_1, n_2, n_3) = (+1,+1,-1)$
the result for a general $n^{ab} = {\rm diag} (n_1, n_2, n_3)$, which reads

\begin{equation}
\label{eq2.92}
T_w^{(0)} (\beta , \chi) =  \frac{1}{8} \,  \left( \frac{ n_1 e^{\beta_2-\beta_1} + n_2 e^{\beta_1-\beta_2}  }{  n_1 e^{\beta_2-\beta_1} - n_2 e^{\beta_1-\beta_2}  } \right)^2  (\Sigma^{12})^2  + {\rm cyclic} \, .
\end{equation}

The Hamiltonian (\ref{eq2.90})--(\ref{eq2.92}) describes the coupled dynamics of the diagonal degrees of freedom of the metric coupled to a spinor. In the next Section we shall discuss its quantization.

\section{Quantum formulation of the coupled spinor-Bianchi-IX system}\label{sec3}
\setcounter{equation}{0}

\subsection{Dependence on the Euler angles}\label{ssec3.1} We shall denote by $\Phi$ the abstract quantum state of the system. Following Dirac, we interpret the various classical constraints, ${\mathcal H} \approx 0$, ${\mathcal H}_a \approx 0$ as constraints on $\Phi$ of the type
\begin{equation}
\label{eq3.1}
\widehat{\mathcal H} \, \Phi = 0 \, ,
\end{equation}
\begin{equation}
\label{eq3.2}
\widehat{\mathcal H}_a \, \Phi = 0 \, ,
\end{equation}
where $\widehat{\mathcal H}$ and $\widehat{\mathcal H}_a$ are suitably defined operatorial versions of ${\mathcal H}$ and ${\mathcal H}_a$. We shall work in the gravity configuration space $\beta_1 , \beta_2 , \beta_3$; $\theta_1 , \theta_2 , \theta_3$, together with a suitable description of the spinor degrees of freedom, labelled, say, by $\sigma$ (see below). This leads to a wave function of the universe described by $\Phi (\beta , \theta , \sigma)$. Let us start by remarking that the classical link (\ref{eq2.86}) naturally suggests an ordering such that Eq.~(\ref{eq3.2}) becomes
\begin{equation}
\label{eq3.3}
\widehat\pi_{w^{ab}} \, \Phi (\beta , \theta , \sigma) = 0 \, .
\end{equation}

The classical angular momenta $\pi_w$ are linear combinations of the canonical angular momenta $p_{\theta_a}$, see Eq.~(\ref{eq2.78}). The canonical quantization $\widehat p_{\theta_a} = -i \, \partial / \partial \, \theta_a$ (with $\hbar = 1$), together with the natural ordering of the $\widehat\pi_w$ as differential operators on the $SO(3)$ space\footnote{Note that the $\theta$'s parametrize $SO(3)$ matrices, rather than the $SU(2)$ group $G$ which constitutes the cosmological space. The ``factor $2$'' difference between $SO(3)$ and $SU(2)$ might have been important, if we had needed to consider situations where $\widehat\pi_w \, \Phi \ne 0$.} naturally leads to concluding that the quantum constraint (\ref{eq3.3}) means that $\Phi (\theta_1 , \theta_2 , \theta_3)$ does not depend on the Euler angles: $\partial_{\theta_a} \Phi = 0$. More directly, we can say that Eq.~(\ref{eq3.3}) means that $\Phi$ is a zero-angular-momentum state on $SO(3)$; {\it i.e.} an ``$s$-wave''. As $\Phi$ does not depend on the $\theta$'s, we can simply ignore them in the following. Note that this is consistent with having started the quantization procedure at the level of the final, reduced type~IX Hamiltonian, namely Eq.~(\ref{eq2.90}), which does not contain the $\theta$'s. From this point of view, the quantum description is cleaner than the classical one. Indeed, in the classical description the Euler angles satisfy  first-order differential equations obtained by setting to zero the l.h.s. of Eq.~(\ref{eq2.75}). This yields a non-trivial dynamics for the $\theta$'s sourced by the spinor bilinears $\propto \Sigma^{ab} (t)$, and influenced by the time-dependence of the $\beta$'s. In the quantum description, the $\theta$'s do not appear at all, which is nicely consistent with the fact that they have the character of gauge parameters.

\subsection{Quantum description of the spinor degrees of freedom}\label{ssec3.2}

After the rescaling (\ref{eq2.55}) of the spinor, the spinor kinetic term is simply
\begin{equation}
\label{eq3.4}
L_s^{\rm kin} = i \, \chi^{\dagger} \dot\chi \equiv i \sum_{\alpha = 1}^4 \chi_{\alpha}^{\dagger} (t) \, \dot \chi_{\alpha} (t) 
\end{equation}
where we made explicit the Dirac-spinor index $\alpha = 1,2,3,4$, which was kept implicit up to now. In the case where $\chi$ is a generic {\it Dirac spinor}, with $4$ independent {\it complex} components, the quantization of (\ref{eq3.4}) is done by the standard anticommutator result (with $\hbar = 1$)
\begin{equation}
\label{eq3.5}
[\chi_{\alpha} , \chi_{\beta}]_+ = 0 = [\chi_{\alpha}^{\dagger} , \chi_{\beta}^{\dagger}]_+ \, ; \quad [\chi_{\alpha} , \chi_{\beta}^{\dagger}]_+ = \delta_{\alpha\beta} \, ,
\end{equation}
where $[A,B]_+ \equiv AB + BA$ denotes an anticommutator, and where all the $\chi$'s are taken at the same instant $t$. In other words, each one of the four $\chi_{\alpha}$'s can be viewed (at each instant $t$) as an independent (complex) fermionic destruction operator, with $\chi_{\alpha}^{\dagger}$ being the corresponding fermionic creation operator. To each pair $(\chi_{\alpha} , \chi_{\alpha}^{\dagger})$ then corresponds  a two-dimensional (complex) Hilbert space, so that to the $4$ mutually anticommuting pairs $(\chi_{\alpha} , \chi_{\alpha}^{\dagger})$, $\alpha = 1,\ldots , 4$ correspond  a fermionic Hilbert space of dimension $2^4 = 16$.

\smallskip

We can, however, simplify the problem by demanding, from the start, that $\chi$ be a {\it Majorana spinor}, {\it i.e.} be restricted to contain only $4$ independent {\it real} components. More precisely, let us use the following explicit Majorana representation of the $4$ Dirac gamma matrices $\gamma^{\widehat\alpha}$ entering the spinor action:
$$
\gamma^{\widehat 0} = \begin{pmatrix} 0 &i \, \sigma_2 \\ i \, \sigma_2 &0 \end{pmatrix} \, , \ \gamma^{\widehat 1} = \begin{pmatrix}\sigma_1 &0 \\ 0 &\sigma_1 \end{pmatrix} \, , \ \gamma^{\widehat 2} = \begin{pmatrix}-\sigma_3 &0 \\ 0 &-\sigma_3 \end{pmatrix} \, , 
$$
\begin{equation}
\label{eq3.6}
\gamma^{\widehat 3} = \begin{pmatrix} 0 &-i \, \sigma_2 \\ i \, \sigma_2 &0 \end{pmatrix} \, ,
\end{equation}
where $\sigma_1 = \begin{pmatrix} 0 &1 \\ 1 & 0 \end{pmatrix}$, $\sigma_2 = \begin{pmatrix} 0 &-i \\ i &0 \end{pmatrix}$, $\sigma_3 = \begin{pmatrix} 1 &0 \\ 0 &-1 \end{pmatrix}$ are the standard Pauli matrices. The four gamma matrices (\ref{eq3.6}) are all real, with the 3 spatial $\gamma$'s being symmetric (and therefore hermitian), while $\gamma^{\hat 0}$ is anti-symmetric ({\it i.e.} anti-hermitian).

\smallskip

When using a real representation such as (\ref{eq3.6}), the reality condition for a Majorana spinor is simply $\chi^* = \chi$, {\it i.e.} the condition that each component $\chi_{\alpha}$ be real ({\it i.e.} hermitian, as an operator). In that case, the quantization of the standard spinor kinetic term (\ref{eq3.4}), {\it i.e.} $i \, \underset{\alpha}{\sum} \,  \chi_{\alpha} \, \dot\chi_{\alpha}$, involves an extra factor $\frac{1}{2}$, namely
\begin{equation}
\label{eq3.7}
[\chi_{\alpha} , \chi_{\beta}]_+ \equiv \chi_{\alpha} \, \chi_{\beta} + \chi_{\beta} \, \chi_{\alpha} = \frac{1}{2} \, \delta_{\alpha\beta} \, .
\end{equation}
The need for the factor $\frac{1}{2}$ in the quantization condition (\ref{eq3.7}) can be viewed in several ways. The most direct way is that the normalization of the basic field commutators should be chosen so that the universal Heisenberg-representation equation of motion for operators (with $\hbar = 1$)
\begin{equation}
\label{eq3.8}
i \ \frac{d}{dt} \, Q = [Q,H]
\end{equation}
(where $[A,B] \equiv AB-BA$ denotes a commutator) hold true, and be consistent with the usual Euler-Lagrange equations of motion. As the Euler-Lagrange variation of the kinetic term $\delta \int dt \, i \, \chi \, \dot\chi$ (for a real, anticommuting $\chi$) involves a factor 2 (after integrating by parts), one needs the factor $\frac{1}{2}$ in (\ref{eq3.7}).

\smallskip

An indirect check consists of decomposing a general, complex Dirac spinor, with kinetic term (\ref{eq3.4}), into real ({\it i.e.} Majorana) and imaginary parts, say $\chi = \chi_R + i \, \chi_I$, $\chi^{\dagger} = \chi_R - i \, \chi_I$, where $\chi_R$ and $\chi_I$ are both Majorana. It is then easily checked that the standard anticommutators (\ref{eq3.5}) imply that $\chi_R$ and $\chi_I$ both satisfy (\ref{eq3.7}), and mutually anti-commute. In addition, it is easily seen that if we insert the decomposition $\chi = \chi_R + i \, \chi_I$ in the full Dirac action (\ref{eq2.56}), it simply decomposes as the sum of two {\it decoupled} Majorana actions, one involving $\chi_R$ and one involving $\chi_I$. [This is true because all the spinor bilinears, including the mass term $\propto \chi^{\dagger} \, \gamma_{\widehat 0} \, \chi$, involve an antisymmetric Clifford-algebra matrix, such as $\gamma^{\widehat 1} \, \gamma^{\widehat 2}$, $\gamma_{\widehat 0} \, \gamma^{\widehat 1 \, \widehat 2 \, \widehat 3}$ or $\gamma_{\widehat 0}$, sandwiched between $\chi^{\dagger}$ and $\chi$.] This shows that considering a complex Dirac spinor source is equivalent to coupling gravity to the sum of two independent Majorana spinor sources. At the Hilbert-space level, the Dirac fermionic state space is just the tensor product of two independent Majorana fermionic state spaces. For simplicity, we shall consider in the following the simpler, single Majorana case, {\it i.e.} $\chi^* = \chi$. Note, in particular, that in that case the fermionic Hilbert space is simply of dimension $2^2 = 4$. Indeed, this fermionic Majorana space is the (irreducible) representation space of the algebra (\ref{eq3.7}), which is simply a {\it Clifford algebra} on the $4$-dimensional Euclidean space. Actually, if we introduce the notation
\begin{equation}
\label{eq3.9}
\Gamma_{\alpha} := 2 \, \chi_{\alpha} \, , \qquad \alpha = 1,\ldots , 4
\end{equation}
we see that the four operators (in Hilbert space) $\Gamma_{\alpha}$, $\alpha = 1,\ldots , 4$, satisfy the standard $O(4)$ Clifford algebra relations
\begin{equation}
\label{eq3.10}
\Gamma_{\alpha} \, \Gamma_{\beta} + \Gamma_{\beta} \, \Gamma_{\alpha} = 2 \, \delta_{\alpha\beta} \, .
\end{equation}
In particular, each $\Gamma_{\alpha}$ has a unit square. In addition, note that the product
\begin{equation}
\label{eq3.11}
\Gamma_5 := \Gamma_1 \, \Gamma_2 \, \Gamma_3 \, \Gamma_4 \equiv 16 \, \chi_1 \, \chi_2 \, \chi_3 \, \chi_4 \, ,
\end{equation}
defines an operator which anticommutes with each $\chi_{\alpha}$, and which has also a unit square: $\Gamma_5^2 = +1$. [This contrasts with the $O(3,1)$ case where $\gamma_5 \equiv \gamma_{\widehat 0} \, \gamma_{\widehat 1} \, \gamma_{\widehat 2} \, \gamma_{\widehat 3}$ squares to {\it minus} $1$, because of the time-like character of $\gamma_{\widehat 0}$.]

\smallskip

Summarizing: when $\chi$ is taken to be a Majorana spinor, the second-quantiza\-tion of $\chi$, Eq.~(\ref{eq3.7}), leads to a $4$-dimensional space of fermionic degrees of freedom. In other words, the wavefunction of the universe, $\Phi (\beta_a , \theta_a , \sigma)$, is labelled, besides the $6$ continuous bosonic  labels $(\beta_1 , \beta_2 , \beta_3 ; \theta_1 , \theta_2 , \theta_3)$ of the gravitational minisuperspace, by a discrete index $\sigma$, which takes $4$ values, and which is algebraically equivalent to the spinor index of the Dirac representation of the $O(4)$ Clifford algebra (\ref{eq3.10}). When $\chi$ is taken to be a complex, Dirac spinor, the second quantization of $\chi$, Eq.~(\ref{eq3.5}), leads to a $16$-dimensional space of fermionic degrees of freedom, which is the tensor product of two independent Majorana-spinor spaces. Note that if one works in a Heisenberg picture, where the $\chi (t)$'s evolve in time according to the equations of motion (\ref{eq3.8}), the corresponding Clifford algebra generators (\ref{eq3.9}) must also be viewed as time-evolving objects, rather than fixed numerical gamma matrices.

\subsection{Explicit form of the Einstein-Majorana Bianchi-IX  quantum Hamiltonian}\label{ssec3.3}

After having taken into account the diffeomorphism constraints (\ref{eq3.2}) in the form (\ref{eq3.3}), so that the wave function $\Phi (\beta , \sigma)$ depends only on the three diagonal metric variables $\beta_1 ,\beta_2 ,\beta_3$ and the discrete spin-space label $\sigma$, it is natural, in the gauge $\widetilde N \equiv N / \sqrt g = 1$, to quantize the gravitational degrees of freedom by replacing the classical $\beta$-momenta $\pi_{\beta}$ [which enter the Hamiltonian (\ref{eq2.90}) only through $T_{\beta} (\pi_{\beta})$, Eq.~(\ref{eq2.72})] by the differential operators $\widehat\pi_{\beta_a} := -i \, \partial / \partial \, \beta_a$. This leads to interpreting the Hamiltonian constraint as the following Klein-Gordon-like Wheeler-DeWitt equation
\begin{equation}
\label{eq3.12}
\widehat{\widetilde{\mathcal H}} { \ }^{\!\!\rm IX} \, \Phi (\beta , \sigma) = 0 \, .
\end{equation}
with
\begin{equation}
\label{eq3.13}
\widehat{\widetilde{\mathcal H}} { \ }^{\!\!\rm IX} := T_{\beta} (\widehat{\pi}_{\beta}) + \widehat{T}_w^{(0)} (\beta , \chi) + V_g^{\rm IX} (\beta) + \widehat V_{s \, {\rm grav}}^{\rm IX} (\beta , \chi) + \widehat V_{s \, {\rm mass}} (\beta , \chi) + c \, ,
\end{equation}
\begin{equation}
\label{eq3.14}
T_{\beta} (\widehat{\pi}_{\beta}) = - \frac{1}{4} \, G^{ab} \, \partial_{\beta_a} \, \partial_{\beta_b} = - \frac{1}{4} \, (\partial_{\beta_1}^2 + \partial_{\beta_2}^2 + \partial_{\beta_3}^2) + \frac{1}{8} \, (\partial_{\beta_1} + \partial_{\beta_2} + \partial_{\beta_3})^2 \, ,
\end{equation}
where $V_g^{\rm IX} (\beta)$ is the usual type-IX potential (\ref{eq2.30}), and where the quantum versions of all the $\beta - \chi$ coupling terms $\widehat T_x^{(0)} (\beta , \chi)$, $\widehat V_{s \, {\rm grav}}^{\rm IX} (\beta , \chi)$, $\widehat V_{s \, {\rm mass}} (\beta , \chi)$, are simply obtained from their classical expressions by interpreting the various spinor bilinears they contain as operators in the fermionic space defined, say in the Majorana case, by Eq.~(\ref{eq3.7}). This yields
\begin{eqnarray}
\label{eq3.15}
\widehat{T}_w^{(0)} (\beta , \chi) &= &\frac{1}{8} \coth^2 (\beta_1 - \beta_2) \, (\widehat\Sigma^{12})^2 + \frac{1}{8} \coth^2 (\beta_2 - \beta_3) \, (\widehat\Sigma^{23})^2 \nonumber \\
&&+ \, \frac{1}{8} \coth^2 (\beta_3 -\beta_1) \, (\widehat\Sigma^{31})^2 \, , 
\end{eqnarray}
where
\begin{equation}
\label{eq3.16}
\widehat\Sigma^{12} := \frac{i}{2} \, \chi^{\dagger} \, \gamma^{\widehat 1 \, \widehat 2} \, \chi \, , \ \widehat\Sigma^{23} := \frac{i}{2} \, \chi^{\dagger} \gamma^{\widehat 2 \, \widehat 3} \, \chi \, , \ \widehat\Sigma^{31} := \frac{i}{2} \, \chi^{\dagger} \gamma^{\widehat 3 \, \widehat 1} \, \chi \, ,
\end{equation}
are operators acting in the $4$-dimensional fermionic space. Similarly, denoting by ``$+$ cyclic'' the addition of two other cyclically permuted terms,
\begin{eqnarray}
\label{eq3.17}
V_{s \, {\rm grav}}^{\rm IX} (\beta , \chi) &= &- \frac{1}{4} \, e^{-2\beta_1} (i \, \chi^{\dagger} \, \gamma_{\widehat 0} \, \gamma^{\widehat 1 \, \widehat 2 \, \widehat 3} \, \chi) + \mbox{cyclic} \nonumber \\
&= &- \frac{1}{4} \, (e^{-2\beta_1} + e^{-2\beta_2} + e^{-2\beta_3}) (i \, \chi^{\dagger} \, \gamma_{\widehat 0} \,\gamma^{\widehat 1 \, \widehat 2 \, \widehat 3} \, \chi) \, ,
\end{eqnarray}
and,
\begin{equation}
\label{eq3.18}
\widehat V_{s \, {\rm mass}} (\beta , \chi) = m \, e^{-(\beta_1 + \beta_2 + \beta_3)} \, (i \, \chi^{\dagger} \, \gamma_{\widehat 0} \, \chi) \, .
\end{equation}
Finally, we have allowed for the addition of an ``ordering constant'' $c$ in Eq.~(\ref{eq3.13}). There are several motivations for allowing for such an additional constant. First, if we were working in a different gauge, {\it e.g.} $N=1$, {\it i.e.} $\widetilde N = g^{-1/2} = e^{(\beta_1 + \beta_2 + \beta_3)}$ instead of $\widetilde N = 1$, we would have had to quantize ${\mathcal H} = \frac{1}{4} \, e^{(\beta_1 + \beta_2 + \beta_3)}$ $G^{ab} \, \pi_{\beta_a} \, \pi_{\beta_b}$ instead of $\frac 14 G^{ab} \, \pi_{\beta_a} \, \pi_{\beta_b}$.  More generally the quantization
of  ${\mathcal H} = \frac{1}{4} \, e^{2 \alpha(\beta)}$ $G^{ab} \, \pi_{\beta_a} \, \pi_{\beta_b}$,
where $\alpha(\beta)= \alpha_a \beta_a$ is some linear form in the $\beta$'s, leads (after
reabsorbing  $e^{\alpha(\beta)}$ in the definition of the wavefunction) 
to an ordering ambiguity which is equivalent to adding a constant 
$c \propto G^{ab} \alpha_a \alpha_b$ in Eq.~(\ref{eq3.13}). Second, the contribution (\ref{eq3.15}) being {\it quartic} in the non-commuting $\chi$'s has also ordering ambiguities (though it is natural to define it as we did, {\it i.e.} by taking the squares of the spinor bilinears (\ref{eq3.16}), which have no ordering ambiguities because $\gamma^{\widehat 1 \, \widehat 2}$, {\it etc.}, are antisymmetric matrices). Finally, there is a basic ambiguity in the gravity-spinor Lagrangian due to the possible use of first-order (Palatini) versus second-order formulations. It has been shown long ago \cite{Weyl29} that this leads to an ambiguity in the action density proportional to
\begin{eqnarray}
\label{eq3.19}
\Delta {\mathcal L} &= & \frac{1}{3!} \, \sqrt{- \, ^4 g} \ (\overline\psi \, \gamma^{\widehat\alpha \, \widehat\beta \, \widehat\gamma} \, \psi) (\overline\psi \, \gamma_{\widehat\alpha \, \widehat\beta \, \widehat\gamma} \, \psi) \nonumber \\
&= &\widetilde N \, [- (i \, \chi^{\dagger} \, \gamma^{\widehat 1 \, \widehat 2} \, \chi)^2 - (i \, \chi^{\dagger} \, \gamma^{\widehat 2 \, \widehat 3} \, \chi)^2 - (i \, \chi^{\dagger} \, \gamma^{\widehat 3 \, \widehat 1} \, \chi)^2 \nonumber \\
&&+  \  (i \, \chi^{\dagger} \, \gamma_{\widehat 0} \, \gamma^{\widehat 1 \, \widehat 2 \, \widehat 3} \, \chi)^2] \, .
\end{eqnarray}
We will come back below to the latter ($\beta$-independent when $\widetilde N = 1$, but $\chi$-dependent) ambiguity. Finally, note that one could also consider adding to the quantum
Dirac action other Lorentz-invariant contributions quadratic in spinor bilinears,
$\Delta {\mathcal L} \sim  (\overline\psi \, \gamma^{A} \, \psi)^2$,
because no symmetry prevents the appearance of such quartic contributions.

\subsection{Properties of the spinor bilinears}\label{ssec3.4}

For the time being, let us note that, in the representation (\ref{eq3.6}), one can compute explicit expressions for the Majorana-spinor bilinears entering the Hamiltonian. Let the $4$ components of the Majorana spinor $\chi$ in the representation (\ref{eq3.6}) be denoted as $\chi_1 , \chi_2 ,  \chi_3 , \chi_4$, and let $\chi^T$ denote the transpose of the column vector $\chi_{\alpha}$, {\it i.e.} the horizontal row $\chi^T = (\chi_1 , \chi_2 ,  \chi_3 , \chi_4)$. We find
\begin{eqnarray}
\label{eq3.20}
\chi^T \, \gamma^1 \, \gamma^2 \, \chi &= &\chi_1 \, \chi_2 - \chi_2 \, \chi_1 + \chi_3 \, \chi_4 - \chi_4 \, \chi_3 \, , \nonumber \\
\chi^T \, \gamma^2 \, \gamma^3 \, \chi &= &\chi_2 \, \chi_3 - \chi_3 \, \chi_2 + \chi_1 \, \chi_4 - \chi_4 \, \chi_1 \, , \nonumber \\
\chi^T \, \gamma^3 \, \gamma^1 \, \chi &= &\chi_3 \, \chi_1 - \chi_1 \, \chi_3 + \chi_2 \, \chi_4 - \chi_4 \, \chi_2 \, , 
\end{eqnarray}
for the bilinears entering $\widehat T_w^{(0)}$ (without the $i/2$ prefactor), while
\begin{equation}
\label{eq3.21}
\chi^T \, \gamma_0 \, \gamma_1 \, \gamma_2 \, \gamma_3 \, \chi = \chi_1 \, \chi_2 - \chi_2 \, \chi_1 - \chi_3 \, \chi_4 + \chi_4 \, \chi_3 \, ,
\end{equation}
and
\begin{equation}
\label{eq3.22}
\chi^T \, \gamma_0 \, \chi = \chi_2 \, \chi_3 - \chi_3 \, \chi_2 - \chi_1 \, \chi_4 + \chi_4 \, \chi_1 \, ,
\end{equation}
are the bilinears entering $\widehat V_{s \, {\rm grav}}$ and $\widehat{V}_{s \, {\rm mass}}$.

\smallskip

In view of (\ref{eq3.20}) it would seem that $\widehat T_w^{(0)}$, which contains the squares of these bilinears, is a hopelessly complicated expression. However, things simplify very much if we use the Clifford-algebra properties of the $\chi$'s. More precisely, using the notation (\ref{eq3.9}), and simple properties such as $\Gamma_1^2 = 1$, $(\Gamma_1 \, \Gamma_2)^2 = -1$, and $\Gamma_1 \, \Gamma_2 \, \Gamma_5 = - \Gamma_3 \, \Gamma_4$ (with the definition (\ref{eq3.11}) of $\Gamma_5$), one finds the following simplified expressions for the spinor bilinears
\begin{eqnarray}
\label{eq3.23}
\chi^T \gamma^{12} \, \chi &= &\Gamma_1 \, \Gamma_2 \, \frac{1-\Gamma_5}{2} \, , \nonumber \\
\chi^T \gamma^{23} \, \chi &= &\Gamma_2 \, \Gamma_3 \, \frac{1-\Gamma_5}{2} \, , \nonumber \\
\chi^T \gamma^{31} \, \chi &= &\Gamma_3 \, \Gamma_1 \, \frac{1-\Gamma_5}{2} \, , \nonumber \\
\chi^T \gamma_{0123} \, \chi &= &\Gamma_1 \, \Gamma_2 \, \frac{1+\Gamma_5}{2} \, , \nonumber \\
\chi^T \gamma_0 \, \chi &= &\Gamma_2 \, \Gamma_3 \, \frac{1+\Gamma_5}{2} \, . 
\end{eqnarray}
Note that these expressions contain the projectors
\begin{equation}
\label{eq3.24}
P_{\pm} := \frac{1 \pm \Gamma_5}{2} \, ,
\end{equation}
which satisfy the usual properties of Dirac's helicity projectors $(1 \pm i \, \gamma_5)/2$ [with $\gamma_5 =  \gamma_0 \, \gamma_1 \, \gamma_2 \, \gamma_3$ so that $(i \, \gamma_5)^2 = +1$], namely: $P_+ + P_- = 1$, $P_+^2 = P_+$, $P_-^2 = P_-$, $P_+ \, P_- = 0$.

\smallskip

Let us also note that the above bilinears satisfy simple commutation relations among themselves. First, as $\Gamma_{ab} := \Gamma_{[a} \Gamma_{b]}$ ({\it i.e.} $\Gamma_{11} = 0$, $\Gamma_{12} = \Gamma_1 \, \Gamma_2$, {\it etc.}) commutes with $\Gamma_5$, and $P_+ \, P_- = 0$, we see that the first three bilinears in (\ref{eq3.23}) commute with the last two. Second, one easily checks that (together with its cyclic analogs)
\begin{equation}
\label{eq3.25}
\left[ \chi^T \, \frac{\gamma^{12}}{2} \, \chi \, , \ \chi^T \, \frac{\gamma^{23}}{2} \, \chi \right] = \chi^T \, \frac{\gamma^{13}}{2} \, \chi \, ,
\end{equation}
{\it i.e.} adding the factor $i$ transforming the anti-hermitian $\frac{1}{2} \, \chi^T \gamma^{ab} \, \chi$ into the hermitian operators $\widehat{\Sigma}^{ab}$ of Eq.~(\ref{eq3.16}),
\begin{equation}
\label{eq3.26}
\left[ \widehat\Sigma^{12} , \widehat\Sigma^{23} \right] = -i \, \widehat\Sigma^{31} \, ,
\end{equation}
together with its cyclic kins. In other words, the second-quantized operators $\widehat\Sigma^{12}$, $\widehat\Sigma^{23}$, $\widehat\Sigma^{31}$ satisfy the (reverse-sign) commutation law of body-frame angular momenta (the classical Poisson bracket $\{ L_1 , L_2 \} = -L_3$ mentioned above becoming $[L_1 , L_2] = -i \, L_3$ at the quantum level).

\smallskip

Let us note that, more generally, one can deduce directly from the basic anticommutation relations (\ref{eq3.7}) that the commutators of two bilinears $\chi^T A \chi \equiv \chi_{\alpha} \, A_{\alpha\beta} \, \chi_{\beta}$ and $\chi^T B \, \chi \equiv \chi_{\alpha} \, B_{\alpha\beta} \, \chi_{\beta}$ (where $A$ and $B$ are some, say antisymmetric, $\chi$-independent matrices) is equal to
\begin{equation}
\label{3.27}
[\chi^T A \, \chi \, , \ \chi^T B \, \chi] = \chi^T \, [A,B] \, \chi \, ,
\end{equation}
where $[A,B]_{\alpha\beta} = A_{\alpha\sigma} \, B_{\sigma\beta} - B_{\alpha\sigma} \, A_{\sigma\beta}$ is the usual matrix commutator.

\smallskip

In addition the squares of the bilinears (\ref{eq3.23}) simplify. Notably, $(\chi^T \gamma^{12} \, \chi)^2 = (\Gamma_1 \, \Gamma_2)^2 \, P_-^2 = -P_-$, {\it i.e.}
\begin{equation}
\label{eq3.28}
\left( \widehat\Sigma^{12} \right)^2 = \left( \widehat\Sigma^{23} \right)^2 = \left( \widehat\Sigma^{31} \right)^2 = \frac{1}{4} \, P_- \, .
\end{equation}

\subsection{``Squared-mass'' in the Wheeler-DeWitt equation}\label{ssec3.5}

The main result of Ref.~\cite{BK} was the finding that a classical spinor source ({\it i.e.} treating the spinor bilinears $\chi^T A \, \chi$ as $c$-numbers) modifies the usual quadratic relation $\underset{a}{\sum} \, p_a^2 - \left( \underset{a}{\sum} \, p_a \right)^2 = 0$ satisfied by Kasner exponents (in the Bianchi type~I case, {\it i.e.} far from any potential wall) into the relation\footnote{We normalize the type~I spatial metric so that $\sqrt g = T$, where $T$ is the proper time.}
\begin{equation}
\label{eq3.29}
\sum_a p_a^2 - \left( \sum_a p_a \right)^2 = - \frac{1}{2} \left[ (\Sigma^{12})^2 + (\Sigma^{23})^2 + (\Sigma^{31})^2 \right] \, ,
\end{equation}
where the $\Sigma^{ab}$ are the spinor bilinears defined above (treated as $c$-numbers). In the language of cosmological billiards (and using the time gauge $N = \sqrt g = T$, i.e. a coordinate time $t=- \ln T$), the result (\ref{eq3.29}) means that the Lorentzian squared velocity in $\beta$-space $\dot\beta^2 := G_{ab} \, \dot\beta_a \, \dot\beta_b = \underset{a}{\sum} \, p_a^2 - \left( \underset{a}{\sum} \, p_a \right)^2$ is negative, $\dot\beta^2 < 0$, {\it i.e. time-like}, rather than light-like ($\dot\beta^2 = 0$) as usual. Belinsky and Khalatnikov \cite{BK} argued that this implied that a classical spinor would (like a scalar field does) ultimately quench the chaotic, oscillatory regime near the singularity, and ultimately turn it into a monotonic Kasner-like, power-law behaviour (because the $\beta$-particle can ultimately move on a time-like geodesic which no longer hits any gravitation wall).

\smallskip

The result (\ref{eq3.29}) was understood via an Hamiltonian approach, and within a more general Kac-Moody coset model approach, by de~Buyl, Henneaux and Paulot \cite{dBHP}. These authors found that, far from the walls associated to a general coset model, the mass-shell condition for the $\beta$-momenta $\pi_{\beta_a} \propto G_{ab} \, \dot\beta_b$ was modified, by the coupling to a classical spinor, from the usual light-like condition $\pi_{\beta}^2 = 0$ (where $\pi_{\beta}^2 := G^{ab} \, \pi_{\beta_a} \, \pi_{\beta_b}$) to a massive-shell condition
\begin{equation}
\label{eq3.30}
\pi_{\beta}^2 + \mu^2 = 0 \, ,
\end{equation}
where the general expression for the ($\beta$-space) squared-mass $\mu^2$ was found to be
\begin{equation}
\label{eq3.31}
\mu_{\rm coset}^2 = \frac{1}{2} \sum_{\alpha} (i \, \chi^{\dagger} J_{\alpha}^s \, \chi)^2 \, .
\end{equation}
Here, $\chi$ is the coset Dirac field [which corresponds to the rescaled Einstein-Dirac field (\ref{eq2.55})], and $\alpha$ labels all the (positive) Kac-Moody roots (including their degeneracy) that enter the bosonic  coset model $G/K$. Each root $\alpha$ is a linear form in the $\beta$'s, and corresponds (in Iwasawa gauge) to a bosonic  wall potential proportional to $\exp (- \, 2 \, \alpha(\beta))$. The object $J_{\alpha}^s$ in (\ref{eq3.31}) is an anti-hermitian generator which represents (in the spinor representation $s$) the ``rotation'' $E_{\alpha} - E_{-\alpha}$ of the maximally compact subgroup $K$ of $G$ associated to the root $\alpha$. The extra factor $i$ in (\ref{eq3.31}) turns $J_{\alpha}^s$ into a hermitian operator, so that $i \, \chi^{\dagger} J_{\alpha}^s \, \chi$ is real and $\mu^2$ is given (for a Kac-Moody coset $G/K$) by an infinite sum of positive quantities.

\smallskip

In the language of the general Kac-Moody result (\ref{eq3.30}), (\ref{eq3.31}) (which uses a normalization where $\pi_{\beta_a} = G_{ab} \, \dot\beta_b$, without the factor $2$ entering our Eq.~(\ref{eq2.71}) above) the original Belinsky-Khalatnikov result (\ref{eq3.29}) corresponds to the squared-mass
\begin{equation}
\label{eq3.32}
\mu_{\rm BK}^2 = \frac{1}{2} \left[ (\Sigma^{12})^2 + (\Sigma^{23})^2 + (\Sigma^{31})^2 \right] \, .
\end{equation}
This squared-mass is of the general form (\ref{eq3.31}) (see below) but includes only three roots, namely the roots corresponding to the so-called ``symmetry walls''\footnote{When considering
potential wall forms $w_A(\beta)$ from a coset perspective we denote them as $\alpha_A(\beta)$.},
$\alpha_{12} (\beta) = \beta^2 - \beta^1$, $\alpha_{23} (\beta) = \beta^3 - \beta^2$ and $\alpha_{13} (\beta) = \beta^3 - \beta^1$. One might wonder whether the presence of only those three symmetry roots, and, in particular, the absence of any contribution coming from the gravitational-wall roots $\alpha^g_{1;23}(\beta)=2 \beta_1$, etc. 
(which are crucial to generate the BKL chaos) is due to the approximate nature of the treatment of Ref.~\cite{BK}, based on a Bianchi type-I model, {\it i.e.} a model which contains all the symmetry walls, but no gravitational walls. Separately from this issue, we wish here to use our results above to clarify the effect of considering, as one should, the spin source as being quantum, rather than classical, as was assumed both in \cite{BK} and in \cite{dBHP}.

\smallskip

One can define the squared-mass term in the Wheeler-DeWitt equation as the term remaining with $\widehat\pi_{\beta}^2$ in a BKL-type limit $\beta \to + \infty$ in which one stays far away from all exponential   walls. We have already mentioned that $\widehat V_{s \, {\rm grav}}^{\rm IX}$, Eq.~(\ref{eq3.17}), contains (half) gravitational walls, while $\widehat V_{s \, {\rm mass}}$, Eq.~(\ref{eq3.18}), contains a (half) cosmological-constant-related exponential wall. All these terms, as well as the usual bosonic  type-IX potential $V^{\rm IX} (\beta)$, Eq.~(\ref{eq2.30}), will tend to zero in this far-wall BKL limit. It remains to consider the contribution $\widehat T_w^{(0)}$, Eq.~(\ref{eq3.15}). Using the hyperbolic-trigonometry identity
$ \coth^2 x = 1 + 1/\sinh^2 (x)$,
we see that we can split $\widehat T_w^{(0)}$ as
\begin{equation}
\label{eq3.34}
\widehat T_w^{(0)} = \frac{1}{4} \, \widehat{\mu}_{\Sigma}^2 + \widehat{V}^{\rm centrif} (\beta , \chi) \, ,
\end{equation}
where
\begin{equation}
\label{eq3.35}
\widehat{\mu}_{\Sigma}^2 = \frac{1}{2} \left[ (\widehat\Sigma^{12})^2 + (\widehat\Sigma^{23})^2 + (\widehat\Sigma^{31})^2 \right] \, ,
\end{equation}
and
\begin{equation}
\label{eq3.36}
\widehat{V}^{\rm centrif} (\beta , \chi) = \frac{1}{8} \, \frac{(\widehat\Sigma^{12})^2}{\sinh^2 (\beta_1 - \beta_2)} + {\rm cyclic} \, .
\end{equation}
One recognizes in (\ref{eq3.36}) the sum of three $\sinh^{-2}$-type centrifugal-wall potentials, which is the form taken by symmetry walls when one uses a Gauss decomposition of the off-diagonal components of the metric rather than an Iwasawa decomposition 
(see \cite{Ryan72,RyanShepley75}).
Far from the symmetry (or centrifugal) walls, {\it i.e.} when $\vert \beta^2 - \beta^1 \vert \gg 1$, {\it etc.}, $\widehat{V}^{\rm centrif}$ becomes a sum of exponentially small terms $\propto \exp (-2 \, \vert \beta^2 - \beta^1 \vert) + {\rm cyclic}$. This leaves as only terms contributing to the dynamics in the far-wall limit 
\begin{equation}
\label{eq3.37}
\widehat{\widetilde{\mathcal H}} { \ }^{\!\!\rm IX} = \frac{1}{4} \, \left[ \widehat\pi_{\beta}^2 + \widehat\mu_{\Sigma}^2 + 4c \right] + \mbox{(exponentially small terms)},
\end{equation}
where $\widehat\pi_{\beta}^2 = - G^{ab} \, \partial_{\beta_a} \, \partial_{\beta_b}$ is the Klein-Gordon operator associated to the $\beta$-space Lorentzian metric.

\smallskip

This result shows that, in Bianchi type-IX (and also, as one easily sees, in type VIII), the coupling of gravity to a quantum spinor generates a $q$-number squared mass in the Klein-Gordon-like Wheeler-DeWitt equation given by 
$$
\widehat\mu^2 = \widehat\mu_{\Sigma}^2 + 4c \, ,
$$
where $\widehat\mu_{\Sigma}^2$ is defined in Eq.~(\ref{eq3.35}), and where $c$ is the ordering ambiguity that we allowed for in Eq.~(\ref{eq3.13}).

\smallskip

Note, first, that if we take a formal classical limit of $\widehat\mu_{\Sigma}^2$, Eq.~(\ref{eq3.35}), we recover the Belinsky-Khalatnikov result $\mu_{\rm BK}^2$, Eq.~(\ref{eq3.32}), without any extra contribution coming from the gravitational walls (which were explicitly taken into account in our calculation, contrary to the type-I computation of Ref.~\cite{BK}). When comparing this result with the coset-model result $\mu_{\rm coset}^2$, Eq.~(\ref{eq3.31}), we see that there is a genuine difference.  Even if one were considering a truncated coset model involving only the couplings to the walls entering the type-IX gravity model, the sum in Eq.~(\ref{eq3.31}) should include as roots both the three symmetry roots $\alpha_{12} (\beta) = \beta_2 - \beta_1$, $\alpha_{23} (\beta) = \beta_3 - \beta_2$ and $\alpha_{13} (\beta) = \beta_3 - \beta_1$, and the three gravitational roots $\alpha_{123}^g (\beta) = \beta_1 - \beta_2 - \beta_3 + \Sigma_a \, \beta_a = 2 \, \beta_1$, $\alpha_{231}^g (\beta) = 2 \, \beta_2$ and $\alpha_{312}^g (\beta) = 2 \, \beta_3$.  The corresponding $K(AE_3)$ generators, $J_{\alpha} = E_{\alpha} - E_{-\alpha}$, within the {\it spinor} representation of $K(AE_3)$, are known to 
be \cite{dBHP,DKN,de Buyl:2005mt,DKN2,DaHi},
\begin{equation}
\label{eq3.38}
J_{\alpha_{12}}^s = \frac{1}{2} \, \gamma^{12} \, , \quad J_{\alpha_{23}}^s = \frac{1}{2} \, \gamma^{23} \, , \quad J_{\alpha_{13}}^s = \frac{1}{2} \, \gamma^{13} \, ,
\end{equation}
for the three symmetry roots, and
\begin{equation}
\label{eq3.39}
J_{\alpha_{123}^g}^s = \frac{1}{2} \, \gamma_0 \, \gamma^{123} \, , \quad J_{\alpha_{231}^g}^s = \frac{1}{2} \, \gamma_0 \, \gamma^{231} \, , \quad J_{\alpha_{312}^g}^s = \frac{1}{2} \, \gamma_0 \, \gamma^{312} \, ,
\end{equation}
for the three gravitational walls. Inserting the expressions (\ref{eq3.38}) into the general coset result (\ref{eq3.31}) yields, in view of the definition (\ref{sigma}) of $\Sigma^{ab}$, precisely the result (\ref{eq3.35}) (including the correct normalization factor). However, a coset model involving the couplings to the walls entering the type-IX gravity model would yield a (quantum) squared-mass of the form
\begin{equation}
\label{eq3.40}
\widehat{\mu}_{\rm coset}^{2 \,({\rm IX}) } = \widehat\mu_{\Sigma}^2 + \frac{3}{2} \left( \frac{i}{2} \, \chi^{\dagger} \gamma_0 \, \gamma^{123} \, \chi \right)^2 \, ,
\end{equation}
where the factor $3$ comes from the fact that the three gravitational-root generators (\ref{eq3.39}) happen to be equal among themselves.

\smallskip

The reason why the coset-results (\ref{eq3.31}) or (\ref{eq3.40}) (for the type-IX truncation) contain more contributions than the original type-IX gravity model, Eq.~(\ref{eq3.35}), is easily understood from the corresponding derivations of these results in \cite{dBHP} and in Section~\ref{sec2} above. Indeed, one sees that each individual contribution $\frac{1}{2} \, (i \, \chi^{\dagger} J_{\alpha}^s \, \chi)^2$ to $\mu^2$ comes for the presence of a corresponding {\it time-derivative} coupling $\sim e \, \dot q \, A$, see Eq.~(\ref{eq2.64}), between gravity and the spinor. For instance, it is because in Eq.~(\ref{eq2.56}) there were (say in the far-wall limit) three Lagrangian-coupling terms,
\begin{equation}
\label{eq3.41}
-\cosh (\beta_1 - \beta_2) \, w^{12} \, \Sigma^{12} + {\rm cyclic} \simeq - \frac{1}{2} \, e^{\alpha_{12} (\beta)} \, w^{12} \, \Sigma^{12} + {\rm cyclic} \, ,
\end{equation}
between the rotational velocities $w^{12} = \dot\varphi \cos \theta + \dot\psi$, {\it etc.} of the off-diagonal metric variables (parametrized by Euler angles), that one ended up with $\widehat{\mu}_{\Sigma}^2$ given by the sum of three symmetry wall contributions $\alpha_{12}$, $\alpha_{23}$, $\alpha_{13}$. By contrast, the couplings corresponding to the three gravitational walls $\alpha_{123}^g$, $\alpha_{231}^g$, $\alpha_{312}^g$, between gravity and the spinor, predicted by the Einstein-Dirac action, were contained in $V_{s \, {\rm grav}}$, Eq.~(\ref{eq2.60}), {\it i.e.} were of the type
\begin{equation}
\label{eq3.42}
V_{s \, {\rm grav}} = - \frac{1}{2} \, C^1 \, _{23} \, e^{-\alpha_{123}^g (\beta)} \left[ \frac{i}{2} \, \chi^{\dagger} \gamma_0 \, \gamma^{123} \, \chi \right] + {\rm cyclic} \, .
\end{equation}

This is very analogous to the coupling (\ref{eq3.41}), with, however, the crucial difference that the rotational velocities $w^{12}$, {\it etc.}, appearing in (\ref{eq3.41}) are replaced in (\ref{eq3.42}) by the structure constants $C^1 \, _{23}$, {\it etc.} In the gravity picture, the structure constants are {\it spatial} derivatives of the metric, and therefore do not contribute a squared-mass term {\it via} the
\begin{equation}
\label{eq3.43}
\frac{(p-e \, A)^2}{2m} = \frac{1}{2} \, \frac{e^2 \, A^2}{m} + \ldots
\end{equation}
mechanism (with $e \, A \propto \frac{i}{2} \, \chi^{\dagger} J_{\alpha}^s \, \chi$) explained above, which was linked to a {\it time-derivative} coupling $\Delta {\mathcal L} = e \, \dot q \, A$. By contrast, the coset action is related to the gravity action by a {\it dualization} of some of the gravity degrees of freedom. In particular, the structure constants $C^a \, _{bc}$ become replaced by the conjugate momenta, {\it i.e.} essentially by the time-derivative, of some dual coset field, namely $\widetilde C^{ab} := \frac 12 \varepsilon^{bcd} \, C^a \, _{cd} \propto \Pi^{ab} \sim \dot\phi_{ab}$ for the $AE_3$ case linked to the $(3+1)$-dimensional Bianchi-IX dynamics (see Section~8 of \cite{DHN2}). This explains why the gravitational roots (and actually all the coset roots) give a contribution to $\mu^2$ within the coset model 
(for the spin 1/2 case), but not within the original gravity model.

\smallskip

Let us now turn to the quantum nature of the Bianchi-IX squared-mass term (\ref{eq3.35}). Though each term $(\Sigma^{ab})^2 = \left( \frac{i}{2} \, \chi^{\dagger} \gamma^{ab} \, \chi \right)^2$ is quartic in the quantized spinor field $\chi$, we have shown above that the Clifford-algebra nature of the quantization conditions for a Majorana spinor led to simple final results. More precisely, we see from (\ref{eq3.28}) that each separate $(\Sigma^{ab})^2$ term yields the same result, so that the total quantum $\mu^2$ reads
\begin{equation}
\label{eq3.44}
\widehat{\mu}_{\Sigma}^2 = \frac{3}{8} \, P_- = \frac{3}{8} \, \frac{1-\Gamma_5}{2} \, ,
\end{equation}
where we recall that $\Gamma_5$ denotes the involutive $(\Gamma_5^2 = 1)$ operator (\ref{eq3.11}). The simple result (\ref{eq3.44}) shows that, in the quantum theory, the squared-mass term [in absence of any additional ordering-constant contribution $4c$, as in Eq.~(\ref{eq3.37})] is an operator which has two possible eigenvalues: the eigenvalue $\mu_{\Sigma}^2 = 0$ in the $2$-dimensional space of ``helicity'' $\Gamma_5 = +1$, or the eigenvalue $\mu_{\Sigma}^2 = 3/8$ in the complementary (orthogonal) $2$-dimensional space of ``helicity'' $\Gamma_5 = -1$. 
[Note that  $\mu_{\Sigma}^2$ vanishes in half of the total ($4$-dimensional) Majorana-spinor Hilbert space.] In terms of $(2j+1)$-dimensional irreducible representations $D_j$ of $SU(2)$, we can view the three components $\widehat\Sigma^{12}, \widehat\Sigma^{23}, \widehat\Sigma^{31}$ of
a single Majorana field as describing the sum $D_0 + D_0 +D_{\frac 12}$.
Let us also note that our results above, also gives the eigenvalues 
of $\widehat\Sigma^{12}, \widehat\Sigma^{23}, \widehat\Sigma^{31}$ (and then of $\mu^2$) when considering a complex Dirac spinor. Indeed, this case is obtained by considering $\chi = \chi_R + i \, \chi_I$ as the complex combination of two, independent ({\it i.e.} anticommuting), Majorana spinors, so that each final $\widehat\Sigma^{12}$ etc., is given by a sum, 
 $\widehat\Sigma^{12}_R +  \widehat\Sigma^{12}_I$,
of two commuting spin operators. The corresponding representation space is therefore
the tensor product $ (D_0 + D_0 +D_{\frac 12}) \times (D_0 + D_0 +D_{\frac 12})$,
i.e. the sum $ 5 D_0 + 4 D_{\frac 12} + D_1$.
This shows that the possible eigenvalues of $\widehat\mu^2$ are $0$, $3/8$ and $1$. Moreover, the eigenvalue $\widehat\mu^2 = 0$ is obtained (when assuming $c=0$)  in a $5$-dimensional subspace of the $16$-dimensional total Hilbert space.
Note that we can extend this result by considering the case where gravity couples to a sum of
$N$ independent (i.e. anticommuting) Majorana spinor fields. In that case  each
spin operator $\widehat\Sigma^{12}$ etc. will be the sum of $N$ independent (commuting) 
contributions belonging to a $D_0 + D_0 +D_{\frac 12}$ representation space.
This implies that the eigenvalues of $\widehat\mu_{\Sigma}^2$ will range between $0$ and $N(N+2)/8$. 
Decomposing the Hilbert space obtained from $N$  tensorial products of $(D_0 + D_0 +D_{\frac 12})$ into a direct sum of irreducible spin $s$ subspaces $D_s$,
one finds that the number $\Delta(s,N)$ of irreducible spaces $D_s$ appearing in the $4^N$ dimensional product space is given by :
$$\Delta(s,N)= \frac{(2\,s +1)}{N+1}\frac{ (2N+2)!}{ (N+2\,s+2)!\, (N-2\,s)!}.$$
For large $N$ and fixed $s$ this behaves as : $\Delta(s,N)\simeq 4^{N+1}\,(1+ 2\,s)/(\sqrt{\pi}\,N^{3/2})$. Applying this result to
the case $s=0$ shows that the eigenvalue $\mu_{\Sigma}^2=0$ will be realized in a rather large fraction of the total Hilbert space,
namely a subspace of dimension $ (2N+2)!/((N+1)! (N+2)!)\simeq 4^{N+1}/(\sqrt{\pi}\,N^{3/2})$ of the total $4^N$-dimensional space. For any $N$, the mean value of $  \mu_{\Sigma}^2 $  (treating all states as equally probable) is  equal to $3 N/16>0$.
The standard deviation of  $  \mu_{\Sigma}^2$ is found to be $\sigma_{\mu_{\Sigma}^2} =\frac 1{16}\sqrt{3(2\,N^2+N)}\approx  0.15 \,N $ (for large $N$).
In the large $N$ limit, one would recover the result of the  classical treatments of Refs \cite{BK,dBHP}.

\smallskip

Let us also note that, if one were considering the coset-type mass (\ref{eq3.40}), including the gravitational-root contributions, the result (\ref{eq3.23}) shows that one would obtain
\begin{equation}
\label{eq3.45}
\widehat\mu_{\rm coset}^{2 \, ({\rm IX})} = \frac{3}{8} \, \frac{1-\Gamma_5}{2} + \frac{3}{8} \, \frac{1+\Gamma_5}{2} = \frac{3}{8} \, ,
\end{equation}
{\it i.e.} a constant, $c$-number result, all over the Hilbert space. This result is reminiscent of the full coset result (\ref{eq3.31}), whose r.h.s., is the quadratic Casimir invariant of the spinor representation of $K(G)$, which is a $c$-number in any irreducible representation.
Note, however, that, though the representation space of the spinor representation of
the Kac-Moody maximally compact subgroup $K(G)$ is finite dimensional, there is an
infinite number of roots $\alpha$, and a corresponding infinite number of generators
$J^s_\alpha$, that one should sum over in the coset result Eq. (\ref{eq3.31}), so that the
value of $\widehat\mu_{\rm coset}^2$ is formally infinite, or, at least, ill defined
as an operator. This indicates the need to renormalize it by some ordering prescription.
This suggests that, in spite of contrary appearances, the renormalized value
$\widehat\mu_{\rm coset}^2=0$ is an allowed possibility.

\smallskip

Let us come back to the result (\ref{eq3.44}) predicted by the usual Einstein-Dirac action, and further discuss its meaning. Let us first note that the non-zero value (\ref{eq3.44}) of $\widehat\mu^2$ is a quantum effect, which directly comes from the anti-commutation relation (\ref{eq3.7}) of the quantized spinor $\chi$. If we reinstate the $\hbar$ on the r.h.s. of (\ref{eq3.7}), we see that the objects $\Gamma_\alpha$ that satisfy the unit-normalized Clifford algebra (\ref{eq3.10}) are actually related to $\chi$ {\it via}: $\chi_{\alpha} = \frac{1}{2} \, \sqrt\hbar \, \Gamma_{\alpha}$. As $\widehat\mu^2$ is quartic in $\chi$ this shows that $\widehat\mu_{\Sigma}^2$ is actually $\frac{3}{8} \, \frac{1-\Gamma_5}{2} \, \hbar^2$, {\it i.e.} of order $\hbar^2$. In turn, this shows that, when one is in a quasi-classical (WKB-type) regime where the gravitational momenta $\pi_{\beta}$ are large compared to $\hbar$, the mass term will have only a sub-leading effect. This is consistent with the finding above that $\mu^2$ has an intrinsic quantum ambiguity due to ordering problems: indeed, the constant $c$ in Eq.~(\ref{eq3.37}) is also seen to be ${\mathcal O} (\hbar^2)$. On the other hand, when considering the Klein-Gordon-type equation defined by  (\ref{eq3.37}) with $\widehat\pi_{\beta} = -i \hbar \, \partial / \partial \beta_a$, we see that a same factor $\hbar^2$ appears in front of the $\beta$-space d'Alembertian $- \, G^{ab} \, \partial_{\beta_a} \partial_{\beta_b}$ and of both the $\Sigma$-generated mass $\widehat\mu_{\Sigma}^2$, and any quantum-ordering contribution $4c$. This means that  the mass term $\mu^2$ will be important when the wave function $\Phi (\beta , \sigma)$ has a characteristic scale of variation in $\beta$-space of order unity.

\smallskip

Let us finally note that the ambiguity in the Einstein-Dirac action linked to using a first-order or second-order formalism is equivalent, in view of Eq.~(\ref{eq3.19}), to adding to $\widehat\mu^2$ a term proportional to $-  \widehat\mu_{\Sigma}^2 -+ \frac{1}{3} \widehat\mu_{\rm grav}^2$, where $\widehat\mu_{\rm grav}^2$ is the contribution to the coset result (\ref{eq3.31}) coming from the three gravitational roots, {\it i.e.} the second term on the r.h.s. of Eq.~(\ref{eq3.40}) (which is the {\it sum} $\widehat\mu_{\Sigma}^2 + \widehat\mu_{\rm grav}^2$).

\smallskip

This shows that the addition of a suitable multiple of the extra contribution (\ref{eq3.19}) to the Einstein-Dirac action can modify the basic result $\mu_{\Sigma}^2$ in a more general result of the type
\begin{equation}
\label{eq3.46}
(1 -k) \widehat\mu_{\Sigma}^2 + \frac{k}{3} \, \widehat\mu_{\rm grav}^2 + 4c \, ,
\end{equation}
where we took also into account a possible ($c$-number) ordering constant. For instance, if $k = \frac{3}{4}$ this would generate a mass term $\frac{1}{4} \,  (\widehat\mu_{\Sigma}^2 + \widehat\mu_{\rm grav}^2) + 4c$. When $c=0$, this would correspond (modulo the factor $\frac{1}{4}$) with the coset-like result (\ref{eq3.45}). On the other hand, if we choose $k = \frac{3}{4}$ and $4c = -3/32$, we end up with a vanishing total squared-mass.

\smallskip

Keeping in mind all these (quantum) ambiguities in the value of $\widehat\mu^2$, we shall now discuss in more detail the quantum dynamics of the Bianchi-IX-spinor system.

\subsection{Quantum dynamics of the Bianchi-IX-spinor system}\label{ssec3.6}

The Wheeler-DeWitt equation for the Bianchi-IX-Majorana-spinor system has the form
\begin{eqnarray}
\label{eq3.47}
\biggl( - \frac{1}{4} \, G^{ab} \, \partial_{\beta_a} \, \partial_{\beta_b} +V_g^{\rm IX} (\beta) + \widehat{V}^{\rm centrif} (\beta , \chi) + \widehat{V}^{\rm IX}_{s \, {\rm grav}} (\beta , \chi) + \nonumber \\
+ \widehat{V}_{s \, {\rm mass}} (\beta , \chi) + \frac{1}{4} \, \widehat{\mu}_{\rm tot}^2 (\chi) \biggl) \Phi (\beta , \sigma) = 0 \, .
\end{eqnarray}
Here, the spin-independent term $V_g^{\rm IX} (\beta)$ is the usual bosonic  type-IX potential (\ref{eq2.30}), while there appears three different spin-dependent potentials. Using our results above, their explicit expressions are (for type IX)
\begin{eqnarray}
\label{eq3.48}
\widehat V^{\rm centrif}_{\rm IX}(\beta , \chi) &= &\frac{1}{32} \left( \frac{1}{\sinh^2 (\beta_1 - \beta_2)} + \frac{1}{\sinh^2 (\beta_2 - \beta_3)} + \frac{1}{\sinh^2 (\beta_3 - \beta_1)} \right) \nonumber \\
&&\times \frac{1-\Gamma_5}{2} \, , \\
\label{eq3.49}
\widehat V_{s \, {\rm grav}}^{\rm IX} (\beta , \chi) &= &- \frac{1}{4} \, (e^{-2\beta_1} + e^{-2\beta_2} + e^{-2\beta_3})(i \, \Gamma_1 \Gamma_2) \, \frac{1+\Gamma_5}{2} \, , \\
\label{eq3.50}
\widehat V_{s \, {\rm mass}}^{\rm IX} (\beta , \chi) &= &+ \, m \, e^{-(\beta_1 + \beta_2 + \beta_3)} \, (i \, \Gamma_2 \Gamma_3) \, \frac{1+\Gamma_5}{2} \, .
\end{eqnarray}
The various possible values of $\widehat{\mu}_{\rm tot}^2$ have been discussed in the previous subsection.

\smallskip

The discrete spin label $\sigma$, indicated as argument of $\Phi$, takes four values, and reminds us that the various spin operators $\Gamma_5$, $i \, \Gamma_1 \Gamma_2$ and $i \, \Gamma_2 \Gamma_3$ act upon a $4$-dimensional Hilbert space. In other words, Eq.~(\ref{eq3.47}) is similar to the second-order form of the Dirac equation coupled to an external electromagnetic field, namely
\begin{equation}
\label{eq3.51}
\left[ (i \, \partial_{\mu} - e \, A_{\mu})^2 + \frac{1}{2} \, e \, \sigma^{\mu\nu} \, F_{\mu\nu} + m^2 \right] \psi = 0
\end{equation}
where $\sigma^{\mu\nu} = \frac{i}{2} \, \gamma^{\mu\nu}$ is the spin generator (in the Clifford algebra).

\smallskip 

In the case of the usual Dirac equation (\ref{eq3.51}), the spin coupling term $\frac{1}{2} \, e \, \sigma^{\mu\nu} \, F_{\mu\nu}$ (linked to the magnetic moment of the electron) couples the $4$-different components of the (first-quantized) Dirac spinor $\psi$, and embodies physical phenomena such as the precession of the electron spin in an external magnetic field. Similarly, we can think of the wave function of the universe $\Phi (\beta)$ as a column vector of $4$ components $\Phi_{\sigma} (\beta)$, which propagate in the Lorentzian $\beta$-space, and are deflected by the spin-independent potential $V_g^{\rm IX} (\beta)$, together with the three spin-dependent potentials (\ref{eq3.48})--(\ref{eq3.50}) which are analogous to the $\frac{1}{2} \, e \, \sigma^{\mu\nu} \, F_{\mu\nu}$ spin-coupling term in Eq.~(\ref{eq3.51}). In addition, the mass term $\widehat{\mu}_{\rm tot}^2 (\chi)$ is also spin-dependent, as discussed in the previous subsection.

\smallskip

In this work, we shall focus on understanding the dynamics of the universe's multi-component wave function $\Phi (\beta)$ in the BKL regime where one approaches the singularity, {\it i.e.} in the limit where the various $\beta_a$'s all tend to $+ \, \infty$. For concreteness, let us first consider the case where $\widehat\mu^2$ is given by the Bianchi-IX prediction (\ref{eq3.44}) together with a possible ordering constant, but without any further contribution of the type of Eq.~(\ref{eq3.19}). In that case we have
\begin{equation}
\label{eq3.52}
\widehat\mu^2 = \frac{3}{8} \, \frac{1-\Gamma_5}{2} + 4c \, .
\end{equation}

Let us start by noting that the helicity operator $\Gamma_5$ commutes with all the terms in the Wheeler-DeWitt equation (\ref{eq3.47}). This means that the $4$ components of $\Phi$ can be viewed as the superposition  of two independent, $2$-component wave functions; say $\Phi_+$ and $\Phi_-$, with
\begin{equation}
\label{eq3.53}
\Phi_{\pm} = \frac{1 \pm \Gamma_5}{2} \, \Phi \, , \quad \Phi = \Phi_+ + \Phi_- \, ,
\end{equation}
and where $\Phi_+$ and $\Phi_-$ undergo two uncoupled evolutions. More precisely $\Phi_+$ (which is a positive-helicity eigenstate $\Gamma_5 \, \Phi_+ = + \Phi_+$) satisfies (with $\Box_{\beta} \equiv G^{ab} \, \partial_{\beta_a} \, \partial_{\beta_b}$)
\begin{eqnarray}
\label{eq3.54}
&&\biggl( - \frac{1}{4} \, \Box_{\beta} + V_g^{\rm IX} (\beta) - \frac{1}{4} \, (e^{-2\beta_1} + e^{-2\beta_2} + e^{-2\beta_3}) \, i \, \Gamma_1 \Gamma_2 \nonumber \\
&&+ \, m \, e^{-(\beta_1 + \beta_2 + \beta_3)} \, i \, \Gamma_2 \Gamma_3 + c \biggl) \Phi_+ (\beta) = 0 \, ,
\end{eqnarray}
while $\Phi_- (\Gamma_5 \, \Phi_- = - \Phi_-)$ satisfies
\begin{eqnarray}
\label{eq3.55}
&&\biggl( - \, \frac{1}{4} \, \Box_{\beta} + V_g^{\rm IX} (\beta) + \frac{1}{32} \Bigl( \frac{1}{\sinh^2 (\beta_1 - \beta_2)} + \frac{1}{\sinh^2 (\beta_2 - \beta_3)} \nonumber \\
&&+ \, \frac{1}{\sinh^2 (\beta_3 - \beta_1)} \Bigl)+ \frac{3}{32} + c \biggl) \Phi_- (\beta) = 0 \, .
\end{eqnarray}

These two sub-equations have rather different structures: (i) they contain a different mass term, (ii) the $\Phi_-$ equation is spin-independent ({\it i.e.} the two independent components of $\Phi_-$ satisfy the same equation), while the two-independent components of $\Phi_+$ are coupled {\it via} the presence of the $i \, \Gamma_1 \Gamma_2$ and $i \, \Gamma_2 \Gamma_3$ spin-coupling terms; and (iii) they contain different spin-averaged potentials (besides the mass term). Concerning the last property, the spin-averaged potential term for $\Phi_+$ (taking into account that $(i \, \Gamma_a \Gamma_b)^2 = + \, 1$, {\it i.e.} that the eigenvalues of $i \, \Gamma_a \Gamma_b$, $a \ne b$, are $\pm \, 1$) is
\begin{eqnarray}
\label{eq3.56}
V_+ (\beta) &= &V_g^{\rm IX} (\beta) = \frac{1}{2} \,  \, (e^{-4\beta_1} + e^{-4\beta_2} + e^{-4 \beta_3}) - \nonumber \\
&&-(e^{-2(\beta_1 + \beta_2)} + e^{-2 (\beta_2 + \beta_3)} + e^{-2 (\beta_3 + \beta_1)}) \, ,
\end{eqnarray}
while the spin-averaged potential for $\Phi_-$ is
\begin{eqnarray}
\label{eq3.57}
V_- (\beta) &= &V_g^{\rm IX} (\beta) + \frac{1}{32} \biggl( \frac{1}{\sinh^2 (\beta_1 - \beta_2)} + \frac{1}{\sinh^2 (\beta_2 - \beta_3)} \nonumber \\
&&+ \frac{1}{\sinh^2 (\beta_3 - \beta_1)} \biggl) \, .
\end{eqnarray}

We note that the spin-averaged potential $V_+ (\beta)$ for $\Phi_+$ is the usual type-IX potential for a {\it diagonal} metric, which contains only gravitational walls, but no symmetry (or centrifugal) walls. Classically, this potential (approximately) confines the motion of the $\beta$ particle (when considering the near-singularity limit) to stay within the Lorentzian billiard chamber defined by the condition that the three gravitational wall forms be positive:
\begin{equation}
\label{eq3.58}
w_{123}^g (\beta) = 2 \, \beta_1 \geq 0 \, , \quad w_{231}^g (\beta) = 2 \, \beta_2 \geq 0 \, , \quad w_{312}^g (\beta) = 2 \, \beta_3 \geq 0 \, .
\end{equation}
At the quantum level, this confinement mechanism will be blurred by tunnelling effects. As is well-known the physics of  tunnelling in quantum cosmology crucially depends on the choice of boundary conditions in configuration space, see, e.g.,  \cite{Vilenkin}. Leaving a discussion of more general states to future work, we shall focus here on wave functions which are quantum analogs of the classical billiard motions within the appropriate billiard chamber ({\it e.g.} the chamber (\ref{eq3.58}) when considering the $\Phi_+$ component). For such states, taking the ordering constant $c$ in (\ref{eq3.54}) to its naive value, $c=0$, we can qualitatively describe the quantum evolution of $\Phi_+$ by looking at the various potential terms in (\ref{eq3.54}). First, we note that, when trying to penetrate the gravitational walls (\ref{eq3.58}), {\it i.e.} when exploring regions where some of the wall forms, {\it e.g.}, $2 \, \beta_1$, become negative, the spin-averaged bosonic  potential $V_+ (\beta) = V_g^{\rm IX} (\beta)$ will grow like $+ \, \frac{1}{2} \, e^{-4\beta_1}$, and will therefore dominate over the corresponding (non positive-definite) growing gravitational spin-dependent potential $\propto e^{-2\beta_1} \, i \, \Gamma_1 \Gamma_2$. In addition, the spin-dependent potential related to the mass $m$ of the spinor becomes exponentially small in the near-singularity limit where the volume of the universe, $\propto \sqrt g = e^{-(\beta_1 + \beta_2 +\beta_3)}$, tends to zero ({\it i.e.} when the sum $\beta_1 + \beta_2 + \beta_3 \to + \, \infty$). This discussion shows that there will exist quasi-classical spinor-like wave functions $\Phi_+ (\beta)$ (with two independent components) consisting of WKB-like solutions approximately bouncing between the walls of the chamber (\ref{eq3.58}), and decaying in the ``forbidden'' domain, $\beta_1 < 0$, $\beta_2 < 0$, $\beta_3 < 0$. Compared to the usual, pure gravity Wheeler-DeWitt equation for Bianchi-IX, which would be
\begin{equation}
\label{eq3.59}
\left( - \, \frac{1}{4} \, \Box_{\beta} + V_g^{\rm IX} (\beta) \right) \Phi_0 (\beta) = 0 \, ,
\end{equation}
for a scalar-valued (one component) $\Phi_0 (\beta)$, the interesting new feature of the gravity-spinor system is the presence of additional spin-dependent couplings. In the case of the $\Phi_+$ equation (\ref{eq3.54}), these are the terms containing $i \, \Gamma_1 \Gamma_2$ and $i \, \Gamma_2 \Gamma_3$. As these two (Clifford algebra) operators do not commute among themselves, their presence means that the two independent components of $\Phi_+$ will continuously mix under the influence of these terms. We thereby end up with the picture of a quantum fermionic billiard where the various polarization state of a spinor-valued wave function mix as the quantum state bounces within the billiard chamber. We shall discuss below the precise link between such a quantum fermionic billiard, and its Grassmann-valued correspondent, studied in Ref.~\cite{DaHi}.

\smallskip

In the BKL limit $\beta_1 + \beta_2 + \beta_3 \to + \, \infty$, the term proportional to $m$ in the evolution equations for $\Phi$ become negligible. Let us then consider the special case where $m=0$, which should also describe the general behaviour when $\beta_1 + \beta_2 + \beta_3 \to + \, \infty$. For the case $m=0$ (and $c=0$), the evolution equation (\ref{eq3.54}) for $\Phi_+$ further simplifies in that it contains only one spin-dependent operator, namely $i \, \Gamma_1 \Gamma_2$. In that case, the idempotent operator $i \, \Gamma_1 \Gamma_2$ {\it commutes} with the Hamiltonian, so that we can further decompose $\Phi_+$ into two (scalar) components, say
\begin{equation}
\label{eq3.60}
\Phi_+^{(\pm)} := \frac{1 \pm i \, \Gamma_1 \Gamma_2}{2} \, \Phi_+ \, ,
\end{equation}
that evolve independently of each other. Namely, these components satisfy
\begin{equation}
\label{eq3.61}
\left( - \, \frac{1}{4} \, \Box_{\beta} + V_g^{\rm IX} (\beta) \mp \frac{1}{4} \, (e^{-2\beta_1} + e^{-2\beta_2} + e^{-2\beta_3}) \right) \Phi_+^{(\pm)} (\beta) = 0 \, .
\end{equation}
In other words, when $m=0$, we can reduce the dynamics of the multi-component wave function $\Phi (\beta , \sigma)$ to the uncoupled dynamics of several separate components satisfying different scalar, Wheeler-DeWitt equations.

\smallskip

Turning to the dynamics of $\Phi_-$, Eq.~(\ref{eq3.55}), we already noted that it did not contain spin-dependent terms. This means that, similarly to the $m=0 \, , \Phi_+$ case discussed above, the two components of $\Phi_-$ satisfy (even when $m \ne 0$) the same scalar, Wheeler-DeWitt equation (\ref{eq3.55}). The latter equation has two interesting new features compared to the $\Phi_+$ one. First, though the off-diagonal, gauge-like Euler-angle degrees of freedom have been eliminated by the diffeomorphism constraints, they have left behind new centrifugal contributions, $\propto 1/\sinh^2 (\beta_1 - \beta_2) + {\rm cyclic}$, to the  potential (see Eq.~(\ref{eq3.57}). These centrifugal terms create infinite potential barriers located at the three symmetry walls: $\beta_1 = \beta_2$, $\beta_2 = \beta_3$ and $\beta_3 = \beta_1$. To examine the effect of these barriers on the quantum wave function, let us consider, say, what happens near the $\beta_1 = \beta_2$ symmetry wall. Locally, we can change coordinates in $\beta$-space and use $x \equiv \beta_1 - \beta_2$ as new coordinate. As the hyperplane $\beta_1 = \beta_2$ is timelike ({\it i.e.} its normal is spacelike), the coordinate $x$ is a spatial coordinate in the Lorentzian $\beta$-space. This means that the $\Phi_-$ equation has, near the $x=0$ hyperplane, the structure
\begin{equation}
\label{eq3.62}
\left( \partial_t^2 - \partial_y^2 - \partial_x^2 + \frac{\alpha^2}{x^2} + \mbox{regular} \right) \Phi_- (t,y,x) = 0 \, ,
\end{equation} where we completed the spatial coordinate $x$ by another spatial one, $y$, together with some time-like variable $t$, and where $\alpha^2$ denotes a positive numerical constant linked to the coefficient of the centrifugal terms. [In the case considered here
we have  $\alpha^2=1/16$, but it is instructive to leave its value arbitrary.]
Here, $t,x,y$ are some linear combinations of the three variables $\beta_1 , \beta_2 , \beta_3$, chosen so that the Lorentzian metric $G_{ab} \, d\beta_a \, d\beta_b = -dt^2 + dx^2 + dy^2$. We can separate the motion w.r.t. $t$ and $y$, {\it i.e.} $\Phi_- = e^{iky - i\omega t} \, \varphi (x)$. Finally, the behaviour of $\varphi (x)$ near the singular point $x=0$ is given by $\partial_x^2 \, \varphi (x) \simeq \alpha^2 \, x^{-2} \, \varphi (x)$. This implies that $\varphi (x) \sim x^s$ with a power $s$ satisfying the indicial equation $s(s-1) = \alpha^2$, so that $s_{\pm} = \frac{1}{2} \pm \sqrt{\frac{1}{4} + \alpha^2}$. The important point is that the two possible solutions satisfy $s_+ > 1$ and $s_- < 0$. If we reject the possibility of a singular wave function at $x=0$, this eliminates $\varphi_- (x) \sim x^{s_-}$. Finally, we conclude that the centrifugal terms force the wave function to vanish at each symmetry wall like $\vert \beta_1 - \beta_2 \vert^{s_+}$, $\vert \beta_2 - \beta_3 \vert^{s_+}$ and $\vert \beta_3 - \beta_1 \vert^{s_+}$, with $s_+ > 1$. In other words, the centrifugal terms confine the evolution of $\Phi_- (\beta)$ to six separate chambers, corresponding to the six possible orderings of $\beta_1 , \beta_2 , \beta_3$: namely, $\beta_1 \leq \beta_2 \leq \beta_3$, $\beta_2 \leq \beta_1 \leq \beta_3$, {\it etc.} In each symmetry chamber, the WKB-like solutions for $\Phi_-$ will bounce between two ``hard'' symmetry walls ({\it e.g.} $\beta_1 = \beta_2 < \beta_3$ and $\beta_1 < \beta_2 = \beta_3$ for the chamber $\beta_1 \leq \beta_2 \leq \beta_3$) and the ``soft'' (exponential) gravitational wall ({\it e.g.} the $2 \, \beta_2 = 0$ wall for the $\beta_1 \leq \beta_2 \leq \beta_3$ chamber).

\smallskip

If we use the naive ordering constant $c=0$ (which led to a ``massless'' Wheeler-DeWitt equation for $\Phi_+$) we see in Eq.~(\ref{eq3.55}) that the $\Phi_-$ components have acquired a positive squared-mass $\mu_-^2 = 3/32$. This mass term will strongly affect the near-singularity behaviour of the $\Phi_-$ wave function. Indeed, as there exist wave packets of the free {\it massive} Klein-Gordon equation $(\Box_{\beta}  - \mu_-^2) \, \Phi = 0$ that are approximately localized (in $\beta$ space) around, say, a time-like line $\beta_a = v^a \, \tau + \beta_a ^{(0)}$, with $G_{ab} \, v^a \, v^b < 0$, such wave packets yield approximate asymptotic solutions of the Wheeler-DeWitt equation (\ref{eq3.55}), because, as $\beta_1 + \beta_2 + \beta_3 \to + \, \infty$ (within, say, the chamber $\beta_1 \leq \beta_2 \leq \beta_3$) the potential terms will become negligible in the domain where the wave packet is approximately localized. This is the quantum version of the result of \cite{BK,dBHP} that a positive $\mu^2$ ultimately quenches the chaotic billiard motion, in the near singularity limit.

However, let us note that, when $c=0$, the fact that the q-number $\widehat{\mu}^2$
admits the eigenvalue zero means that there will always exist a part of the quantum wavefunction
(namely the one described by its $\Phi_+$ component) which will
exhibit a chaotic behavior near the singularity. In other words, at the quantum level,
the classical result of \cite{BK,dBHP} does not prevent part of the quantum reality to
behave chaotically.

\section{Quantum versus Grassmannian fermionic billiards}\label{sec4}

The two crucial new features brought by coupling gravity to a spinor are: (i) the appearance
of new spin-dependent potential terms; and (ii) the possible presence of a squared-mass term
in the Klein-Gordon-Wheeler-DeWitt equation.  As discussed above, though the feature (ii) 
is important for knowing whether the mixmaster chaos is affected or not by the coupling
to a quantum spinor, it is also a delicate quantum effect, which is sensitive to the choice
of ordering in the Wheeler-DeWitt  equation, as well as to other choices. We shall assume
in this Section either that we have made choices so that the relevant 
squared-mass term vanishes\footnote{This would, for instance, be the case for the $\Phi_+$ components of $\Phi$, with the
naive ordering constant $c=0$, see Eq. (\ref{eq3.54}).}, or that we are considering situations
where one can distinguish the effect of the spin-dependent potential from other effects.
Under this assumption, we can focus on the effect of the spin-dependent potential terms,
and contrast a {\it quantum} treatment of these spin-dependent terms, to a treatment
where the spinor variables are considered as Grassmann-valued (G-numbers).

The recent work \cite{DaHi}  has studied in detail the fermionic billiards defined
by the chaotic dynamics of G-valued  (spin-3/2) spinor fields,
as they arise in the near-singularity limit of supergravity (both in dimension eleven,
and dimension four).  The two main results of Ref. \cite{DaHi} that are relevant to our
present study are: (1)  the spin-3/2 billiard defined by supergravity factorizes into
a spin-1 (vector)  billiard, and a spin-1/2 (Majorana-spinor) one; in view of the results
of Ref.  \cite{dBHP},  it is the latter spin-1/2
billiard which is of relevance for us here; and (2) the spin-1/2 billiard consists of a succession
of generalized Weyl reflections, defined in a kind of spin-covering $\mathcal{W}^s$ 
of the standard Weyl group 
of the Kac-Moody algebra associated with the considered supergravity model. 
More precisely, it was found that each collision of the universe on a wall labelled
by a Kac-Moody root $\alpha$ (i.e. so that the corresponding wall form is
simply $w(\beta) =\alpha(\beta)$), corresponding to a  spin-dependent potential proportional to 
$ \exp (- \alpha(\beta) ) \,  i  \, \chi^{\dagger} J_{\alpha}^s \, \chi$,  causes
the G-valued spin-1/2 field $\chi$ to ``rotate'', in the spin-1/2 representation space, 
by an angle of $\pi/4$ along the axis defined by $ J_{\alpha}^s$. More precisely,
the value of the Dirac spinor $\chi$ after the collision on the wall $\alpha$ differs
from its incident value by the matrix

\begin{equation}
\label{Ralpha}
\mathcal{R}_{\alpha} = e^{\varepsilon_\alpha  \frac{\pi}{2}  J_{\alpha}^s} \, .
\end{equation}
Here, $\varepsilon_\alpha = \pm 1$ is a sign which will be defined below, and
the object $J_{\alpha}^s$, which was introduced above, is an anti-hermitian generator which represents (in the spinor representation $s$) the ``rotation'' $E_{\alpha} - E_{-\alpha}$ of the maximally compact subgroup $K$ of $G$ associated to the root $\alpha$.  For instance, in the 
($3+1$)-dimensional case of relevance here, the generators $J_{\alpha}^s$ associated,
respectively, to the symmetry walls and the leading gravitational walls, were
written down in Eqs.   (\ref{eq3.38}),   (\ref{eq3.39}). Note that the generators $ J_{\alpha}^s$
all include a factor $1/2$, so that their eigenvalues are $\pm \frac{1}{2}$. 
[In conjunction with the
$\frac{\pi}{2}$ prefactor, this implies, as stated, a rotation angle of $\pm \frac{\pi}{4}$.]
Note also that all the generators $ J_{\alpha}^s$ belong to the Clifford algebra defined
by the usual ($3+1$)-dimensional gamma matrices, so that the spinor-wall-reflection matrices
(\ref{Ralpha}) are $4 \times 4$ matrices acting on the usual Dirac spinor space. [More precisely,
they are real, and act on the real Majorana spinor space.]

The picture of fermionic billiards emerging from Ref.  \cite{DaHi} is a  growing, chaotic
succession of spinor reflections 

\begin{equation}
\label{Rproduct}
\mathcal{R}_{\alpha_n}  .  \mathcal{R}_{\alpha_{n -1}} \cdots \mathcal{R}_{\alpha_2} . \mathcal{R}_{\alpha_1}
\end{equation}

acting on the spinor index of the G-valued spinor $\chi$.
It was found in Ref.  \cite{DaHi} that the multiplicative group  $\mathcal{W}^s$ 
defined by spin-1/2
billiard, i.e. by all the matrix products (\ref{Rproduct}), is of finite cardinality,
both in the eleven-dimensional case, and the four-dimensional one. For example,
in the 4-dimensional case of relevance here, the group of products (\ref{Rproduct})
of the basic reflection generators, Eqs.   (\ref{eq3.38}),   (\ref{eq3.39}), associated with
the symmetry and gravitational walls is a finite group of cardinality $4 \times 48$.
Actually, it happens that the 4-dimensional case is quite special in that the
symmetry generators happen to {\it commute} with the gravitational wall ones
(which, themselves, reduce to only one element as the three generators (\ref{eq3.39})
differ only by a cyclic permutation of 123, which does not change the value of
$\gamma^{123}$). Due to this,  the group  $\mathcal{W}^s$ is the direct product
of two separate groups.

Having recalled the results concerning the near-singularity limit of the dynamics
of G-valued spinors coupled to a chaotic bosonic  cosmology, we wish now to
clarify the connection of these results to the case of quantum spinors, sourcing
a cosmological model.  In that case, we must replace the classical evolution
of a G-valued spinor $\chi$, considered along the classical evolution of the
bosonic billiard described by the $\beta$-particle bouncing between symmetry
and gravitational walls, by a solution of the coupled multi-component Wheeler-DeWitt
equation.  However, in order to be able to compare the two treatments we must consider
a case where the Wheeler-DeWitt equation do correspond to the spinor coupling
terms studied in Refs. \cite{dBHP,DaHi}. Indeed, the latter works considered
classical Hamiltonians of the form

\begin{equation} \label{Hspincoupling}
H= \frac{1}{4} \, G^{ab} \, \pi_{\beta_a} \, \pi_{\beta_b} + \sum_{\alpha} \frac{1}{2} 
e^{-2 \alpha(\beta)} \Pi_\alpha^2 + \sum_{\alpha} \frac{1}{2} 
e^{- \alpha(\beta)} \Pi_\alpha \Sigma_\alpha + c_{\rm tot}
\end{equation}
Here, the $\Pi_\alpha$'s are dynamical variables whose meaning is different if one
considers a gravity model or a coset model, the $ \Sigma_\alpha$'s are the  spin-coupling
terms corresponding to a general root, namely

\begin{equation} \label{sigmaalpha}
\Sigma_\alpha = i \, \chi^{\dagger} J_{\alpha}^s \, \chi
\end{equation}
where the $J_{\alpha}^s$'s are the products of usual gamma matrices defined
in Eqs. (\ref{eq3.38}), (\ref{eq3.39}) for symmetry and gravitational roots, and
where the constant $c_{\rm tot}$ (defining the squared mass $\mu^2= 4 c_{\rm tot}$)
includes various types of contributions, which depend on the model considered (as discussed above).

The important feature of the spin-couplings included in (\ref{Hspincoupling}) is that
the {\it same} quantity $\Pi_\alpha$ enters the bosonic  potential wall
$\frac{1}{2}  e^{-2 \alpha(\beta)} \Pi_\alpha^2$, and the corresponding  spin-dependent
potential term $\frac{1}{2} e^{- \alpha(\beta)} \Pi_\alpha \Sigma_\alpha $.
One can verify that this is indeed the case for the general Hamiltonian (\ref{eq2.69}),
both for the terms linked to the dominant gravitational walls (with, e.g., $\alpha_{123} = 2 \beta_1$,
$\Pi_{\alpha_{123}}= -C^1_{\   23}$, and $J_{\alpha_{123}}$ given by (\ref{eq3.39}) ), and to the symmetry walls. However,
in the latter case, as Eq. (\ref{Hspincoupling}) refers to an Iwasawa decomposition of the metric
one must reconsider the far-wall limit of the Gauss-decomposition-based Hamiltonian (\ref{eq2.69}).
In that limit (e.g. when $\beta_2- \beta_1 \gg 1$) one finds that the various hyperbolic
functions in Eq. (\ref{eq2.76}) do yield a structure compatible with the general Iwasawa
result  (\ref{Hspincoupling}) (with, e.g., $\Pi_{\alpha_{12}} =  \pi_{w^{12}}$).

In order to see more clearly the differences between the Grassmannian treatment
of spinor couplings, and its quantum treatment, let us consider the quantization
of the simplest version of the general Hamiltonian  (\ref{Hspincoupling}), namely
the case where there is only {\it one wall}, corresponding to one root $\alpha$. 
In that case, the dynamical variable $\Pi_\alpha$ is clearly seen to be a constant
of motion, and can therefore be considered as being a c-number both in a
classical and a quantum treatment. [In the quantum treatment, we consider
eigenstates of the operator $\Pi_\alpha$.]  It then remains to quantize $\beta$
and $\chi$.  Canonical quantization of the
$\beta$ dynamics yields $ \pi_\beta = - i \partial_\beta$, while the quantization
of $\chi$ is done according to the anticommutation relations (\ref{eq3.5}), or
(\ref{eq3.7}) in the Majorana case that we shall consider here. This yields a quantum
Hamiltonian of the form

\begin{equation} \label{Hquantumalpha}
H= -\frac{1}{4} \, G^{ab} \, \partial_{\beta_a} \, \partial_{\beta_b} +  \frac{1}{2} 
e^{-2 \alpha(\beta)} \Pi_\alpha^2 +  \frac{1}{2} 
e^{- \alpha(\beta)} \Pi_\alpha \widehat{\Sigma}_\alpha + c_{\rm tot}
\end{equation}

where $\widehat{\Sigma}_\alpha$ is the quantum operator defined by replacing
$\chi$ in Eq. (\ref{sigmaalpha}) by its quantized version, submitted to (\ref{eq3.7}).

As explained above, the quantum wavefunction $\Phi(\beta, \sigma)$ has both
continuous indices related to the  $\beta$ dynamics, and a discrete one, $\sigma$,
related to the Clifford algebra (\ref{eq3.7}) satisfied by the quantum $\chi$.
Let us consider solutions of $ H \Phi(\beta, \sigma)=0$ describing wavepackets
colliding on the $\alpha$ potential wall. One can write such solutions explicitly
by  separating the variables in Eq.  (\ref{Hquantumalpha}). Namely, as the potential
terms depend only on the combination $\alpha(\beta)$ of the $\beta$'s, we can look
for wavefunctions $\Phi$ of the form

\begin{equation}
\Phi(\beta, \sigma) = e^{i k_{\parallel}(\beta)} F(\beta_{\perp}, \sigma)
\end{equation}

where the linear form $k^{\parallel}$  describes the $\beta$-space momentum
parallel to the wall $\alpha(\beta)=0$  (i.e. $G^{a b} k^{\parallel}_a  \alpha_b =0$),
while $ F(\beta_{\perp}, \sigma)$, where $\beta_{\perp} := \alpha(\beta)$,
describes the motion perpendicular to the $\alpha$ wall
(as well as the spin degrees of freedom).  Inserting this separated wavefunction in the
Wheeler-DeWitt equation $ H \Phi(\beta, \sigma)=0$ (and considering a root $\alpha(\beta)$
of norm  $G^{ab} \alpha_a \alpha_b = 2$, as is the case for all the roots considered here)
yields the following equation for $F$:

\begin{equation} \label{Morse1}
\partial_{\beta_{\perp}}^2 F(\beta_{\perp},\sigma) = (e^{- 2  \beta_{\perp}} \Pi_\alpha^2 + e^{- \beta_{\perp}} \Pi_\alpha \widehat{\Sigma}_\alpha  - {\cal Q} ) F(\beta_{\perp},\sigma)
\end{equation}
where $\cal Q$ is a separation constant which involves both $ (k^{\parallel})^2 :=G^{a b} k^{\parallel}_a  k^{\parallel}_b$ and the constant $c_{\rm tot}$ in  Eq.  (\ref{Hquantumalpha}).
If $ \widehat{\Sigma}_\alpha$ were a c-number, this would be the Schr\"odinger equation in a Morse potential, which is well-known to be reducible
to the general confluent hypergeometric equation.  In our case, $ \widehat{\Sigma}_\alpha$
is an operator in spin space (i.e. acting on the index $\sigma$ labelling the four
components of the wavefunction $\Phi$). However,  we can reduce our problem to a
c-number-valued $ \widehat{\Sigma}_\alpha$ by first considering wavefunctions $\Phi(\sigma)$ 
that are eigenstates of  this operator: say $ \widehat{\Sigma}_\alpha \Phi(\sigma) = {\Sigma}_\alpha  \Phi(\sigma)$.  We can also simplify the resulting equation by 
shifting $\beta_{\perp} $ by a constant so as to absorb the $\Pi_\alpha^2$ factor multiplying
$ e^{- 2  \beta_{\perp}} $. Namely, one defines $\beta'_{\perp}$ such that
$ e^{- 2  \beta'_{\perp}} = \Pi_\alpha^2  e^{- 2  \beta_{\perp}} $
This makes also  $\Pi_\alpha$ disappear from the second term,
except for its sign: $\varepsilon_\alpha := \text{sgn}(\Pi_\alpha) = \pm 1$. 
When $\cQ$ is positive, there is a solution of the resulting equation

\begin{equation} \label{Morse2}
\partial_{\beta'_{\perp}}^2 F(\beta'_{\perp},\sigma) = (e^{- 2  \beta'_{\perp}}  + e^{- \beta'_{\perp}} \varepsilon_\alpha {\Sigma}_\alpha -  {\cal Q}) F(\beta'_{\perp},\sigma)
\end{equation}

which starts as an incident plane wave 
$F(\beta'_{\perp},\sigma) \propto \exp - i  \sqrt{ {\cal Q}} \beta'_{\perp} $  faraway 
from the wall, i.e. when $\beta_{\perp} \gg +1 $,  reflects on the wall located
around $\beta'_{\perp} =0$ (with an evanescent wave in the `forbidden' region 
$\beta'_{\perp} < 0$), and ends up, with some dephasing, as an outgoing
plane wave  $F(\beta'_{\perp},\sigma) \propto \exp + i  \sqrt{ {\cal Q}} \beta'_{\perp} $ 
when  $\beta'_{\perp} \to + \infty $. Using the notation $U(a,b,z)$ for the second
Kummer function  (see \cite{AS}, chapter 13), the exact solution of Eq. (\ref{Morse2})
describing this scattering process on the combined bosonic+spinorial wall $\alpha(\beta)$
is given by

\begin{equation}
F [ \beta'_{\perp}]=\exp[-e^{-\,\beta'_{\perp}}]\,e^{-\,i\,\sqrt{{\cal Q}}\,\beta'_{\perp}}U[\frac 12+ \frac 12 \varepsilon_\alpha \Sigma_\alpha +i\,\sqrt{{\cal Q}},1+2\,i\,\sqrt{{\cal Q}},2 \, e^{-\,\beta'_{\perp}}]\\
\end{equation}

We can then extract from this exact expression (using the expansion of the Kummer
$U(a,b,z)$ function near $z=0$) the phase factor $e^{i \delta_\alpha}$
between the incident wave and the outgoing one. We find

\begin{equation} \label{q-phase}
e^{i\, \delta_\alpha(\Sigma_\alpha)} = -\frac{\Gamma(\frac 12+ \frac 12 \varepsilon_\alpha \Sigma_\alpha -i\,\sqrt{{\cal Q}})\,\Gamma(\frac 12+2\,i\,\sqrt{{\cal Q}})}{\Gamma(\frac 12+ \frac 12 \varepsilon_\alpha \Sigma_\alpha+i\,\sqrt{{\cal Q}})\,\Gamma(\frac 12 -2\,i\,\sqrt{{\cal Q}})}
\end{equation}

Let us now compare this quantum dephasing with the Grassmannian  
description, recalled above, of the reflection of  the spinor $\chi_m$ on the
wall $\alpha(\beta)$. [Here, we denote by $m,n, \ldots$
spinor indices, to avoid the confusion with the use of $\alpha$ as a label for the root].
This description was that the value $\chi'$ of the spinor after the interaction
with the wall $\alpha$ is obtained from its value $\chi$ before the interaction via
the following matrix transformation:

\begin{equation} \label{c-phase}
\chi'_m  =  (\mathcal{R}_{\alpha})_{mn}  \chi_n = \left[ e^{\varepsilon_\alpha \frac{\pi}{2}  J_{\alpha}^s} \right]_{mn} \chi_n
\end{equation}
where $\varepsilon_\alpha := \text{sgn}(\Pi_\alpha) = \pm 1$ as above.

Similarly to what happens for the usual Dirac equation, we can consider that
the various components, labelled by $m$, of the spinor $\chi_m$ encode
the classical polarization state of $\chi$. This polarization state can then be also  encoded
by decomposing the spinor $\chi_m$ onto a basis of eigenstates of the
first-quantized hermitian spin generator $i \,  J_{\alpha}^s$. [We recall that  $i \,  J_{\alpha}^s$ is
$i$ times the product of an  even number of ordinary gamma matrices; e.g. for the
symmetry root $\alpha_{12} = \beta_2 -\beta_1$ it is the ordinary first-quantized
spin generator $i \,\frac 12 \gamma^{12}$, whose eigenvalues are $\pm \frac 12$, and
whose eigen-spinors are often used to decompose a general spinor into various spin states.]
We then see that Eq. (\ref{c-phase}) is saying that the classical dephasing, upon
reflection on the wall $\alpha$, of a classical spinor $\chi_m$ polarized so as to
be an eigenstate of $i  J_{\alpha}^s$, with eigenvalue $  \sigma_\alpha = \pm \frac 12$ 
is given by the phase factor $\exp - i \varepsilon_\alpha \frac{\pi}{2} \sigma_\alpha$.

The quantum dephasing  (\ref{q-phase}) looks {\sl a priori} quite different from the classical
dephasing $\exp - i \varepsilon_\alpha  \frac{\pi}{2}  \sigma_{\alpha}$. Let us, however, check that they
agree, as they should, in the quasi-classical, WKB limit. This limit corresponds to
considering a large value of $\sqrt{ {\cal Q} }$, i.e.  high-frequency
wavepackets $\sim   \exp \pm  i  \sqrt{ {\cal Q}} \beta'_{\perp} $ . In addition, we need
to decompose the quantum phase  $ \delta_\alpha(\Sigma_\alpha)$ in  (\ref{q-phase}) into two separate contributions:
(i) a spin-independent part,  $ \delta_\alpha(0)$, which can be mathematically defined by 
replacing  $\Sigma_\alpha$ by zero on the r.h.s. of Eq.  (\ref{q-phase}), and (ii) 
the spin-dependent contribution, $ \delta_\alpha(\Sigma_\alpha) - \delta_\alpha(0)$.
This yields a spin-dependent dephasing factor given by

\begin{equation} \label{q-spinphase}
e^{i\, (\delta_\alpha(\Sigma_\alpha) -  \delta_\alpha(0))} = 
\frac{\Gamma(\frac 12+ \frac 12 \varepsilon_\alpha \Sigma_\alpha -i\,\sqrt{{\cal Q}})}
{\Gamma(\frac 12 -i\,\sqrt{{\cal Q}})}
\frac{\Gamma(\frac 12 +i\,\sqrt{{\cal Q}})}{\Gamma(\frac 12+ \frac 12 \varepsilon_\alpha \Sigma_\alpha +i\,\sqrt{{\cal Q}})}
\end{equation}
Using now the fact that, for large values of $z$, one has 
$\Gamma(z+a)/\Gamma(z) = z^a (1 + O(1/z))$, we see that, modulo fractional
corrections of order $1/\sqrt{{\cal Q}}$, the spin-dependent part of the quantum phase,
i.e. $\delta_\alpha(\Sigma_\alpha) -  \delta_\alpha(0)$, is equal to 
$ - \frac{\pi}{2} \varepsilon_\alpha \Sigma_\alpha$, in perfect agreement  with the classical result  $ - \frac{\pi}{2} \varepsilon_\alpha \sigma_\alpha$.
This result also shows that the  generalized spin operators
$\widehat{\Sigma}_\alpha = i \chi^{\dagger}  J_{\alpha}^s \chi$ are  the {\it second-quantized}
versions of the first-quantized gamma-algebra generators $i  J_{\alpha}^s$, and that 
classical spinors $\chi_m$ that are eigenstates of $i  J_{\alpha}^s$ do correspond, 
after quantizing $\chi$ according to Eq. (\ref{eq3.5}), to quantum eigenstates $\Phi$
of $\widehat{\Sigma}_\alpha $. Note also that the fact that the quantum phase
(\ref{q-phase}) agrees, in the quasi-classical limit, with the result obtained
by integrating the classical equation of motion of $\chi$, namely
$$
\partial_t \chi = \frac 12 e^{- \alpha(\beta)} \Pi_\alpha  J_\alpha^s \chi
$$
finally rests on the fact that, according to the general result (\ref{eq3.8}),
the  Heisenberg-picture quantum operator $\hat{\chi}$ satisfies the same
equation of motion.

We have compared here the quantum and classical descriptions of the
`collision' of the universe on a single (bosonic+spinorial) potential wall $\alpha(\beta)$.
We could discuss, exactly along the same lines, the dynamics of the two $\Phi_+$
components of the wavefunction $\Phi$ in the Bianchi-IX case. Indeed,  $\Phi_+$
satisfies the (two-component) equation (\ref{eq3.54}) which (in the case $m=0$)
is closely similar to  the Wheeler-DeWitt equation associated to Eq. (\ref{Hquantumalpha}),
except that we now have spin-dependent collisions on three different gravitational
walls. Note, however, that the spin evolution is very simple in this case, as the
three gravitational-wall spin operators $J_{\alpha_{123}}^s, J_{\alpha_{231}}^s, J_{\alpha_{312}}^s$ all coincide.  The spin eigenstates being the same for the three
different walls, there is no room for any interesting chaotic behaviour of the
spin polarization, as could happen if one had had three different (and non commuting)
spin operators. As for the two remaining $\Phi_-$ components of the Bianchi-IX $\Phi$
they satisfy the spin-independent equation  (\ref{eq3.55}) so that we cannot
compare its dynamics to the discussion of \cite{DaHi}. 

There is, however, another Bianchi model for which we can compare the
classical and quantum dynamics, it is the Bianchi-II model, defined by structure constants
$C^a \, _{bc} = \varepsilon_{bcd} \, n^{ad}$ with $n^{ab} = {\rm diag} (1,0,0)$. 
The classical dynamics of this model is discussed in Appendix C, using an
Iwasawa fixing of the dreibein.  The diagonal metric variables are denoted $\beta_1, \beta_2, \beta_3$, while the Iwasawa off-diagonal degrees of freedom are denoted
$\nu_{12}, \nu_{23}$ and $\nu_{13}$. [Note that indices are not naturally cyclically permuted
when working in an Iwasawa representation.]
The quantization of $\chi$ is done as above,
while the quantization of the metric degrees of freedom is conveniently done
in a $\beta, \nu$ representation, i.e. with momenta conjugated to the three $\beta$'s
defined as $ \widehat\varpi^a = -i \partial_{\beta_a}$, 
and with momenta conjugated to the three $\nu$'s defined as
$\widehat\varpi^{ab} = -i \partial_{\nu_{ab}}$.

In the Bianchi-II model, there are only two non trivial diffeomorphism constraints
(classically given by Eqs.  (\ref{BIISconst})), in addition to the Hamiltonian constraints (Eq.  (\ref{BIILconst})).
The quantum mechanical versions of these two constraints will then be the following
constraints on the quantum state $\Phi$:

\begin{equation}
\widehat\varpi^{12} \Phi=\widehat\varpi^{13} \Phi=0\label{QSconst} 
\end{equation}

In the $\beta, \nu$ representation, i.e. for a wavefunction $\Phi(\beta, \nu, \sigma)$
(where, as above, the discrete index $\sigma$ labels the spin degrees of freedom),
the constraints  (\ref{QSconst})) imply that $\Phi(\beta, \nu, \sigma)$ does not depend
on $\nu_{12}$ and $\nu_{13}$, but only (besides $\sigma$) 
on  $\beta_1$, $\beta_2$, $\beta_3$ and $\nu_{23}$.  

The remaining constraint is the Hamiltonian one, of the form

\begin{equation}  \label{BIIWDW1}
\widehat{\cal H}\Phi=0
\end{equation}

Assuming for simplicity a vanishing Dirac mass, $m=0$, but allowing, as above, for
an ordering constant $c$, the Hamiltonian operator 
[whose classical version is  (Eq.  (\ref{BIILconst})) ] may be written as
\begin{eqnarray} \label{qHII}
\widehat {\cal H}&=&  \frac 14\left[\sum_i(\widehat\varpi^i)^2-\frac 12 (\sum_i\widehat\varpi^i)^2\right]+ \frac 12e^{-4\beta_1} \nonumber\\
&&
+ \frac18 \left[ \widehat \Sigma^{[\hat 1\hat 2]}\strut^2+ \widehat \Sigma^{[\hat 3\hat 1]}\strut^2+(\widehat \Sigma^{[\hat 2\hat 3]}+ 2\,e^{(\beta_2-\beta_3)} \widehat{\varpi}^{23})^2  \right]     -\frac {e^{-2\beta_1}}4  i \chi^\dagger\gamma_0 \gamma_{123} \chi \nonumber\\
&&+ c
\end{eqnarray}

It is amusing to note that the quantum dynamics of this Bianchi-II model is
simpler to analyze than its classical dynamics. Indeed, as discussed in Appendix C,
it is difficult to solve generically the classical (or Grassmannian) spinor evolution equation
(\ref{SWeq}). By contrast, in the Schr\"odinger-Wheeler-DeWitt picture of  Eq. (\ref{BIIWDW1})
there are no  fermionic equations to solve.  In addition, we have that: (1)
the spin dependence of the Hamiltonian is analog to that of a {\it symmetric} top, i.e. involving the spin operators $\widehat \Sigma^{ab}$
only through $\widehat \Sigma^{[\hat 2\hat 3]}$, and   the combination $\widehat \Sigma^{[\hat 1\hat 2]}\strut^2+\widehat \Sigma^{[\hat 1\hat 3]}\strut^2$;  and (2) the additional spin-dependent term linked to
the gravitational wall, i.e. $ i \chi^\dagger\gamma_0 \gamma_{123} \chi$, commutes
with the  $\widehat \Sigma^{ab}$-dependent terms (see the subsection \ref{ssec3.4}). [Note that this commutation property (2) is valid in the case where one would consider a complex,
Dirac spinor $\chi$, because of the independence of the two real Majorana
components of such a general $\chi$.] 

In other words, we can diagonalize the Hamiltonian by imposing that the state $\Phi$
is a simultaneous eigenstate of the following three operators:
\begin{eqnarray}
\widehat \Sigma^{[\hat 2\hat 3]} \Phi = \sigma_{23} \Phi, \nonumber \\
\left( \widehat \Sigma^{[\hat 1\hat 2]}\strut^2+\widehat \Sigma^{[\hat 1\hat 3]}\strut^2+\widehat \Sigma^{[\hat 2\hat 3]}\strut^2 \right) \Phi = \cS^2 \Phi,  \nonumber \\
\frac{i}{4} \chi^\dagger\gamma_0 \gamma_{123} \chi \Phi = \cC_g \Phi
\end{eqnarray}

On each $\sigma$ component of such a state, the Hamiltonian constraint equation
leads to an equation of the form
\begin{equation}
\left[-\frac 14\square_{\beta}+ \frac 12e^{-4\beta_1} -\frac 12 e^{2(\beta_2-\beta_3)}\partial^2_{\nu_{23}}-\frac i2\,{ \sigma_{2 3} }\,e^{\beta_2-\beta_3}\partial_{\nu_{23}}+{\cal C}_g\,e^{-2\beta_1}+\frac 18 \cS^2 +c \right]\Phi=0\, .\label{QGenEq}
\end{equation}

By using the results of subsection \ref{ssec3.4} the allowed values of the quantum numbers ${ s_{2 3} }$, ${\cal C}_g$ and $ \cS^2$, and their multiplicity, for a Majorana spinor are,
$$
[\sigma_{23},\, {\cal C}_g,\, \cS^2]_{\rm mult.}: [0,\pm 1/4,0]_1,[\pm 1/2,0,3/4]_1\, .
$$ 

while, for a Dirac spinor they are :
$$
[\sigma_{23},\, {\cal C}_g,\,\cS^2]_{\rm mult.}: [0,0,0]_3, \, [0,\pm 1/2,0], \,
[\pm 1/2,\pm 1/4,3/4]_2, \, [0,0, 2], \, [\pm 1,0,2] \, .
$$
Note that in the Majorana case $\sigma_{23}$ and $\cC_g$ cannot be both non zero, and we
have the simple link $\cS^2 = 3\, \sigma_{23}^2$. These links are relaxed in the Dirac-spinor case.

Eq. (\ref{QGenEq}) can be solved by separation of variables. Indeed, the wave operator
$\square_{\beta}$ depends on three  $\beta$ variables, while the potential walls
entering the equation involve only two combinations of the three $\beta$'s,
namely $\alpha_{123}(\beta) = 2 \beta_1$ and $\alpha_{23}(\beta) = \beta_3 - \beta_2$.
Therefore there exists a linear combination of the three $\beta$'s which will be `orthogonal'
(in the Lorentz-$\beta$-space sense) to the two combinations $\alpha_{123}(\beta)$
and  $\alpha_{23}(\beta)$. It is easy to see that 
$\alpha_0(\beta) := 2 \beta_1 + \beta_2 + \beta_3$ is
such a combination. Actually, one can easily check that the three variables
$\alpha_0, \alpha_{23}, \alpha_{123}$ define an orthogonal coordinate system
in Lorentzian $\beta$-space, with $\alpha_0$ being a time-like coordinate, and
$ \alpha_{23}$, and $ \alpha_{123}$ two space-like coordinates. Finally, a generic
solution  can always be expressed by superposing separated solutions of the form:

\begin{equation}\label{mode } 
\Phi(\beta, \nu_{23})=e^{i\,p\,\nu_{23}}e^{i\,k\,(2\beta_1+\beta_2+\beta_3)} F_1(2 \beta_1)F_2(\beta_3-\beta_2)
\end{equation}
where the two functions $F_1$ and $F_2$ satisfy exactly the same Morse-potential-type
Schr\"odinger equation that we encountered above (in the single-wall case)  namely:
\begin{eqnarray}
F_1^{\prime\prime}[2 \beta_1]&=&\left(e^{-4\,\beta_1}- 2\, {\cal C}_g\,e^{-2\,\beta_1}-{\cal Q}_1\right) F_1[2\beta_1]\\
F_2^{\prime\prime}[ \beta_3-\beta_2]&=&\left(p^2\,e^{2\,(\beta_2-\beta_3)}+   { \sigma_{2 3} }\,p\,e^{(\beta_2-\beta_3)} - \cQ_2 \right) F_2[\beta_3-\beta_2]
\end{eqnarray}

Here the two separation constants $\cQ_1$ and $\cQ_2$ must satisfy the following
mass-shell condition
\begin{equation}  \label{mass-shell1}
-k^2 + \cQ_1 + \cQ_2 = -\frac 14 \cS^2 - 2 c = - \frac 12 \mu^2
\end{equation}
where $\mu^2$ is the squared-mass of the Wheeler-DeWitt equation.
If we look, as above, for wave functions that vanish behind the $ \alpha_{23}$, and 
$ \alpha_{123}$ walls, we must choose  Kummer's $U$-type solution for $F_1$
and $F_2$.

When ${\cal Q}_1 $ and $\cQ_2$ are positive, the physically relevant solutions are 
(when assuming $p>0$ for definiteness):
\begin{eqnarray}
&&F_1 [2 \beta_1]=\exp[-e^{-2\,\beta_1}]\,e^{-2\,i\,\sqrt{{\cal Q}_1}\,\beta_1}U[\frac 12- {\cal C}_g+i\,\sqrt{{\cal Q}_1},1+2\,i\,\sqrt{{\cal Q}_1},2 e^{-2\,\beta_1}] \, , \nonumber \\
&&\\
&&F_2 [ \beta_3-\beta_2]= \exp[-p e^{-(\beta_3-\beta_2)}]\,e^{- i\,\sqrt{\cQ_2}(\beta_3-\beta_2)}\quad \nonumber \\
&&\phantom{F_2 [ \beta_3-\beta_2]= } \times U[\frac 12(1+ {\sigma_{2 3} })+i\,\sqrt{\cQ_2},1+2\,i\,\sqrt{\cQ_2},\, 2 \,p\,e^{-(\beta_3-\beta_2)}] \nonumber\ .\\&&
\end{eqnarray}
Far from the two walls, these modes propagate as plane waves in all the variables, with $\beta$-space momenta  $\varpi_a$ of the form
$$
\varpi_a {[\varepsilon_G,\varepsilon_S]}=\{2(k+\varepsilon_G\, \sqrt{{\cal Q}_1}),k+\varepsilon_S\,\sqrt{\cQ_2},k-\varepsilon_S\,\sqrt{\cQ_2}\} 
$$
satisfying the mass-shell condition
\begin{equation}  \label{mass-shell2}
G^{ab} \varpi_a \varpi_b = - \mu^2 = - \frac 12 \cS^2 - 4 c
\end{equation}
The quantities $\varepsilon_G$ and $\varepsilon_S$ in the momenta $\varpi_a$
denote some  $\pm 1$ signs, that flip upon collisions on the walls. 

As above, we can also compute the phase shifts of these modes as they reflect on a wall.
More precisely we find that, for given quantum numbers $\cC_g$, $\cQ_1$,
$ { \sigma_{2 3} }$  and ${\cal Q}_2$,
the phase shift $\alpha_g$ of $F_1$ as it reflects on a gravitational wall, and the
phase shift $\alpha_s$ of $F_2$ as it reflects on a symmetry wall, are respectively given by:
\begin{eqnarray}
e^{i\,\alpha_g}&=&-\frac{\Gamma(1/2-{\cal C}_g-i\,\sqrt{{\cal Q}_1})\,\Gamma(1/2+2\,i\,\sqrt{{\cal Q}_1})}{\Gamma(1/2-{\cal C}_g+i\,\sqrt{{\cal Q}_1})\,\Gamma(1/2-2\,i\,\sqrt{{\cal Q}_1})}\\
e^{i\,\alpha_s}&=&-\frac{\Gamma(1/2+{ \sigma_{2 3}/2 }-i\,\sqrt{\cQ_2})\,\Gamma(1/2+2\,i\,\sqrt{\cQ_2})}{\Gamma(1/2+{ \sigma_{2 3} /2}+i\,\sqrt{\cQ_2})\,\Gamma(1/2-2\,i\,\sqrt{\cQ_2})}\\
\end{eqnarray}
As above, we can deduce from these results the intrinsic phase shifts due to the
spin dependence of the walls  by subtracting  the phase shift of the spin zero mode. 

Let us finally note that the existence of a non-zero mass term 
$\mu^2 =   \cS^2/2 + 4 c$ in the Wheeler-DeWitt equation can lead
to an interesting phenomenon (whose classical analogue is discussed in Appendix C).
Indeed, a  strictly positive $\mu^2$   (e.g. corresponding to $c=0$ and $\cS^2 \neq 0$)
forces the classical trajectory of the wavepacket to stay time-like in $\beta$-space,
i.e. prevents it to reach the $\beta$-space light-cone. Therefore, such a mass term
constitutes a kind of potential wall that prevents the wavepacket to reach the light-cone.
We can therefore think of the dynamics described by the Bianchi-II Hamiltonian above
as that of a quantum particle moving in a three-dimensional Lorentzian space,
and confined by {\it three} different walls: the two spacelike walls $\alpha_{123}$
and $\alpha_{23}$ (that prevent the particle from going on the negative sides
of those spacelike walls), and a third effective $\mu^2$ wall that prevents the
particle, after it has bounced on the spacelike walls and aims towards
the light-cone, to reach the light-cone. These three walls thereby define
a kind of waveguide that oblige the particle to move in a time-like direction,
which is somewhere midway between the spacelike walls and the light-cone.
The interesting consequence of this waveguide phenomenon is that it can trap
the particle in a bound state, confined between all these walls.
This happens when both $\cQ_1$ and $\cQ_2$ are negative (in view
of the mass-shell condition (\ref{mass-shell1}) this happens only when
$\mu^2> 2 k^2 >0$).  In that case, one should no longer consider 
scattering states of the Morse-potential equations above, but rather
{\it bound states} in the Morse potentials.  It is well known 
(see, e.g., \cite{Landau-LifchitzMQ}) that these
bound states occur for the following quantization conditions
\begin{equation}
\sqrt{-{\cal Q}_1}=-n_1-\frac 12 + {\cal C}_g\, , 
\end{equation}
and 
\begin{equation}
\sqrt{- \cQ_2}=-n_2-\frac 12(1 +{ \sigma_{2 3} })\, .
\end{equation}
where $n_1$ and $n_2$ must be natural integers (starting with 0). 
For instance,  in the minimal case where the ordering constant is simply the
naive value $c=0$, there will exist only one such bound state, namely
the ground state  $n_1=n_2=0$, with $\cC_g = -\sigma_{23}/2 =  1/4$, $\cQ_1=\cQ_2=-1/16$,
and $k=\pm 1/4$. This solution furnishes an interesting example of a quantum
cosmological ground state associated to a Bianchi-II billiard.
Note that we have discussed here a state which is bound simultaneously within
the two separate Morse-potential equations associated to the gravitational
and symmetry walls. There can also exist semi-bound states, i.e. states
which are bound w.r.t, say,  the gravitational-wall Morse-potential, but
which represent scattering states w.r.t. the symmetry-wall potential.

\section{Conclusions}\label{Conc}

We have studied the minisuperspace quantization of spatially homogeneous (Bianchi)
cosmological universes sourced by a Dirac (or Majorana) spinor field. In the main text
we used a formulation of the spinor dynamics in which the local $SO(3,1)$ local Lorentz
symmetry of the vielbein is gauge-fixed from the start. [Appendix A compares this
approach to the one where one does not initially fix the vielbein.] In the Bianchi types
IX and VIII (corresponding to simple homogeneity groups $G$) we fixed the $SO(3)$
freedom in the dreibein by using the existence of a three-dimensional automorphism group
of the Lie algebra of $G$. In the Bianchi type II case, we fixed the $SO(3)$
freedom of the dreibein by using an Iwasawa decomposition (which happens to be
compatible with the automorphism group of the corresponding $G$). 

In the Bianchi types IX and VIII, the quantum version of the three diffeomorphism constraints
means that the wavefunction does not depend on the three Euler angles parametrizing
the (pseudo-)orthogonal matrix $S \in SO(3) $ (for type IX) or $S \in SO(1,2)$ (for type VIII),
entering the  Gauss decomposition $g = S^T {\rm diag} S$ of the metric $g$. This is the
quantum version of the classical possibility of restricting $g$ to a diagonal form,
$ {\rm diag} =  {\rm diag}( e^{-2 \beta_1},   e^{-2 \beta_2}, e^{-2 \beta_3} )$  .

The quantization of the homogeneous spinor (denoted $\chi$ after a rescaling)
leads to a finite-dimensional fermionic Hilbert
space, which means that the wavefunction of the universe, which, in the bosonic  case, has only one component, becomes multi-component in presence of a spinor field.  In addition, in the
Majorana case, the four components of the wavefunction can be identified with the four
components of  a spinor of an Euclidean $O(4)$.

The multi-component Wheeler-DeWitt equation satisfied by the wavefunction $\Phi$ is
similar to the second-order form of the Dirac equation coupled
to an external electromagnetic field ($ \left[ (i \, \partial_{\mu} - e \, A_{\mu})^2 + \frac{1}{2} \, e \, \sigma^{\mu\nu} \, F_{\mu\nu} + m^2 \right] \psi = 0$), namely it has a structure
of the form (for types IX and VIII)

\begin{equation} \label{WDWconcl}
\biggl( - \frac{1}{4} \, G^{ab} \, \partial_{\beta_a} \, \partial_{\beta_b} +V_0(\beta) + \sum_\alpha V_\alpha(\beta)  \widehat{\Sigma}_\alpha( \chi) + \sum_{\alpha'} V_{2 \alpha'}(\beta)  (\widehat{\Sigma}_{\alpha'}( \chi))^2+ \frac{1}{4} \, \widehat{\mu}_{\rm tot}^2 (\chi) \biggl) \Phi (\beta , \sigma) = 0 \, .
\end{equation}
where  $\widehat{\Sigma}_\alpha( \chi) = i \chi^{\dagger} J^s_\alpha \chi$,
 and $\widehat{\Sigma}_{\alpha'}( \chi)=  i \chi^{\dagger} J^s_{\alpha'} \chi   $ are 
some bilinears in $\chi$, and where $\alpha$ and $\alpha'$ run over some sets of
linear forms in $\beta$ (or `roots'). There exist a limit in $\beta$ space (the far-wall, or BKL, limit) where all the potentials $V_A(\beta)$  tend exponentially towards zero. [The existence
of this limit defines the separation between the $V_A(\beta)$-terms and the squared-mass term $ \frac{1}{4} \, \widehat{\mu}_{\rm tot}^2 (\chi)$.] 
The  main features of this multi-component Wheeler-DeWitt equation are the following.

The squared-mass term $\widehat{\mu}_{\rm tot}^2 (\chi)$ is a quantum effect, which is of
order $\hbar^2$, and which is affected by several sorts of quantum ordering ambiguities.
We discussed the fact that it was different in the original minisuperspace Einstein-Dirac theory compared to the spin-1/2 Kac-Moody coset proposed in \cite{dBHP}. This suggests that
we should choose an ordering (and additional terms, such as (\ref{eq3.19})) such that
the total squared-mass term vanishes. On the other hand, if we do not do so, but use
instead the naive ordering that looks natural in the quasi-Gaussian spacetime gauge 
$N=\sqrt{g}$, one gets a specific prediction for $\widehat{\mu}_{\rm tot}^2 (\chi)$.
One then finds that this quantum (spin-dependent) operator admits the eigenvalue zero
in a fraction of the total Hilbert space. This ensures that a part of the total wavefunction,
i.e. a part of the total quantum reality, will formally continue to behave chaotically
near the singularity, in contrast with the case where the spinor source is treated
classically, where $\mu^2$ is a strictly positive c-number \cite{BK} (see (\ref{eq3.32})).

We discussed in some detail the physical effects linked to the other terms in
the multi-component Wheeler-DeWitt equation (\ref{WDWconcl}). We studied in
particular the spin-dependent terms of the form $\sum_\alpha V_\alpha(\beta)  \widehat{\Sigma}_\alpha( \chi) $. 
Such terms appear both in the Bianchi IX and VIII cases, and in the Bianchi II one. 
In the Bianchi IX and VIII cases the set of roots
$\alpha$ entering these terms are the three gravitational roots, while, in the
Bianchi II case there appear both a gravitational root and a symmetry root
(see (\ref{QGenEq})). When combining these spin-dependent terms with the
corresponding potential terms $\sim e^{- 2 \alpha(\beta)}$ of the spinless potential
$V_0(\beta)$ involving the same root $\alpha$, we found that they lead to a
Schr\"odinger equation in a Morse potential. By studying the quantum scattering
on such spin-dependent Morse potentials we could relate the quantum dynamics
of  wavepackets reflecting on them to a previous study of fermionic billiards,
which used Grassmann-valued spinor fields.  The quantum spin dynamics
of Bianchi IX and Bianchi VIII happens to be rather trivial because all the corresponding 
spin-dependent couplings $\widehat{\Sigma}_\alpha( \chi) $ can be simultaneously
diagonalized. A more interesting spin dynamics would, however, be obtained in more
complicated models (e.g. in higher- dimensions) where the various
$\widehat{\Sigma}_\alpha( \chi) $'s do not commute among themselves.

We also studied in detail the Bianchi II model.  For this case we could provide the
exact general solution of the quantum dynamics. It can indeed be decomposed
in separated modes, which can all be expressed in terms of confluent hypergeometric
functions. Some of these solutions describe wavepackets reflecting on the gravitational
and symmetry wall forms $\alpha_{123}(\beta)$ and $\alpha_{23}(\beta)$ of the Bianchi II model,
while other solutions (present when $\mu^2 > 0$) can describe interesting bound
states, trapped between the walls $\alpha_{123}(\beta)$ and $\alpha_{23}(\beta)$,
and the effective wall generated by the positive squared-mass term.

Note finally that the Appendices provide more details about several formal aspects
of the gravity-spinor Hamiltonian dynamics, as well as a study of the classical
limit of this dynamics.

\section*{Acknowledgments}
We  thank Marc Henneaux for a clarifying discussion. Philippe Spindel is grateful to IHES,
where an important part of this work was elaborated, for its kind hospitality.
This work was also partially supported in part by IISN-Belgium (convention 
4.4511.06), and by ``Communaut\'e fran\c caise de Belgique - Actions de Recherche
Concert\'ees''.

\appendix 
\setcounter{equation}{0}
\section*{Appendices}

These Appendices are devoted to some aspects of the Hamiltonian formalism applied to the dynamical problem constituted by the Einstein-Dirac equations considered in the framework of homogeneous Bianchi (class A) cosmological models. As described in the main text, we use an adapted tetrad constituted, at each point, by a time-like vector orthogonal to the slices of homogeneity and a dreibein tangent to these slices: thus these dreibeins are defined up to local $SO(3)$ rotations. In the first Appendix we do not fix the local $SO(3)$ freedom in the dreibeins, and provide all the constraints and Hamiltonian evolution equations of the corresponding dynamical variables. In the second Appendix we make use of an Iwasawa decomposition to fix the $SO(3)$ gauge freedom and display the Hamiltonian and momentum constraints and general constants of motion,  bilinear in the spinorial variables. The third Appendix consists of a sketch  of the resolution of the classical equations for the particular case of the Bianchi II cosmological model. The main aim of this discussion is  to provide the elements needed to compare the classical and quantum dynamics of a billiard collision near the cosmological singularity.
\section
{ Locally $SO(3)$-invariant approach to the Einstein-Dirac dynamics}
\label{A}

In the approach \cite{HenPRD,HenIHP} where one fixes the time-like vector of the vielbein but leaves an $SO(3)$ rotational gauge symmetry in the choice of the spatial dreibein, the dynamical variables (in an Hamiltonian formalism) are $N$, $N^a$, $h_{\widehat k a}$, $\Psi^A$, $\Psi_A^{\dagger}$, together with the conjugate momenta to the dreibein and the spinor, say, respectively, $\Pi^{\widehat k a}$, $\Pi_A$ and $\Pi^A$. Starting from the Lagrangian action $L = L_{EH} + L_D$, where
$$
L_{EH}:= N\,\sqrt{g}\left (\,^3\hspace{-0.6mm}R+K_{ab}K^{ab}-(K^a_a)^2\right)\, ,
$$
with $K_{ab}$ denoting the second fundamental form, Eq.~(\ref{eq2.24}), and
$$
L_D=N\,\sqrt{g}\left(  \Psib\gamma^{\hat 0}\nabla_{\hat 0}\Psi+\Psib\gamma^{\hat k}\nabla_{\hat k}\Psi -m\,\Psib\Psi\right)\, , 
$$
one computes the two sets of momenta. First the gravitational momenta conjugated to the dreibein : 
\begin{equation}
\Pi^{\hat k \,a}:=\partial L/\partial \dot h_{\hat k\,a}=2\,\PG^{\hat k \,a}+2\,P^{\hat k \,a}-   S^{\hat k \,a}
\end{equation}
where we defined
\begin{equation}
\PG^{ab} :=+\sqrt{g}\left[K^{ab}-g^{ab}\,K\right ]=\PG^{ba}\,  ,
\end {equation}
\begin{equation}
P^{ab}:=-\sqrt{g}\,\Psib\gamma^{\hat p} \frac{\gamma^{\hat 0\hat q}}{4}\Psi\, h_{\hat p}^{(a}h_{\hat q}^{b)}=\frac 14 \sqrt{g}\,\Psib\gamma^{\hat 0}\Psi\,g^{ab}=P^{ba}\, ,
\end {equation}
\begin{equation}
S^{{\hat a}{\hat b}}:= \frac{1}{4} \, \sqrt{g} \Psib\gamma^{\hat 0}\gamma^{{\hat a}{\hat b}}\Psi=- S^{{\hat b}{\hat a}}\, ,
\end {equation}
as well as
\begin{equation}
\PG^{\hat k \,b} :=h^{\hat k}_a\PG^{ab}\quad,\quad P^{\hat k \,b}:=h^{\hat k}_aP^{ab}\quad,\quad  S^{\hat k \,b} :=S ^{\hat k{\hat a}}h_{{\hat a}}^b \,.
\end {equation}

Note that, for convenience, we follow Ref. \cite{HenIHP} in working in the Appendices
with the spinor bilinear $S^{{\hat a}{\hat b}}$, which differs by a factor one-half
from the one used in the main text: namely $S^{{\hat a}{\hat b}} = \frac{1}{2} \Sigma^{{\hat a}{\hat b}}$, with the definition (\ref{sigma}).

We also have the fermionic momenta that lead to primary constraints :
\begin{eqnarray}
\Pi_A&:=&\partial^L L/\partial\dot \Psi ^A=-\sqrt{g}\,\left(\Psib\,\gamma^{\hat 0}\right)_A=-i\sqrt{g}\,\,\Psi_A^\dagger\, ,\label{CA}\\
\Pi^A&:=&\partial^L L/\partial\dot \Psi ^\dagger_A=0\, .\label{CdA}
\end{eqnarray}

The Cartan one--form $\varpi = pdq$ is given by:
\begin{equation}\label{Cartform}
\varpi=\Pi^{\hat k\alpha}dh_{\hat k\alpha}+ \Pi_Ad\Psi^A+\Pi^Ad\Psi^\dagger_A\end{equation}
and the total Hamiltonian is given by
\begin{equation}\label{TotH}H=N\,{\cal H}+N_{k}\,{\cal H}^k+ \Omega_{[ab]}\,{\cal H}^{[ab]}
\end{equation}
where (using when necessary the definition $\Pi^{rs} :=h^r_{\hat k}\Pi^{\hat k s}$):
\begin{equation}
{{\cal H}}=\frac 1{\sqrt{g}}\left(\PG^{r s}\PG\hspace{-1mm}\strut_{r s}-\frac12 (\PG^{ r}\hspace{-2.7mm}\strut_r)^2\right)-\sqrt{g}\,^3\hspace{-0.6mm}R -\sqrt{g}\Psib\gamma^{\hat p}D_{\hat p}\Psi+m\,\sqrt{g}\,\Psib\Psi+\frac12\,\sqrt{g}\,D_{{\hat \rho}}(\Psib\gamma^{\hat \rho}\Psi)
\end{equation}
and
\begin{equation}
{\cal H}^s=-D_r\Pi^{sr}+\sqrt{g}\Psib\gamma^{\hat 0} D^s\Psi\quad ,\quad {\cal H}^{[ab]}=\frac12(h_{{\hat k}}^a\Pi^{{\hat k} b}-h_{{\hat k}}^b\Pi^{{\hat k} a})+ S^{[ab]}\, .
\end{equation}

The coefficients $\Omega_{[ab]}$ in Eq.~\ref{TotH} are Lagrange multipliers measuring the arbitrariness in the local rotation rate of the dreibein. We will fix them in Appendix \ref{Iwas}. To the three Lagrange multipliers $N$, $N_k$ and $\Omega_{[ab]}$ are associated the three constraints:
\begin{equation}
{\mathcal H} \approx 0 \, , \quad {\mathcal H}^s \approx 0 \, , \quad {\mathcal H}^{[ab]} \approx 0 \, .
\end{equation}

It is useful to use slightly different canonical spinorial variables: $\chi^A:=g^{1/4}\,\Psi^A$ and $\phi^\dagger_A=i\,g^{1/4}\,\Psi_A^\dagger$. Accordingly we obtain: 
\begin{equation}
\Pi_Ad\Psi^A=\phi^\dagger_Ad\chi^A-\frac 14 \phi^\dagger\chi\,\frac 1g\,dg=\phi^\dagger_Ad\chi^A-\frac 12 \phi^\dagger\chi\,h^{\hat ka}\,dh_{\hat k a} .
\end{equation}
This redefinition of the spinorial variable entails a corresponding redefinition of the
canonical  momentum  conjugated to the dreibein $ h_{\hat ka}$.  Instead of $\Pi^{\hat ka}$
the new momentum conjugated to  $h_{\hat ka}$ becomes
$\pi^{\hat k a}=\Pi^{\hat ka}-\frac 12 \phi^\dagger\chi\,h^{\hat ka}$. This shift in the
definition of the gravitational momentum has the effect of suppressing the second ,
$2\, P^{\hat k \,a}$, contribution in the above expression for  $\Pi^{\hat ka}$,
so that one finally obtains

\begin{equation}\label{piasPG}
\pi^{\hat k a}=2\,\PG^{\hat k \,a}- S^{\hat k \,a} .
\end{equation}
When appropriately moving the indices by means of the dreibein $h_{\hat k}^a$ or its inverse,
the two contributions to the redefined gravitational momentum read:
\begin{equation}
\PG^{ab}=\frac14 (h_{\hat k}^a\pi^{\hat k b}+h_{\hat k}^b\pi^{\hat k a})\,  , \, 
S^{\hat a  \hat b}  = \frac 14 \phi^\dagger\gamma^{{\hat a}{\hat b}}\chi
\end{equation}

Let us now apply the above general formalism to homogeneous spaces of Bianchi class A, {\it i.e.} with structure constants, Eq.~(\ref{eq2.3}),  of the form
\begin{equation}\label{Cn} 
C^a_{\  cd}:=n^{ab}\varepsilon_{bcd}\, .
\end{equation}
In such a framework we have 
\begin{equation}
\strut^3\hspace{-0.6mm}R=-\frac 1g\left(n^{ab}n_{ab}-\frac 12 n^2\right)
\end{equation}
where the indices on $n_{ab}$ have been lowered with $g_{ab}$, and where $n := g_{ab} \, n^{ab}$. The spatial Dirac derivative of $\Psi$ has the general form
\begin{equation}
\slash \hspace{-2.8mm}{D}\Psi=i\,\slash{\hspace{-2.2mm}\lambda}\,\Psi+\frac 14\frac n {\sqrt{g}}\gamma^{\hat 0}\gamma_5\Psi\, ,
\end{equation}
where we have taken into account the possibility to introduce a spatial variation of $\Psi$ via constants $\lambda_a$ such that the Lie derivative of the spinor field along the generators $\xi _a$ of the isometry group of the homogeneous slices verify $\pounds_{\xi_a}\Psi=i\,\lambda_a\,\Psi $. Integrability of these equations implies that $\lambda_a\,n^{ab}=0$ (see Ref. \cite{HenIHP}). When $\lambda_a \ne 0$ the fermionic degrees of freedom are expressed as a product of a time dependent dynamical factor multiplied by a space dependent phase factor.
However, this spatial dependence disappears in the stress-energy tensor $T^{\mu \nu}(\Psi)$.
As indicated by the explicit $i$ entering the condition $\pounds_{\xi_a}\Psi=i\,\lambda_a\,\Psi $ such generalized homogeneous solutions are only possible in the complex, Dirac-spinor
case. Seen from the viewpoint where one decomposes a Dirac field $\Psi$ into two real
Majorana fields $\Psi_R, \Psi_I$, these solutions correspond to having spatial oscillations turning
one Majorana mode into the other one (but keeping quadratic expressions $\sim \Psi_R^2 + \Psi_I^2$ constant in space).

From all these results we obtain the various terms defining the total Hamiltonian (\ref{TotH}) in the framework of Bianchi cosmologies:
\begin{equation}\label{Ham}
{{\cal H}}=\frac 1{\sqrt{g}}\left(\PG^{rs}\PG\hspace{-1mm}\strut_{rs}-\frac12 (\PG^r\hspace{-2.7mm}\strut_r)^2\right)+\frac 1{\sqrt{g}}\left(n^{ab}n_{ab}-\frac 12 n^2\right) +i\,\phi^\dagger\gamma^{\hat 0}\slash{\hspace{-2.2mm}\lambda}\,\chi-\frac n{4\sqrt{g}}\,\phi^\dagger\gamma_5\chi-m\,\phi^\dagger\gamma^{\hat 0}\chi
\end{equation}
where $\gamma_5= \gamma_0 \gamma_1 \gamma_2 \gamma_3$ is the common 
Clifford algebra generator linked to spinorial gravitational walls, see Eqs. (\ref{eq3.39}),
while
\begin{eqnarray}
{\cal H}_{\hat s}&=&2\,\PG^{\hat m\hat n}C_{\hat m{\hat s}{\hat n}}+ \frac 14 \phi^\dagger \gamma^{[{\hat m}{\hat n}]}\chi\,C_{{\hat m}{\hat n}{\hat s}}+i\,\lambda_{\hat s}\,\phi^\dagger\chi \nonumber \\
&=&\pi^{\hat m\hat n}C_{\hat m{\hat s}{\hat n}} +i\,\lambda_{\hat s}\,\phi^\dagger\chi\, ,\label{Ha}\\
{\cal H}^{[{\hat a}{\hat b}]}&=& \pi^{[{\hat a}{\hat b}]}+ \frac 14 \phi^\dagger\gamma^{[{\hat a}{\hat b}]}\chi =   \pi^{[{\hat a}{\hat b}]} +  S^{\hat a  \hat b}    \, .\label{Hab}
\end{eqnarray}
Here the hatted indices on $\pi^{\hat m\hat n}$ and $C_{{\hat m}{\hat n}{\hat s}}$ are 
obtained from their original forms $\pi^{\hat k a}$ and $C^a_{\ b c}$ by suitable uses
of the dreibein $ h_{\hat ka}$ or its inverse.

The Einstein--Dirac Lagrangian leads to a set of  second class constraints that oblige us to consider Dirac brackets instead of Poisson brackets. Denoting Dirac brackets as $\{ \ , \ \}^*$ we obtain from the constraint equations (\ref{CA}, \ref{CdA}) the canonical brackets  :
\begin{equation}
\left\{h_{{\hat m} a},\pi^{{\hat n} b}\right\}^* =\delta^{\hat n}_{\hat m}\delta^b_a\qquad\left\{\chi^A,\phi^{\dagger}_B\right\}^*= \delta^A_B=i\left\{\chi^A,\chi^{\dagger}_B\right\}^*\label{ClassDB}
\end{equation}
with corresponding classical equations of motion  
\begin{equation}
\dot h_{{\hat k} a}\approx \left\{h_{{\hat k} a},H\right\}^* \quad,\quad\dot \pi^{{\hat k} a}\approx \left\{\pi^{{\hat k} a},H\right\}^*\quad,\quad\dot \chi^A\approx\left\{\chi^A,H\right\}^*
\end{equation}
completed by the first class constraints:
\begin{equation}
{\cal H}\approx 0\qquad{\cal H}_{\hat a}\approx 0\qquad{\cal H}_{[{\hat m}{\hat n}]}\approx 0\, .
\end{equation} 

So that, in the gauge $N_a=0$, we obtain (with $h:=\sqrt{g}$ and $\pi_{a{\hat k}} := h_{ \hat m  a} h_{\hat k b} \pi^{\hat m b}$):
\begin{equation}
\dot h_{{\hat k} a}=\frac N{4\,h}\left(\left(\pi_{{\hat k} a}+\pi_{a{\hat k}}\right)-\left( \pi^{{\hat m} b}h_{{\hat m} b}\right)h_{{\hat k} a}\right)+ \Omega_{[{\hat k}{\hat m}]}\,h^{\hat m}_a
\end{equation}
from which we deduce
\begin{equation}
\frac 12\left(\dot h_{{\hat a} q}h^q_{{\hat b}}-\dot h_{{\hat b} q}h^q_{{\hat a}}\right)=\Omega_{[{\hat a}{\hat b}]}
\end{equation}
\begin{equation}
\pi_{({\hat a}{\hat b})}=\frac h{N}\left(\dot h_{{\hat a} q}h^q_{{\hat b}}+\dot h_{{\hat b} q}h^q_{{\hat a}}\right)-\frac{2\,\dot h}N\delta_{{\hat a}{\hat b}}
\end{equation}
and
\begin{eqnarray}
\dot \pi^{{\hat k} a}&=&N\left\{h^{{\hat k} a}  \frac 1{4\,h}\left(\pi^{{\hat r}{\hat s}}\pi_{{\hat r}{\hat s}}-\frac12 (\pi^{\hat r}_{\hat r})^2-4\,n^{rs}n_{rs}+2 n^2\right)\right.\nonumber\\
&&+i\lambda^{\hat k}\,\phi^\dagger\gamma^{\hat 0}\gamma^a\chi+ \frac 1{2\,h}n^{ab}h^{{\hat k} }_b\phi^\dagger\gamma_5\chi\nonumber\\
&&-h_{{\hat q} b}\frac 1{4\,h}\left(\pi^{{\hat q} a}\pi^{{\hat k} b}+\pi^{{\hat q} a}\pi^{b{\hat k}}-\pi^{{\hat q} b}\pi^{{\hat k} a} \right)\nonumber\\
&&\left.+h_{ b}^{\hat k}\frac 1{h}\left(4\,n^{ac}n^{ b}_c-2\,n\,n^{ab} \right)\right\}\nonumber\\
&&-\Omega^{\hat k}_{\ {\hat p}}\pi^{{\hat p} a}\, .
\end{eqnarray}
To be complete let us also display the Dirac equation :
\begin{equation}\label{Digeneq}
\dot \chi =N\left(i\,\gamma^{\hat 0}\lambda\hspace{-2.5mm}\slash -\frac n{4\,h}\gamma_5-m\gamma^{\hat 0} \right)\chi+ \frac 14 \Omega_{[{\hat a}{\hat b}]}\gamma^{[{\hat a}{\hat b}]}\chi\, .
\end{equation}

\setcounter{equation}{0}
\section
{ Approach using an Iwasawa representation to fix the local $SO(3)$ gauge  }\label{Iwas}

In   Appendix \ref{A}, we displayed an Hamiltonian formalism in which the local rotation rate of the spatial dreibein is left unfixed. Starting from this formalism, we can then gauge fix the local rotation of the dreibein by choosing a specific way of defining a dreibein $h_{\widehat k a}$ from the metric components $g_{ab}$. In the main text, we did this by using the form (\ref{eq2.5}), with a matrix $S$ belonging to the automorphism group of the Bianchi structure constants $C^a \, _{bc}$. Here, we show the form of the Hamiltonian obtained when fixing the definition of the dreibein by means of an Iwasawa representation, namely
\begin{equation}\label{Iwasrep}
h_{{\hat k} a}=
\left(
\begin{array}{ccc}
e^{-\beta_1}  &0   & 0  \\
0 & e^{-\beta_2}  &  0 \\
0  & 0  &   e^{-\beta_3}
\end{array}
\right)\left(
\begin{array}{ccc}
1  &\nu_{12}   &\nu_{13}   \\
0 & 1  &\nu_{23}   \\
0  & 0  &   1
\end{array}
\right)
\end{equation}
{\sl i.e.}
\begin{equation}
h_{{\hat k} a}=e^{-\beta_k}\nu_{ka}\qquad  {\text{without summation on $k$.}}
\end{equation}

Let us emphasize that there are (only) three (non-vanishing) off-diagonal $\nu$ variables, namely
$\nu_{12}, \nu_{23}, \nu_{13}$, i.e. the $\nu_{ka}$'s such that $k < a$. [The $\nu_{ka}$'s
with $k \geq a$ are  equal either to 0 or to 1.] These three metric off-diagonal
variables are the Iwasawa-decomposition analogs of the three Gauss-decomposition
Euler-angle variables. 
The momenta conjugated to the Iwasawa diagonal and off-diagonal variables, $\beta_k$ and $\nu_{ka} \, (k<a)$ will be denoted by $\varpi^k$ and $\varpi^{ka} \, (k<a)$: ($\varpi^1,\varpi^2,\varpi^3,\varpi^{12},\varpi^{13},\varpi^{23}$).  Note that $\varpi^{ka} $ is {\it only}
defined for $ k<a$. This contrasts with the momenta  $\pi^{{\hat k} a}$ 
conjugated to the dreibein  $h_{{\hat k} a}$ which are defined for all indices: $k<a$,
$k=a$ and $k>a$.  Among these, the ones with $k<a$ and $k=a$ are easily
related to the $\varpi^{ka} \, (k<a)$'s and the $\beta$'s by considering the
Cartan one--form (\ref{Cartform}). Specifically, one finds : 
\begin{eqnarray*}
&&  {\rm when} \,  {\hat k}<a \, : \,\pi^{{\hat k} a}=e^{\beta_k}\varpi^{ka}\, ,\\
&&\pi^{{\hat k} k}=-e^{\beta_k}(\varpi^k+\sum_{k<a}\varpi^{ka}\nu_{ka})\, .
\end{eqnarray*}
On the other hand, the calculation of the remaining components of the gravitational
momenta, i.e $\pi^{{\hat k} a}$ for $k>a$ is less immediate and must be based on the
fact that, 
when viewed within the non-gauge-fixed formalism of Appendix A the Iwasawa gauge (\ref{Iwasrep}) introduces three {\it new constraints:} $h_{\hat k a}=0$ for $k>a$, that do not commute with the constraints ${\cal H}^{[{\hat a} {\hat b}]}$. The presence of these new constraints oblige us either to apply the Dirac procedure or to solve them explicitly. Using this
last option allows one to compute, recursively,  the remaining momenta 
$\pi^{{\hat k} a}$ with $k>a$, i.e. 
$\pi^{\hat 2 1}$, $\pi^{\hat 3 1}$ and $\pi^{\hat 3 2}$, by using the relation
\begin{equation}\label{pirec}
\pi^{{\hat k} a}=h^a_{\hat p}\,\pi^{\hat p d}\,h_d^{\hat k}-2\, S^{[{\hat k}{  a}]}  ,
\end{equation}
starting  from $({\hat k},k-1)$ down to $(\hat 2,1)$. Indeed, noticing that $h^a_{\hat p}=0$  if $p<a$ and $h^{\hat k}_d=0$ if $k>d$, we see that for given values of $k>a$, the right hand side of Eq.(\ref{pirec}) involves only already known momenta $\pi^{\hat p d}$ with either $p>k$ or $k\geq p\geq a$ but $d\geq p$. In three dimensions, this yields the
explicit results
\begin{eqnarray}
\pi^{\hat32}&=& e^{ (2\beta_2-\beta_3)}\varpi^{23} +e^{\beta_3}\nu_{23}\varpi^3+2\,e^{\beta_2}S^{[\hat 2\hat 3]}\nonumber\\
\pi^{\hat31}&=&e^{-\beta_3}[e^{2\beta_3}(\nu_{13}-\nu_{12}\nu_{23})\varpi^3+e^{ 2\beta_1}\varpi^{13}-e^{2\beta_2}\nu_{12}\varpi^{23}]+2[e^{\beta_1}S^{[\hat 1\hat 3]}-e^{\beta_2}\nu_{12}S^{[\hat 2\hat 3]}]\nonumber\\
\pi^{\hat21}&=&e^{\beta_2}[\nu_{12}\varpi^2+e^{-2(\beta_2-\beta_1)}(\varpi^{12}+\nu_{23}\varpi^{13})+(\nu_{23}\nu_{12}-\nu_{13})\varpi^{23}]+2\,e^{\beta_1}S^{[\hat 1\hat 2]} \nonumber\\ &&\label{pivarpi}
\end{eqnarray}

Now we have in hands all the ingredients needed (\ref{piasPG}) to write the remaining Hamiltonian equations and constraints (\ref{Ham},\ref{Ha}) in terms of the Iwasawa variables. 

The Iwasawa Hamiltonian density (of weight $+1$) reads, 
when using a diagonal form\footnote{As is well known, the matrix $n^{ab}$ can always be diagonalized; we  assume it has been.} $n^{ab} = {\rm diag} \, \{ n_1, n_2 , n_3 \}$ for the symmetric matrix $n^{ab}$ entering the structure constants $C^a \, _{cd} = n^{ab} \, \varepsilon_{bcd}$,
\begin{eqnarray}
{\cal H}&=&
{ e^{\sum_i\beta_i}}\left\{\frac 14(\pX)^2+(\pY)^2+(\pZ)^2-\frac 18(\pX+\pY+\pZ)^2 \right.\nonumber\\
&&+ \frac12 \left[(e^{(\beta_1-\beta_2)}(\varpi^{12}+\nu_{23}\varpi^{13})+S^{[\hat 1\hat 2]})^2+(e^{(\beta_1-\beta_3)} \varpi^{13}+S^{[\hat 1\hat 3]})^2+(e^{(\beta_2-\beta_3)} \varpi^{23}+S^{[\hat 2\hat 3]})^2\right]  \nonumber\\
&& +\frac {{n_1}^2} 2  e^{-4\beta_1}+\frac {{n_2}^2} 2 [e^{-2\beta_2}+e^{-2\beta_1}\nz^2]^2
+\frac {{n_3}^2}2[e^{-2\beta_3}+e^{-2\beta_1}\ny^2+e^{-2\beta_2}\nx^2]^2
\nonumber\\
&&-n_1n_2[e^{-2(\beta_1+\beta_2)}-e^{-4\beta_1}\nz^2]\nonumber\\
&&-n_1n_3[e^{-2(\beta_1+\beta_3)}-e^{-4\beta_1}\ny^2+e^{-2(\beta_1+\beta_2)}\nx^2]\nonumber\\
&&-n_2n_3[e^{-2(\beta_2+\beta_3)}+e^{-2(\beta_1+\beta_3)}\nz^2-e^{-4\beta_2}\nx^2-e^{-4\beta_1}\nz^2\ny^2 \nonumber\\
&&\qquad \quad  +e^{-2(\beta_1+\beta_2)}(\ny^2-4\,\nz\,\ny\,\nx+\nz^2\,\nx^2)]\nonumber\\
&&\left.-\frac 14[n_1e^{-2\beta_1}+n_2(e^{-2\beta_2}+e^{-2\beta_1}\,\nz^2)+n_3(e^{-2\beta_3}+e^{-2\beta_1}\ny^2+e^{-2\beta_2}\nx^2)]  \phi^\dagger\gamma_5\chi\right\}\nonumber\\
&& -m \,\phi^\dagger\gamma^{\hat 0}\chi
\end{eqnarray}
while  the momentum constraints reduce to
\begin{eqnarray}
{\cal H}^1&=& e^{\beta_1}\left\{
n_3\left[e^{2(\beta_2-\beta_3)}\varpi^{23}+\nu_{13}\varpi^{12}-\nu_{23}(\varpi^2-\varpi^3+\nu_{23}\varpi^{23})+2 e^{(\beta_2-\beta_3)}S^{[23]}
\right]
\right.\nonumber\\
&&\qquad\left.-n_2\left(\varpi^{23}+\nu_{12}\varpi^{13} \right)\right\}\\
{\cal H}^2&=&e^{\beta_2}\left\{n_1\,\py +n_2\,\nz\,(\px+\nz\py)\right.\nonumber\\
&&\qquad -n_3\left[e^{2(\beta_1-\beta_3) }\,\py+e^{2(\beta_1-\beta_2)}\,\nx\,(\pz+\nx\,\py)\right.\nonumber\\
&&\qquad\qquad \left. \left. +\ny\,(\pX-\pY+\ny\,\py+\nx\,\px)+2\,e^{(\beta_1-\beta_2)}\nx\,\Sz-2\,e^{(\beta_1 -\beta_3)}\,\Sy\right]\right\}\\
{\cal H}^3&=&e^{\beta_3}  
\left\{ -n_1\,[\pz+\nx\,\py]\right .\nonumber\\
&&\qquad +n_2\,[e^{2(\beta_1-\beta_2) }(\pz+\nx\,\py)  -\nz[\pX-\pY+\nz(\pz+\nx\,\pz)]+2\,e^{\beta_1-\beta_2}\Sz ]\nonumber\\
&& \qquad -n_3[ \ny(e^{2(\beta_2-\beta_3)}\px+(\pX-\pY)+2\, e^{(\beta_2-\beta_3)}\,\Sx)+\ny^2(\pz+\nx\,\py)\nonumber\\
&&\qquad \qquad \left.-\nx\,(e^{2(\beta_1-\beta_3)}\py+2\,e^{\beta_1-\beta_3}\,\Sy)]\right\}
\end{eqnarray}

The constraint algebra reduces to
\begin{equation}
\{{\cal H},{\cal H}_a\}=0\qquad \{{\cal H}_a,{\cal H}_b\}= -C_{\ ab}^c{\cal H}_c\, .
\end{equation}

Obviously bosonic  and fermionic equations decouple in the sense that the bosonic  sector   only involve quadratic expressions in the fermionic variables. In particular, using the previous equations we obtain (under the simplifying condition $\lambda_a=0$)  :
\begin{equation}
(\phi^\dagger\gamma_5\chi)\dot {\vphantom{)}}=2\,m\,\phi^\dagger\gamma^{\hat 0}\gamma_5\chi
\end{equation}
which constitute a first integral when the Dirac mass $m$ vanishes (chirality conservation), and
\begin{equation}
\dot S^{\hat 1\hat 2}=\frac 12\left(e^{\beta_2-\beta_3}S^{\hat 1\hat 3}\varpi^{23}-e^{\beta_1-\beta_3}S^{\hat 2\hat 3}\varpi^{13}\right)
\end{equation}
plus two similar equations obtained by circular permutation of the indices $1,2$ and $3$. It is worthwhile to notice that these spin  equations are independent from the Bianchi type considered.  They lead to the conservation law of the norm of the spin   tensor :
\begin{equation}
(S^{\hat 1\hat 2})^2+(S^{\hat 1\hat 3})^2+(S^{\hat 2\hat 3})^2\equiv {\mathbb S}^2 =C^{st}\qquad.\label{TotS2}
\end{equation}

\setcounter{equation}{0}
\section{ Classical dynamics of the Bianchi-II-Dirac system}
\label{C}

To analyze the behaviour of the spinor variables under a collision, it suffices to restrict ourselves to the (almost) simplest case: the Bianchi type II homogeneous model. The relevant structure constants are obtained by using  $n^{1}=1$, $n^2=n^3=0$ in Eq. (\ref{Cn}). An important remark is that (see Ref.\cite{HenIHP}, in which various special solutions of the Einstein-Dirac system are discussed) the subgroup $SInv$ of unimodular $SL(3,{\mathbb R})$ transformations that preserves these structure constants is a five dimensional group whose generators will provide, {\sl a priori},  five constants of motion [see eqs (\ref{Hconst1}--\ref{Hconst3})].

The   Hamiltonian constraint is:
\begin{eqnarray}
{\cal H}&=& { e^{\sum_i\beta_i}}\left(\frac 14\left[\sum_i(\varpi^i)^2-\frac 12 (\sum_i\varpi^i)^2\right]+ \frac 12e^{-4\beta_1}\right.\nonumber \\
&&+\left.\frac12 \left[(e^{(\beta_1-\beta_2)}(\varpi^{12}+\nu^{23}\varpi^{13})+S^{[\hat 1\hat 2]})^2+(e^{(\beta_1-\beta_3)} \varpi^{13}+S^{[\hat 1\hat 3]})^2+(e^{(\beta_2-\beta_3)} \varpi^{23}+S^{[\hat 2\hat 3]})^2\right]\right)\nonumber\\
&& -m \,\phi^\dagger\gamma^{\hat 0}\chi-\frac {e^{(\beta_2+\beta_3-\beta_1)}}4 \phi^\dagger\gamma_5\chi \, , \label{BIILconst}
\end{eqnarray}
while the three diffeomorphism constraints reduce (upon using $n^{1}=1$, $n^2=n^3=0$)
to only two non-empty constraints, namely
\begin{equation}\label{BIISconst}
\varpi^{12}\approx 0\qquad,\qquad \varpi^{13}\approx 0
\end{equation}
Let us notice that  the Hamiltonian does not depend on the variables $\nu_{12}$ and $\nu_{13}$, 
so that there will be no ordering ambiguities when  quantizing  this system.

When replacing the constraints (\ref{BIISconst}) within the Hamiltonian, and using as
lapse and shift $N =e^{-(\beta_1+\beta_2+\beta_3)}$, $ N^k=0$ ,
one ends up with a Hamiltonian of the form
\begin{eqnarray}  \label{HBII}
H&=& \frac 14\left[\sum_i(\varpi^i)^2-\frac 12 (\sum_i\varpi^i)^2\right]+ \frac 12e^{-4\beta_1}\nonumber\\
&&+ \frac12 \left[ (S^{[\hat 1\hat 2]})^2+(S^{[\hat 1\hat 3]})^2+(e^{(\beta_2-\beta_3)} \varpi^{23}+S^{[\hat 2\hat 3]})^2\right]\nonumber\\
&& -m \,  e^{- (\beta_1 +\beta_2 +\beta_3)} \phi^\dagger\gamma^{\hat 0}\chi-\frac 14 e^{-2\beta_1} \phi^\dagger\gamma_5\chi \, .
\end{eqnarray}

The non vanishing extra constants, taking into account the momentum constraints, are
\begin{eqnarray}
&&\varpi^{23} \approx { p\, }\, ,\label{Hconst1}\\ &&\varpi^3-\varpi^2-2\,\nu_{23}\,\varpi^{23} \approx {\cal C}_1\label{Hconst2}\, ,\\ && \nu_{23}(\varpi^3-\varpi^2)+(e^{2(\beta_2-\beta_3)} -\nu_{23}^2)\,\varpi^{23}+2\,e^{(\beta_2-\beta_3)}S^{[\hat 2 \hat 3]}\approx{\cal C}_2\,  .\label{Hconst3}
\end{eqnarray}
Moreover we also obtain that 
\begin{equation}\label{S23}
S^{[\hat 2\hat 3]}\approx{ s_{2 3} }
\end{equation}
is a constant of motion, as well as the norm of the spinor field 
\begin{equation}
\chi^\dagger\chi\approx{ \cal C }\label{kdk}
\end{equation}
and, if we assume that the Dirac mass $m$ is zero, 
\begin{equation}
(i/4)\,\chi^\dagger\gamma_5\chi\approx{\cal C}_g\label{kdg5k}
\end{equation}
constitutes one more constant of motion.

As usual in this framework we adopt as lapse and shift functions:
\begin{equation}
N =e^{-(\beta_1+\beta_2+\beta_3)}\qquad,\qquad N^k=0\, .
\end{equation}
The equations of motion reduce to $\dot F=N\,[F,{\cal H}]$. When combined with the energy and momentum constraints, and the previous constants of motion, they lead to:
\begin{eqnarray}\label{Eqspin12}
\dot S^{[\hat 1\hat 2]}&=&+\frac 12 { p\, }e^{(\beta_2-\beta_3)}S^{[\hat 1\hat 3]}\\
\dot S^{[\hat 1\hat 3]}&=&-\frac 12 { p\, }e^{(\beta_2-\beta_3)}S^{[\hat 1\hat 2]} \label{Eqspin13}
\end{eqnarray}
from which we deduce another constant of motion:  $(S^{[\hat 1\hat 2]})^2+(S^{[\hat 1\hat 3]})^2=L^2$, in accordance with eqs (\ref{TotS2}) and (\ref{S23}).[ Note the links $s_{23}=\sigma_{23}/2$ and
$L^2 + s_{23}^2 ={\mathbb S}^2= \cS^2/4$ between these constants of the motion, and the other constants introduced above; in the following discussion of the integration of the classical equations of motion we denote some constants of integration by the same symbols as their corresponding quantum numbers introduced at the end of section \ref{sec4}.]

We also obtain :
\begin{eqnarray}
\dot \beta_2-\dot\beta_3&=&\frac 12 (\varpi^2-\varpi^3)\label{eqb23}\, ,\\
\dot \varpi^3-\dot\varpi^2&=&2\,e^{(\beta_2- \beta_3)} { p\, }(e^{(\beta_2- \beta_3)} { p\, }+{ s_{2 3} })\label{eqpb23}\, .
\end{eqnarray}
Obviously this system may be integrated by two quadratures. However it will furnish only one new constant of integration as from eqs (\ref{Hconst1}--\ref{Hconst3}) we obtain:
\begin{equation}\label{Intprem}
\frac 14 (\varpi^2-\varpi^3)^2={ p\, }{\cal C}_2+\frac 14 {\cal C}_1^2+{ s_{2 3} }^2-({ p\, }e^{(\beta_2- \beta_3)}+{ s_{2 3} })^2\, .
\end{equation}
{\sl i.e.} equivalently 
\begin{equation}
(\dot\beta_2-\dot\beta_3)^2={\cal K}^2-({ p\, }\,e^{(\beta_2-\beta_3)}+{ s_{2 3} })^2\label {db12sq}
\end{equation}
where we have put 
${\cal K}^2:={ p\, }{\cal C}_2+ {\cal C}_1^2/4 +{ s_{2 3} }^2 = \cQ_2+s_{23}^2$.
Let us notice that this equation implies that $\beta_2-\beta_3$ is bounded from above if there is effectively a symmetry wall, {\sl i.e.} unless ${ p\, }=0$. More precisely the domain of variation of $\beta_2-\beta_3$ are the following, depending on the relative values of ${\cal K} (>0)$ and $ s_{2 3}$ i.e. of $\cQ_2$ :
\begin{eqnarray}
\cQ_2>0\  ({\cal K}^2>{s_{2 3}}^2)&&\label{K2gs}\\
\text{if }p\,{s_{2 3}}>0\quad :&&\beta_2-\beta_3\in \left]-\infty,\ln [({\cal K}-\vert {s_{2 3}} \vert)/\vert p\vert ]\right]\nonumber\\
\text{if }p\,{s_{2 3}}<0\quad :&&\beta_2-\beta_3\in \left]-\infty,\ln [({\cal K}+\vert {s_{2 3}} \vert)/\vert p\vert]\right]\nonumber\\
\cQ_2<0\ ({\cal K}^2<{s_{2 3}}^2)&&\label{K2ls}\\
\text{if }p\,{s_{2 3}}<0\quad :&&\beta_2-\beta_3\in \left]\ln [(\vert {s_{2 3}} \vert - {\cal K})/\vert p\vert],\ln [(\vert {s_{2 3}} \vert + {\cal K})/\vert p\vert]\right]\nonumber
\end{eqnarray}
The situation corresponding to integration constants obeying the inequality (\ref{K2gs}) corresponds to a situation where $\beta_3$ is (almost) always\footnote{Always if  the ratio $({\cal K}\mp\vert {s_{2 3}} \vert)/\vert p\vert <1$, otherwise there is a short period of time during which  $\beta_2>\beta_3$.} greater than $\beta_2$. When ${\cal K}^2<{s_{2 3}}^2$, the difference between $\beta_2$ and $\beta_3$ could be of constant (positive or negative) sign or oscillate around zero.
Once integrated, eqs (\ref{Eqspin12}) and (\ref{Eqspin13}) furnish the spin components $ S^{[\hat 1\hat 2]}$ and $ S^{[\hat 1\hat 3]}$ after a single quadrature while the metric coefficient $\nu_{23}$ is directly obtained thanks to equations (\ref{Hconst2}) and (\ref{eqb23}) : 
\begin{equation}\label{Soln23}
\nu_{23}=\frac 1p(\dot \beta_3-\dot\beta_2)-\frac{ {\cal C}_1}{2\,p}\, .
\end{equation}

Another constant of motion is obtained from the equation
\begin{equation}
\dot \varpi^2+\dot\varpi^3=- 2\,N \,{\cal H}=0\, .
\end{equation}
We denote it by $\varpi^2+ \varpi^3=2\,{ k\, }$.

\par\noindent The remaining bosonic  equations, taking into account the previous ones, reduce to :
\begin{eqnarray}
\dot \beta_1&=&\frac 14 \varpi^1- \frac 12 { k\, }\, ,\label{Eqb1}\\
\dot \beta_2+\dot \beta_3&=&-\frac 12 \varpi^1\, ,\label{Eqb23}\\
\dot \varpi^1&=&2\,e^{-4\,\beta_1}-2\,{\cal C}_g\,e^{-2\,\beta_1}\, ,\label{Eqpb1}\\
\dot \nu_{12}&=&  S^{[\hat 1\hat 2]}e^{(\beta_1-\beta_2)}\, ,\label{Eqn12}\\
\dot \nu_{13}&=&  S^{[\hat 1\hat 3]}e^{(\beta_1-\beta_3)} + e^{(\beta_1-\beta_2)}S^{[\hat 1\hat 2]} \nu_{23}\, .\label{Eqn13} 
\end{eqnarray}
Equations (\ref{Eqb1}) and (\ref{Eqb23}) immediately provide:
\begin{equation}\label{2b1b2b3}
2\beta_1+\beta_2+\beta_3= { k\, }(t_0-t)\, .
\end{equation}
Just as equations (\ref{eqb23}) and (\ref{eqpb23}), equations (\ref{Eqb1}) and (\ref{Eqpb1}) are integrated by two quadratures, defining by the way the constant of motion ${ \cal Q }_1$:
\begin{equation}\label{dotb1sq}
\dot\beta_1^2=\frac 14\left({ \cal Q }_1+2\,{\cal C}_g\,e^{-2\,\beta_1}-e^{-4\,\beta_1}\right)\, .
\end{equation}
For $\beta_1$ to be classically defined we need ${\cal C}_g^2+{ \cal Q }_1>0$ and ${\cal C}_g>0$ or ${ \cal Q }_1>0$. The domain of variation of $e^{-2\,\beta_1}$ is 
\begin{eqnarray}
e^{-2\,\beta_1}\in \left ]0,\sqrt{{\cal C}_g^2+{\cal Q}_1}+{\cal C}_g\right ]&& \text {if }{ \cal Q }_1>0\, ,\\
e^{-2\,\beta_1} \in \left [ {\cal C}_g-\sqrt{{\cal C}_g^2+{\cal Q}_1},{\cal C}_g+\sqrt{{\cal C}_g^2+{\cal Q}_1}\right ]&& \text {if }{ \cal Q }_1<0 \text{ and  }{\cal C}_g>0\, .
\end{eqnarray}
The first case corresponds to a bounce defining asymptotic Kasner's exponents; the second one to an oscillatory behaviour of the metric component.

So we obtain the expression of the three variables $\beta_1$, $\beta_2$ and $\beta_3$.  Then the remaining three equations can   be integrated by quadratures. For completeness we display them hereafter:

\subsubsection*{The solutions:}
After elementary integrations, we obtain when ${\cal K}^2-{ s_{2 3} }^2>0$ :
\begin{eqnarray}
  \beta_3-\beta_2&=&  \ln \left\{\frac{ {\cal K}\,\vert p \vert \cosh[\sqrt{{\cal K}^2-{ s_{2 3} }^2}(t-t_-)]+p\,{ s_{2 3} }}{{\cal K}^2-{ s_{2 3} }^2}\right\}\label{b3b2}\, ,\\
  {\cal B}(t)&:=&\frac {{ p\, }}2 \int e^{(\beta_2-\beta_3)}\,dt\nonumber\\
  &=&
\text{sgn}(p)\,\arctan\left[ \sqrt{
\frac{{\cal K}-\text{sgn}(p)\,{ s_{2 3}}}{{\cal K}+\text{sgn}(p)\,{ s_{2 3}}}
}
\tanh[\sqrt{{\cal K}^2-{ s_{2 3} }^2}(t-t_-)/2] \right]\, ,\nonumber
\end{eqnarray}
and when ${\cal K}^2-{ s_{2 3} }^2<0$ and $p\,{ s_{2 3} }<0$ :
\begin{eqnarray}\label{b3b2bis}
\beta_3-\beta_2&=&  \ln \left\{\frac{ {\cal K}\,\vert p \vert \cos[\sqrt{{ s_{2 3} }^2-{\cal K}^2}(t-t_-)]+p\,{ s_{2 3} }}{{\cal K}^2-{ s_{2 3} }^2}\right\}\label{b3b2bis}\, ,\\
{\cal B}(t)& =&
\text{sgn}(p)\, \arctan \left[ \sqrt{
\frac{{ s_{2 3}}-\text{sgn}(p)\,{\cal K}}{{ s_{2 3} }+\text{sgn}(p)\,{\cal K}}
}
\tan[\sqrt{{ s_{2 3} }^2-{\cal K}^2}(t-t_-)/2]
\right]
\nonumber\, .
\end{eqnarray}
The function ${\cal B}(t)$ furnishes the time evolution of the non constant spin tensor components:
\begin{eqnarray}
S^{[\hat 1\hat 2]}&=&L\,\cos[  {\cal B}(t)]\, ,\label{S12}\\
S^{[\hat 1\hat 3]}&=&L\,\sin[  {\cal B}(t)]\, .\label{S13}
\end{eqnarray}
Let us notice that, by fixing appropriately the initial values of the spinorial mode, we also have : ${ p\, }e^{(\beta_2-\beta_3)}+{ s_{2 3} }={\cal K}\,\sin [2\,  {\cal B}(t)]$.

In the same way we obtain the expression of $\beta_1$. When ${\cal Q}_1>0$  it reads as:
\begin{equation}\label{b1}
      \beta_1=\frac 12\ln\left[\frac{\sqrt{{\cal C}_g^2+{ \cal Q }_1}\cosh[\sqrt{{ \cal Q }_1} (t-t_\ast)]-{\cal C}_g }{{\cal Q}_1} \right]\, ;
\end{equation}
when ${\cal Q}_1<0$ and thus ${\cal C}_g>\vert {\cal Q}_1\vert$, we obtain :
\begin{equation}\label{b1bis}
      \beta_1=\frac 12\ln\left[\frac{{\cal C}_g+\sqrt{{\cal C}_g^2+{ \cal Q }_1}\cos[\sqrt{{ -\cal Q }_1} (t-t_\ast)] }{{\vert {\cal  Q}_1\vert}} \right]\, .
\end{equation}
Combining these equations with eqs (\ref{2b1b2b3}) and eqs (\ref{b3b2} or \ref{b3b2bis}) we obtain the explicit expression of $\beta_2$ and $\beta_3$. The functions $\nu_{12}$ and $\nu_{13}$ are then obtained by integration of eqs (\ref{Eqn12}), \ref{Eqn13}) where the spin tensor components are expressed thanks eqs (\ref{S12}, \ref{S13}).
The various integration constants ${ p\, },\dots,{ \cal Q },L$ are not all independent. The Hamiltonian constraint imposes:
\begin{equation}\label{Q}
{ k\, }^2-{ p\, }{\cal C}_2-\frac 14{\cal C}_1^2-{ s_{2 3} }^2-L^2=  { k\, }^2-{\cal K}^2-L^2\, ,
\end{equation}
which is the classical analogs of the mass shell condition  (\ref{mass-shell1}).
Obviously there are different qualitative behaviors of the metric components, according to the relative values and signs of some integration constants. From Eq. (\ref{2b1b2b3})  and eqs (\ref{b3b2}, \ref{b1}) we immediately obtain   Kasner exponents (see next section). The function ${\cal B}(t)$, obtained from Eq. (\ref{b3b2}), varies of a finite amount, and the spin tensor components $S^{[\hat 1\hat 2]}$ and $S^{[\hat 1\hat 3]}$ rotate by a finite angle. But contrary to what happens for gravity coupled to bosonic  fields, the non-diagonal metric coefficients $\nu_{12}$ and $\nu_{13}$, which are driven by these spin tensor components blow up exponentially.

The  solutions given by eqs  (\ref{b3b2bis}) and (\ref{b1bis}) are of a different nature. Here the spin tensor components $S^{[\hat 1\hat 2]}$ and $S^{[\hat 1\hat 3]}$ oscillate. The metric component  $\exp[-2\beta_1]$ also oscillate between two positive values. The coefficients $\exp[-2\beta_2]$ and $\exp[-2\beta_3]$ are given by the product of a non-vanishing oscillatory factor and the exponential $\exp[2\,k\,t]$. Accordingly they blow up when $k\,t\rightarrow + \infty$ and collapse to a singularity when
$k\,t\rightarrow - \infty$. The evolution of the non-diagonal terms are different according to the subspace they concern. The coefficient  $\nu_{23}$ follows the derivative of $\beta_3-\beta_2$, and remains bounded, oscillating. The coefficients $\nu_{12}$ and $\nu_{13}$  grow up like an exponential, but modulated by an oscillating factor.
Notice that this last solution   absolutely requires a non vanishing spinor field.

\subsubsection*{Collision rules}
The collision rules, i.e. the  transformation law of the Kasner exponents of the metric, when they exist ({\sl i.e.} for ${\cal Q}_1>0$),  are directly read from the asymptotic behaviour of the solutions displayed in the previous section.
These solutions  allow one to discuss both collisions on a gravitational wall or on a symmetry wall. The arbitrariness  of the constants $t_\ast$ and $t_-$ reflects the independence of the order of the collision processes. 
\par\noindent The kinetic matrix is, with respect to the velocities $\{\dot \beta_1,\dot \beta_2,\dot \beta_3,\dot \nu_{12},\dot \nu_{13},\dot \nu_{23}\}$,:
\begin{equation}\left(
\begin{array}{cccccc}
0 & -1 & -1 & 0 & 0 & 0 \\
-1 & 0 & -1 & 0 & 0 & 0 \\
-1 & -1 & 0 & 0 & 0 & 0 \\
0 & 0 & 0 & \frac{1}{2} \left(e^{2 \text{$\beta_3 $}-2 \text{$\beta_1 $}} \text{$\nu_{23}
  $}^2+e^{2 \text{$\beta _2$}-2 \text{$\beta _1$}}\right) & -\frac{1}{2} e^{2
  \text{$\beta _3$}-2 \text{$\beta _1$}} \text{$\nu _{23}$} & 0 \\
0 & 0 & 0 & -\frac{1}{2} e^{2 \text{$\beta _3$}-2 \text{$\beta _1$}} \text{$\nu _{23}$}
  & \frac{1}{2} e^{2 \text{$\beta _3$}-2 \text{$\beta _1$}} & 0 \\
0 & 0 & 0 & 0 & 0 & \frac{1}{2} e^{2 \text{$\beta _3$}-2 \text{$\beta _2$}}
\end{array} 
\right)
\end{equation}
\par\noindent The Hamiltonian constraint implies that asymptotically the trajectories in the $\beta$-space are non-spacelike:
\begin{equation}
\dot {\boldsymbol \beta}\cdot\dot{\boldsymbol \beta}=\frac 14 (\varpi_1^2+\varpi_2^2+,\varpi_3^2)-\frac 18(\varpi_1+\varpi_2+\varpi_3)^2=-\frac 14 (L^2+{ s_{2 3} }^2)\, . \label{betamass} 
\end{equation}
with a mass term in agreement with \cite{BK}.
In terms of the velocities $\{\dot\beta_i\}$, the collision rules are, for a collision on a gravitational wall $\{\dot\beta_1,\dot\beta_2,\dot\beta_3\}\mapsto \{ -\dot\beta_1,\dot\beta_2 + 2\,\dot\beta_1,\dot\beta_3 + 2\,\dot\beta_1\}$
and on a symmetry wall $\{\dot\beta_1,\dot\beta_2,\dot\beta_3\}\mapsto \{ \dot\beta_1,\dot\beta_3,\dot\beta_2 \}$. In terms of momenta we obtain :
\begin{equation}\label{picoll}
\{\varpi_1,\varpi_2,\varpi_3\}\mapsto \{ -\varpi_1 + 2(\varpi_2+\varpi_3),\varpi_2,\varpi_3\}\qquad\text{and}\qquad  \{\varpi_1,\varpi_2,\varpi_3\}\mapsto \{ \varpi_1, \varpi_3,\varpi_2\}
 \end{equation}
These mappings conserve the value of the "mass'' (\ref{betamass}). As is well known, the timelike nature of the trajectories makes that, after a finite number of collisions, the $\beta$
``billiard ball''  will end up on a  worldline which does not catch up anymore the receding walls.

\subsubsection*{The Dirac sector}
It remains to solve the Dirac equation (\ref{Digeneq}), taking into account the expression of the metric and spin components obtained. It is given by:
\begin{equation}\label{BIIDieq}
\dot \chi=N(t)\{\chi,{\cal H}\}= \frac 14\left(  S^{[\hat 1\hat 2]}\, \gamma^{\hat 1\hat 2}+   S^{[\hat 1\hat 3]}\,\gamma^{\hat 1\hat 3}+   (S^{[\hat 2 \hat 3]}+e^{\beta_2-\beta_3}\,\varpi^{23})\,\gamma^{\hat 2\hat 3}-  e^{-2\,\beta_1} \gamma_5\right)\chi
\end{equation}
This classical equation for the rotation of the spinor $\chi$ is more complicated to solve than the quantum problem posed by the
quantum type II Hamiltonian (\ref{qHII}) because it represents the rotation of the body-frame object $\chi$ w.r.t. the space frame.
This rotation is the combination of several different precessions and rotations corresponding to the various terms in the equation above. 
We can simplify a bit the study of this combined rotation
by ``encoding" the last term (linked to the interaction with the gravitational wall) in a change of spinor variable.
Indeed, using the commutation of $\gamma_5$ with the spin matrices $\gamma^{\hat j\hat k}$, its  effect is taken into account by replacing, for ${\cal Q}_1 >0$, $\chi$ by the variable $\psi$, with:
\begin{equation}
\chi=\exp\left[-\frac 14\int e^{-2\,\beta_1}\,dt\,\gamma_5\right]\,\psi=\exp\left[-\frac 12 \arctan\left(\frac{\sqrt{{\cal C}_g^2 +{ \cal Q }_1} +{\cal C}_g}
{\sqrt{{ \cal Q }_1}}\tanh[\sqrt{{ \cal Q }_1}\,(t-t_*)/2]\right)\,\gamma_5\right]\psi\, ,
\end{equation}
or
\begin{equation}
\chi=\exp\left[-\frac 12 \arctan\left(\frac{\sqrt{{\cal C}_g^2 +{ \cal Q }_1}+{\cal C}_g}
{ \sqrt{{ -\cal Q }_1}}\tan ({\sqrt{{- \cal Q }_1}\,(t-t_*)/2})\right)\,\gamma_5\right]\psi
\end{equation}
when ${\cal Q}_1 <0$.

Note that, after a collision on the gravitational wall, the spinor undergoes a total rotation around the ``direction" $\gamma_5 = \gamma_0 \gamma^{123}$
(associated to the gravitational wall)  given by the matrix : 
\begin{equation}\label{ThetaG}
\Theta_G:= \exp\left[\left(  \pi /2- \frac 12 \arctan( \sqrt{{ \cal Q }_1}/{{ {\cal C}_g}})\right)\gamma_5\right]\, .
\end{equation}
In particular, in the limit where $ \sqrt { \cal Q }_1 \gg  {\cal C}_g$, we get a rotation by an angle of $\frac{\pi}{4}$
around $\gamma_5 = \gamma_0 \gamma^{123}$, in agreement with the sharp (gravitational) wall limit studied \cite{DaHi}.

Let us finally discuss the effects of the other terms in the rotational evolution of the spinor, which, in terms of the object $\psi$, are contained in 
the equation:
\begin{equation}\label{SWeq}
\dot \psi= \frac 14 \left( S^{[\hat 1 \hat 2]}\,\gamma^{\hat 1\hat 2}+S^{[\hat 1 \hat 3]}\,\gamma^{\hat 1\hat 3}+(S^{[\hat 2 \hat 3]}+e^{\beta_2-\beta_3}\,\varpi^{23})\,\gamma^{\hat 2\hat 3}\right)\psi
\end{equation}
For arbitrary values of $L$, it seems difficult to express simply the spinorial mode but assuming  $L=0$ we obtain:
\begin{equation}\label{Lzerospinor}
\psi[t]=\left(\cos[\frac 14{ s_{2 3} } \,t+\frac 12\, {\cal B}(t)] +\sin[\frac 14{ s_{2 3} } \,t+\frac 12\, {\cal B}(t)] \, \gamma^{\hat 2 \hat 3}\right)\psi_0 \, .
\end{equation} 
This result exhibits a rotation in spinor space which is the sum of a uniform, continuous precession (with angular velocity ${ s_{2 3} }/4$),
and of the rotation  ${\cal B}(t)$ (which spans a finite angle in the infinite time $ t \in (- \infty, + \infty)$).
If ${ s_{2 3} }=0$ things further simplify. [Note that the two conditions $L=0$ and  ${ s_{2 3} }=0$ cancel the rotation of the body frame w.r.t.
the space frame, and leave only the effects of the two dynamical walls: the gravitational wall $2 \beta_1$ (studied above),
and the symmetry wall $\beta_3 - \beta_2$ that we consider next.]  We obtain   ${\cal B}(t)=\arctan \{\exp[{\cal K}(t-t_-)]\}$ and after the collision the spinor will be rotated
as :
\begin{equation}
\psi[t=+\infty]= \exp\left[ \frac 14 \pi \,  \gamma^{\hat 2  \hat 3} \right]\psi[t=-\infty]:=\Theta_S\,\psi[t=-\infty]\, .
\end{equation}
in accordance with the collision law on a symmetry wall obtained in  Ref.  \cite{DaHi}.

Thus to conclude this section let us remark that the bosonic  part of the $m=0$ and $\lambda_k=0$ Einstein-Dirac system, 
under the assumption of a cosmological model of Bianchi type II, is completely integrable, in terms of elementary functions. For the fermionic part we have
obtained explicit solutions only when simplifying the problem by, essentially, restricting the continuous rotation between the space frame
and the body frame (to which the spinor is attached). However,  
as mentioned earlier it is not necessary to know precisely the spinor dynamics to elucidate the gravitational one.
As discussed in the main text, the quantum dynamics of this system is simpler
and can be solved in an exact manner.

\end{document}